\documentclass{aa}  

\usepackage{graphicx}
\usepackage{array}
\usepackage{tabularx}
\usepackage{amsmath}
\usepackage{multirow}
\usepackage{chngcntr}
\usepackage{hyperref}
\usepackage{txfonts}
\usepackage{natbib}
\bibpunct{(}{)}{;}{a}{}{,} 
\usepackage{units}
\usepackage{comment}
\usepackage{color}
\usepackage[normalem]{ulem}

\newcommand{\sinc}{\text{sinc}}

\begin{document} 
  
  \titlerunning{SPHYNX: an accurate DBSPH method for astrophysical applications}
  \title{SPHYNX: an accurate density-based SPH method\\for astrophysical applications}

   \author{R.M. Cabez\'on\inst{1,2} \and D. Garc\'ia-Senz\inst{3,4} \and J. Figueira\inst{3}}

   \institute{Departement Physik, Universit\"at Basel, Klingelbergstrasse 82, 4056 Basel, Switzerland\\
         \email{ruben.cabezon@unibas.ch}
         \and
             Scientific Computing Core, sciCORE, Universit\"at Basel, Klingelbergstrasse, 61, 4056 Basel, Switzerland
         \and
             Departament de F\'isica, Universitat Polit\`ecnica de Catalunya, EEBE, Eduard Maristany 10-14 m\`odul C2, 08019 Barcelona, Spain\\
         \email{domingo.garcia@upc.edu, joana.figueira@upc.edu}
         \and
             Institut d'Estudis Espacials de Catalunya, Gran Capit\`a 2-4, 08034 Barcelona, Spain
         }

   \date{Received December 7, 2016;accepted xxx}

 
  \abstract
   {}
   {Hydrodynamical instabilities and shocks are ubiquitous in astrophysical scenarios. Therefore, an accurate numerical simulation of these phenomena is mandatory to correctly model and understand many astrophysical events, such as Supernovas, stellar collisions, or planetary formation. In this work, we attempt to address many of the problems that a commonly used technique, smoothed particle hydrodynamics (SPH), has when dealing with subsonic hydrodynamical instabilities or shocks. To that aim we built a new SPH code named SPHYNX, that includes many of the recent advances in the SPH technique and some other new ones, which we present here.}
   {SPHYNX is of Newtonian type and grounded in the Euler-Lagrange formulation of the smoothed-particle hydrodynamics technique. Its distinctive features are: the use of an integral approach to estimating the gradients; the use of a flexible family of interpolators called $sinc$ kernels, which suppress pairing instability; and the incorporation of a new type of volume element which provides a better partition of the unity. Unlike other modern formulations, which consider volume elements linked to pressure, our volume element choice relies on density. SPHYNX is, therefore, a density-based SPH code.}
   {A novel computational hydrodynamic code oriented to Astrophysical applications is described, discussed, and validated in the following pages. The ensuing code conserves mass, linear and angular momentum, energy, entropy, and preserves kernel normalization even in strong shocks. In our proposal, the estimation of gradients is enhanced using an integral approach. Additionally, we introduce a new family of volume elements which reduce the so-called tensile instability. Both features help to suppress the damp which often prevents the growth of hydrodynamic instabilities in regular SPH codes.}
   {On the whole, SPHYNX has passed the verification tests described below. For identical particle setting and initial conditions the results were similar (or better in some particular cases) than those obtained with other  SPH schemes such as GADGET-2, PSPH or with the recent density-independent formulation (DISPH) and conservative reproducing kernel (CRKSPH) techniques.}


   \maketitle

\section{Introduction}
\label{sec:introduction}
Many interesting problems in Astrophysics involve the evolution of fluids and plasmas coupled with complex physics. For example, in core collapse Supernova, magnetohydrodynamics meets with general relativity, nuclear processes and radiation transport. Other scenarios, such as neutron star mergers, Type Ia Supernova, and planet- or star formation, face similar challenges in terms of complexity. Besides that, these phenomena often have a strong dependence on the dimensionality and they must be studied in three dimensions. This requires accurate numerical tools which translate to rather sophisticated hydrodynamic codes. Because of its adaptability to complex geometries and good conservation properties, the Smoothed Particle Hydrodynamics (SPH) method is a popular alternative to grid-based codes in the astrophysics community. SPH is a fully Lagrangian method, born forty years ago \citep{luc77,gin77}, that since then has undergone sustained development \citep{mon92,mon05,ros15,spr10,pri12}. Recent years have witnessed a large range of improvements specially aimed at reducing the numerical errors inherent to the technique. These errors are known as $E_0$ errors \citep{rea10} and mainly appear due to the conversion of the integrals, representing local-averaged magnitudes of the fluid, into finite summations. The most simple and naive way to get rid of them would be to work closer to the continuum limit, which implies working with a number of particles and neighbors as big as possible (ideally, $N\to \infty$ and $N_{nb}\to \infty$, see \citealp{zhu15}). Unfortunately, this is not feasible in common applications of the technique because the total number of particles is limited by the computing power (both by speed and storage). Moreover, the number of neighbors of a given particle cannot be arbitrarily increased without suffering pairing-instability \citep{sch81}. This is a numerical instability that acts as an attractive force which appears at scales slightly shorter than the smoothing length $h$, provoking artificial particle clumping and effectively decreasing the quality of the discretization, which eventually leads to unrealistic results.

In order to reduce the $E_0$~errors, another more practical possibility has been studied during recent years: finding interpolating functions that are less prone to particle clustering than the widely used $M_4$ or cubic-spline kernel \citep{mon85}. Among the various candidates, the most used (pairing-resistant) kernels come either from an extension of the $M_n$ family to higher-order polynomials \citep{sch46} or those based on the Wendland functions \citep{wendland1995}. In particular, the Wendland family is specially suited to cope with pairing instability \citep{deh12}. Another possibility is the $sinc$ family of kernels \citep{cabezon2008}, which are functions of type $S(x)=C(n)(\sin x/x)^n$, and add the capability of dynamically modifying their shape simply by changing the exponent $n$. That adaptability of the $sinc$ kernels makes the SPH technique even more flexible and can be used, in particular, to prevent particle clustering, as shown in Sect.~\ref{Sec.sinc}.

Historically, the growth of subsonic hydrodynamical instabilities has been problematic for SPH simulations, as they damp them significantly. The Rayleigh-Taylor (RT) instability is a ubiquitous phenomenon which serves as a paradigmatic example. It appears wherever there is cold and dense fluid on top of a hot and diluted one in the presence of gravity (or any inertial force, in virtue of the principle of equivalence). The entropy inversion leads to the rapid overturn of the fluid layers. In the real world, the overturn is triggered by small perturbations at the separation layer between the light and dense fluids. The RT instability is one of the most important agents driving the thermonuclear explosion of a white dwarf, which gives rise to the Type Ia supernova (SNIa) explosions. Their correct numerical description is also crucial to understanding the structure of the supernova remnants (SNR) and to model core collapse supernova (CCSN) explosions. Additionally, the RT instability is also the source of other interesting phenomena such as the KH instability or turbulence. The numerical simulation of the Rayleigh-Taylor instability using SPH has traditionally been a drawback for the technique, especially for low-amplitude initial perturbations in the presence of a weak gravitational force. At present, for a similar level of resolution, the best SPH codes cannot yet compete with the state-of-art grid-based methods. For example, the finite-volume/difference Godunov methods such as ATHENA and PLUTO, AMR codes such as FLASH, the Meshless Finite Mass (MFM) and Volume methods (MFV), and especially the moving mesh methods based on Voronoi tessellations, as in the code AREPO, provide a good approach to the RT instability. This problem is partially overcome using a large number of neighbors and by adding an artificial heat diffusion term to the energy equation, as in the PSPH proposal by \cite{sai13} and \cite{hop13}. However, these problems still persist when either the size of the initial perturbation or the gravity value are reduced \citep{val12}, meaning that they are a symptom of another source of numerical error in SPH named {\it tensile instability}. This is an artificial surface tension that appears at contact discontinuities because of an insufficient smoothness of pressure between both sides of the discontinuity \citep{mon00}. As a consequence, the integration of the momentum equation gives incorrect results. An excess of that tension provokes the damping of fluid instabilities, especially those with short wavelengths. Several techniques have been proposed to treat this problem, like averaging the pressure by means of the interpolating kernel itself, scheme PSPH \citep{hop15}, volume element estimation bounded to pressure, the density independent scheme (DISPH) \citep{hop13,sai13,sai16}, or by adding an artificial diffusion of heat to the energy equation, which helps to balance the pressure across the discontinuity \citep{pri07}. They have paved the road that led the SPH technique to a new standard within the last few years, and have helped to overcome this long lasting inconvenience. In particular, it has been proved that it is fundamental to increase the accuracy of the gradient estimation across the contact discontinuities via reducing its numerical noise. To achieve that, \cite{garciasenz2012} used an integral scheme to calculate spatial derivatives, which proved to be especially efficient at handling fluid instabilities. The validity of that approach was assessed in subsequent works by \cite{cabezon2012} and \cite{ros15}. In this last case, the integral approach to the derivatives (IAD) was used to extend the SPH scheme to the special-relativistic regime (see also \cite{ros15b}). See also the recent work of \cite{val16}, where the efficiency of the IAD scheme to reduce the $E_0$~errors is studied in detail.

Finally, a recent breakthrough in SPH was the emergence of the concept of generalized volume elements \citep{rit01,sai13,hop13}. In these works, it was shown that a clever choice of the volume element (VE) can reduce the tensile-instability, leading to a better description of hydrodynamic instabilities. In this paper, we present a novel estimator to the VE that preserves the normalization of the kernel. We show in this work that having a good partition of the unity is also connected to the tensile-instability problem. The calculations of the growth of the Kelvin-Helmholtz and Rayleigh-Taylor instabilities using these new VE are encouraging.

In this work, we also introduce the hydrodynamics code SPHYNX, that gathers together the latest advances in the SPH technique, including those new ones presented here. SPHYNX has already been used in production runs simulating type Ia and core collapse supernova and is publicly accessible\footnote{\url{astro.physik.unibas.ch/sphynx}}.

The organization of this paper is as follows. In Section~\ref{sec:generalities} we review the main properties of the $sinc$~kernels as well as the integral approach to the derivatives, which are at the heart of SPHYNX. Section~\ref{sec:preconditioning} is devoted to the choice of the optimal volume element and to the update of the smoothing-length $h$ and of the kernel index $n$. Section~\ref{sec:sphynx} describes the structure of the hydrodynamics code SPHYNX: moment and energy equations and included physics. Sections~\ref{sec:2Dtests} and \ref{sec:3Dtests} are devoted to describing and analyzing a variety of tests carried out in two-dimensions and three-dimensions, respectively. Finally, we present our main conclusions and prospects for the future in Section~\ref{sec:conclusions}.

\section{Interpolating kernels and gradients evaluation}
\label{sec:generalities}
\subsection{The Sinc kernels}
\label{Sec.sinc}
Interpolating kernels in SPH must fulfill several basic requirements in order to be considered suitable interpolators. Some of these features are automatically satisfied if the functional form of the kernel is spherically symmetric and approaches the Dirac$-\delta$ in the continuum limit. A functional representation of the Dirac-$\delta$ is \citep{cou97}

\begin{equation}
\delta = \lim_{\epsilon\to 0}\frac{1}{\pi x}\sin\left(\frac{x}{\epsilon}\right)
\label{deltasinc}
.\end{equation}

\noindent
Writing $\epsilon=\frac{2h}{\pi}$ the function above becomes

\begin{equation}
W_1^S = \frac{1}{2h}\left[\frac{\sin\left(\frac{\pi}{2}\frac{x}{h}\right)}{\frac{\pi}{2}\frac{x}{h}}\right]= \frac{1}{2h}~ \sinc\left(\frac{\pi}{2}\frac{x}{h}\right)
\label{kernelsinc1}
,\end{equation}

\noindent
where the magnitude $\sinc(\xi)=\frac{\sin(\xi)}{\xi}$ is a widely known function used in signal analysis. Nevertheless, the expression given by Eq.~(\ref{kernelsinc1}) does not have compact support, limiting its practical applications as SPH interpolation kernel. Moreover, for $x/h> 2$, the $sinc$ function oscillates and eventually produces negative values. An obvious solution to this problem is to restrict the domain to $\vert x/h\vert\le 2$, but in that case the first derivative of this function does not go to zero at the limits of the interval. To fix that, we define the $W^S_n$ kernel set as:

\begin{equation}
W^S_n(q,h,n) = \frac{B_n}{h^d}~S_n\left(\frac{\pi}{2}q\right)\qquad 0\le q\le 2\,,
\label{kernelsincn}
\end{equation}

\noindent where $S_n(\cdot)= \sinc^n(\cdot)$, $q= |\mathbf{r}|/h$, and $d$ is the spatial dimension. $B_n$ is a normalization constant, whereas $n$ is a real number with $n\geq 2$ to guarantee the nullity of the first derivative at $q=2$.
We generically refer to the $W^S_n$ set as $sinc$ kernels; see \cite{cabezon2008} (where they were called $harmonic$ kernels) and \cite{garciasenz2014}. These interpolators fulfill the regular desirable characteristics for SPH interpolators. Namely, tending to a $\delta$-function in the continuum limit, having compact support, and being spherically symmetric. But they have a number of interesting additional features: a) they are connected by definition to the $\delta$-function; b) they add flexibility to the SPH calculations as the kernel profile can be changed by varying the kernel index $n$; c) working with a high index $(n\ge 5)$ not only ensures that high-order derivatives are well behaved, but also overcomes particle clustering (see Sect.~\ref{pairing}); and d) they tend to fulfill separability when the index $n$ increases (see App.~\ref{App:separability}). 

Unfortunately, there is not a general analytic expression for the normalization constant $B_n$, but this small problem can be circumvented by fitting $B_n$, numerically calculated for a series of values of $n$. A fitting formula for $B_n$, valid in the range $3\le n\le 12$, was provided in \cite{garciasenz2014}. We reproduce it here for the sake of completeness, 

\begin{equation}
B_n=
\begin{cases}
b_0+b_1 n^{1/2}+b_2 n+b_3 n^{-1/2} & \mathrm{1D}\\
b_0+b_1 n+b_2 n^{-1}+b_3 n^{-2} & \mathrm{2D}\\
b_0+b_1 n^{1/2}+b_2 n+b_3 n^{3/2} & \mathrm{3D}
\end{cases}
\label{normalization}
,\end{equation}

\noindent where the values of coefficients $b_0, b_1, b_2, b_3$ as a function of the dimensionality are provided in Table~\ref{table1}. 

\begin{table*}
\centering
\begin{tabular}{@{}crrrrr@{}}
\hline
Dimensions& \multicolumn{1}{c}{$b_0$} & \multicolumn{1}{c}{$b_1$} & \multicolumn{1}{c}{$b_2$} & \multicolumn{1}{c}{$b_3$}  \\
\hline
\hline
1D& $-1.5404568~10^{-2}$ & $3.6632876~10^{-1}$ & $-4.6519576~10^{-4}$ &$7.3658324~10^{-2}$ \\
2D& $5.2245027~10^{-2}$  & $1.3090245~10^{-1}$ & $1.9358485~10^{-2}$  &$-6.1642906~10^{-3}$ \\
3D& $2.7012593~10^{-2}$  & $2.0410827~10^{-2}$ & $3.7451957~10^{-3}$  &$4.7013839~10^{-2}$ \\
\hline
\end{tabular}
\caption{Coefficients for calculating the normalization constant $B_n$ in Eq.~(\ref{normalization}).}
\label{table1}
\end{table*}

In the limit of large $n$, the $sinc$~kernels also display an interesting feature: separability, which has not been sufficiently emphasized in the extant literature concerning SPH. Standard kernels used in SPH are spherically symmetric functions which naturally lead to a second-order accurate scheme. Interestingly, the majority of these kernels do not fulfill the identity:

\begin{equation}
W_{3d}({\vert\mathbf{r}\vert}, h) = W_{1d}(x,h)\cdot W_{1d}(y,h)\cdot W_{1d}(z,h)\,.
\label{separability}
\end{equation}

This property guarantees the consistency of simulations involving planar symmetry, which should render identical results when calculated in 1D, 2D, or 3D, if the resolution and the interpolating kernel are the same in all three cases. Exploring in detail the applications of this property is beyond the scope of this paper. Nevertheless, we added a short discussion in Appendix~\ref{App:separability}.

Additionally, some particular values of the kernel index $n$ can mimic the profile of the most notable kernels used in SPH. In Table~\ref{table2} we show the correspondence between the $sinc$ and both, the $M_n$ and $C_m$ families of interpolators. The value of $n$ shown in Table~\ref{table2} was obtained by minimizing the metric distance $d=\sum_{k=1}^N \vert W1_k-W2_k\vert$, where $W1$ and $W2$ are the interpolators to be compared\footnote{This is not the only way to compare the kernels as one could, for instance, compare the logarithms of the functions. Alternatively, the minimization of the metric distance $d$ between the first derivative of the kernels can be also of interest. All these procedures lead to only slight variations of the kernel exponents with respect to those given in Table~\ref{table2}.}. It is remarkable that $n\simeq 5$ is able to approach both the quintic $M_6$ B-spline and the $C_4$ Wendland kernel. Thus, the choice $n=5$ seems a suitable default value when the number of neighbors is moderate, $n_b\simeq 60-120$ in 3D (see f.e. Figs.~4 and 5 of \citealt{ros15}). For $n_b\ge 120$ it may be advisable to raise the index of the $sinc$~kernel to avoid the pairing instability. Interestingly, the choice $n=6.315$ provides a very similar profile to that of the Wendland $C_6$~kernel. Nevertheless, a similarity in the real space is not the only characteristic to take into account. As proved by \cite{deh12}, having a positive Fourier transform for a large range of modes is of utmost importance when dealing with pairing instability (see Sect.~\ref{pairing}).

It is also worth noting that the $W_n^S, n \in \Re (+)$ family forms a continuous set, making it possible to locally change the index $n$ during the runtime of the simulation so that one can, for example, raise the value of $n$ in presence of sharp density gradients, (see Sect.~\ref{updating n}).

Finally, the evaluation of a $\sinc$ function is indeed computationally more costly than the regular spline kernels. Nevertheless, it is very easy to circumvent this problem by interpolating the value of the $\sinc$ function from a relatively small pre-calculated table. In practical applications, though, the choice of the kernel has a negligible impact on the computational burden, being completely masked by efficient cache usage and other sections of the code, as neighbor search or gravity calculation.

Although SPHYNX uses the $sinc$~kernels by default, it also incorporates the Wendland $C_2, C_4$ and $C_6$~kernels which can easily be selected by the user whenever necessary. On another note, the impact of the $C_m$~kernels has recently been studied by other authors \citep{deh12,ros15} so we focus here almost exclusively on the $W_n^S$ set. Nevertheless, a comparison between the performance of the $C_6$ and $sinc$ interpolators is provided in Section~\ref{gresho}, where the evolution of the Gresho-Chan vortex is simulated and discussed.

\begin{table*}
\centering
\begin{tabular}{@{}crrrrrrr@{}}
\hline
Kernel& \multicolumn{1}{c}{$M_4$} & \multicolumn{1}{c}{$M_5$} & \multicolumn{1}{c}{$M_6$} & \multicolumn{1}{c}{$C_2$}&\multicolumn{1}{c}{$C_4$}&\multicolumn{1}{c}{$C_6$}  \\
\hline
\hline
n & $3.002$ & $3.934$ & $4.85$ &$3.67$&$4.98$&$6.315$ \\
\hline
\end{tabular}
\caption{Values of the exponent $n$ of the $sinc$~kernels that provide the best match (i.e., minimum metric distance $d=\sum_{k=1}^N \vert W1_k-W2_k\vert$~) to the profile of known interpolators, such as members of the B-splines, $M_n$, and Wendland, $C_m$ families.}
\label{table2}
\end{table*}

\subsubsection{Suppressing pairing instability}
\label{pairing}

Convergence to the continuum limit is only achieved when $h\to 0$. According to \cite{zhu15}, it can only be reached using {\sl both} a large number of particles $N$, and a large number of neighbors $n_b$. The value of $N$ chiefly depends on the available computer resources and on algorithmic details. However, working with a large number of neighbors is not only expensive but it may be totally impractical due to the tendency of the mass-particles to gather in clusters when $n_b$ is high, a phenomenon known as \textit{pairing instability}. A leap forward to alleviate this problem was the proposal of using Wendland functions \citep{wendland1995} as interpolators by \cite{deh12}, which are much less sensitive to this instability than the low-order members of the $M_n$ B-spline family. Another possible option to avoid particle clustering is to choose a $sinc$~kernel with a large enough exponent, \cite{garciasenz2014}. 

As mentioned above, having a Fourier transform with positive modes is necessary for a kernel to be stable. In Fig.~(\ref{figure1.1}) we show the Fourier transform of several B-spline, Wendland, and $sinc$ kernels. As expected, Wendland kernels show positive Fourier transform at very large modes. Nevertheless, it is clear that the $sinc$ family systematically becomes negative at higher modes and their negative regions decrease in size very quickly as the exponent $n$ increases. Therefore, particle clustering can be totally suppressed, in practice, simply by raising $n$. It is also worth noting that, in order to improve the performance of Wendland kernels, it is advisable to considerably increase the number of neighbors up to $n_b\simeq 400$ in 3D \citep{val16} which has a substantial increase on the computational burden of actual simulations\footnote{In fact, much larger than computing high-order kernels like the $sinc$-family with large exponents ($n\sim 7-10$).}.

We tested the impact of the kernel and the number of neighbors on the emergence of pairing instability in a dynamical simulation. Results can be found in Section~\ref{pairingtest}.

\subsection{The integral approach to derivatives}
\label{IAD}

In SPH, gradients are usually calculated simply by applying the nabla operator to the kernel function \citep{gin77}. Nevertheless, the same procedure is not applied to calculating higher-order derivatives, such as the Laplace operator, needed, for example, to evaluate the heat transport contribution to the energy equation. For these cases, an integral approach is preferred \citep{brookshaw1985} because it is less noisy and provides excellent results. Working in that direction, an integral approach to calculating first derivatives, called \emph{IAD}, was recently introduced by \cite{garciasenz2012}. The \emph{IAD} approach provides a more accurate estimation of the first derivative, thus improving the overall quality of the simulations. This was demonstrated through the study of several test cases in 1D, 2D, and 3D by \cite{cabezon2012}, \cite{ros15}, and \cite{val16}, hence we refer the reader to these papers for technical details of the method. In \cite{garciasenz2012} it was shown that a restriction of the full integral approach (called \emph{IAD}$_0$) leads to a conservative formulation of the SPH equations  in the same way as the standard scheme{\footnote{For this work, we consider the term "standard codes" defined as those SPH codes fulfilling that: a) they estimate gradients using the analytical derivative of the kernel; and b) the VE is linked to the local density value~$\frac{m}{\rho}$.} does. This, however is not without cost. Some accuracy is lost in the gradient calculation, which is exact for linear functions when \emph{IAD} is used. In his work, \cite{ros15} evaluated this accuracy loss in several tests, and the outcome was that even if \emph{IAD}$_0$ was used, the accuracy loss, in comparison with \emph{IAD}, was much smaller than the accuracy gain when comparing with standard gradient estimators. Furthermore, \cite{val16} presents a series of tests that show the effectiveness of the \emph{IAD}$_0$ method at removing sampling errors in sub-sonic flows. Additionally, the SPH equations linked to the \emph{IAD}$_0$ scheme are formally similar to those of the standard method provided that the kernel gradient is computed as follows (for the sake of clarity, from now on we drop the dependence of the kernel on the particle inter-distance, $\mathbf{r}_b-\mathbf{r}_a$),

\begin{equation}
\frac{\partial W_{ab}(h_a,n_a)}{\partial x_{i,a}}= {\mathcal A}_{i,ab}(h_a,n_a)\,; i=1,d\,,
\label{gradk}
\end{equation}
                 
\noindent
where, 

\begin{align}
{\mathcal A}_{i,ab}(h_a,n_a) &=\sum_{j=1}^{d} c_{ij,a}(h_a) (x_{j,b}-x_{j,a}) W_{ab}(h_a,n_a)\,,\label{lagr7}
\end{align}

\noindent $c_{ij,a}$ being the coefficients of the inverse of matrix $\mathcal T$ for particle $a$, whose elements are, 

\begin{equation}
\tau_{ij,a}=\sum_b \frac{m_b}{\rho_b}(x_{i,b}-x_{i,a})(x_{j,b}-x_{j,a})W_{ab}(h_a,n_a)\,; i,j=1,d
\label{tauijsph2}
.\end{equation}

In \cite{garciasenz2012}, it is shown that for the Gaussian kernel the \emph{IAD}$_0$ scheme is equivalent to the gradient estimation with the analytical derivative of the kernel function, this being a particular case of the integral approach. It was also proven that \emph{IAD}$_0$, in fact, exactly reproduces the gradient of linear functions provided that,

\begin{equation}
\langle\Delta\mathbf{r}\rangle_a=\sum_b\frac{m_b}{\rho_b} (\mathbf{r}_b-\mathbf{r}_a) W(h_a,n_a)=0\,.
\label{approxI}
\end{equation}

Equation~(\ref{approxI}) is, in general, fulfilled only approximately, and this is the main difference between \emph{IAD} and \emph{IAD}$_0$ accuracies. The best gradient estimation is therefore attained by \emph{IAD}$_0$ only when $\langle\Delta\mathbf{r}\rangle\simeq 0$. In this respect, the initial model should be built so that Eq.~(\ref{approxI}) is approached as much as possible. As we see in the following section, it turns out that getting a small $\langle\Delta\mathbf{r}\rangle$ is closely related to the choice of the adequate volume element for the integration. An analytical proof showing that an improvement in the partition of the unit translates to a better estimation of gradients within the $IAD_0$~paradigm is provided in the Appendix~\ref{App:demo}.

A collateral, albeit positive, feature of \emph{IAD}$_0$ is that it hinders the emergence of the pairing stability. This is shown in Section~\ref{pairingtest}.
     
\section{Preconditioning the calculation}
\label{sec:preconditioning}
\subsection{The choice of the volume element}
\label{choice_volel}

A recent breakthrough in SPH has been the development of a general method to assign the volume element to a particle \citep{sai13,hop13,hop15,ros15}. In SPH, the volume element of particle $a$ has traditionally been $V_a=m_a/\rho_a$ but, as noted in the seminal proposal by \cite{rit01}, other options could be more convenient for handling specific problems. SPHYNX makes use of the scheme developed by \cite{hop13}, where an estimator $X_a$ is introduced so that the particle volume is

\begin{equation}
V_a = \frac{X_a}{\sum_b X_b W_{ab}} 
\label{estimatorX}
.\end{equation}

The density of the particle is then calculated as $\rho_a=m_a/V_a$. Current choices for the estimator are $X_a=1$, $m_a$, and $P_a^k$ $(k\le 1)$, where $P_a$ is the pressure. Taking $X_a=m_a$ leads to the common choice $V_a=m_a/\rho_a$. Actually, the same volume element is obtained with $X_a=1$ when the mass of the particles is the same. On the other hand, the choice $X_a = P_a^k$ helps to suppress the surface tension that appears at contact discontinuities.

Here we want to discuss another option for $X_a$, not considered in the aforementioned papers, which could also be of value for SPH users. We choose

\begin{equation}
X_a=(m_a/\rho_a)^p\qquad p\le 1
\label{estimator}
,\end{equation}

\noindent and put it into Eq.(\ref{estimatorX}):

\begin{equation}
V_a = \frac{(\frac{m_a}{\rho_a})^p}{\sum_b (\frac{m_b}{\rho_b})^p W_{ab}}\,, 
\label{estimatorXrho}
\end{equation}

\noindent which for $p=1$ becomes,

\begin{equation}
V_a = \frac{\frac{m_a}{\rho_a}}{\sum_b \frac{m_b}{\rho_b} W_{ab}} 
\label{estimatorXrho1}
,\end{equation}

\noindent
where the summation underneath is simply the kernel normalization condition for the standard volume choice $m_a/\rho_a$. Therefore, the volume element $V_a$ given by Eq.~(\ref{estimatorXrho1}) comes after re-normalizing $m_a/\rho_a$. We note that if $\sum_b (\frac{m_b}{\rho_b}) W_{ab}=1$, then Eq.~(\ref{estimatorXrho1}) leads to the identity $\rho_a=\rho_a$, as expected. Furthermore, one can take Eq.~(\ref{estimatorXrho}) as a recursive equation, which, jointly with $\rho_a=m_a/V_a$, allows to find the optimal $\rho_a$ leading to an almost perfect kernel normalization after several iterations. That is, starting from an initial guess of the density, for example $\rho_a=\sum_b m_b W_{ab}$, the value of $V_a$ is computed in an explicit way using Eq.(\ref{estimatorXrho}). This gives a new density $\rho_a=m_a/V_a$ to be used in the next integration step, and so on. The consistency and robustness of this procedure has been checked with the tests described in Sections~\ref{sec:2Dtests}, \ref{sec:3Dtests}, and in Appendix \ref{App:convergence}.

The impact of changing the volume element is highlighted in the hydrostatic square test described by \cite{sai13} and that we reproduce in Sect.~\ref{squaretest}.

\subsection{Choosing smoothing length and kernel exponent: equalization}
\label{updating n}
The use of the $sinc$ kernels allows us to work with continuum adaptive index $n(\mathbf{r}, t)$. Interestingly, the smoothing length $h(\mathbf{r}, t)$ and kernel index $n(\mathbf{r}, t)$ can be calculated jointly with the density estimation. Because of the large number of involved variables, $\rho_a$, $h_a$, $n_a$, $V_a$, and $\Omega_a$, this point deserves some discussion.

Firstly, the volume estimator $X_a$ is updated only when the global iteration has finished. Therefore, $X_a$ is handled explicitly so that the coupling with other variables during the current iteration is avoided. Secondly, self-consistent values of $h_a$, $n_a$, $\rho_a$, and $\Omega_a$ are simultaneously calculated with a Newton-Raphson iterative scheme, that we named $equalization$~(because it equalizes the resolution at the post-shock tail of shock waves) and is described in \cite{garciasenz2014}. We summarize it here for the sake of completeness:

\begin{enumerate}
 \item Choose a trial value of $h_a$, as well as the baseline kernel index $n=n_0$, at the beginning of the integration step. 
 \item Calculate the density of each particle and its logarithm average over neighbors: $\ln\bar\rho_a=\frac{1}{\mathcal N_a}\sum_{b=1}^{\mathcal N_a} \ln\rho_b$.
 \item Evaluate $\lambda_a$, the ratio between $\rho_a$ and $\bar\rho_a$, which is taken as a local indicator of linearity and is quantified with the following:
\end{enumerate}

\begin{equation}
\lambda_a=
\begin{cases}
\left(\frac{\bar\rho_a}{\rho_a}\right) & \text{for~~$\bar\rho_a\geq\rho_a$}\\
\left(\frac{\rho_a}{\bar\rho_a}\right) & \text{for~~$\bar\rho_a <\rho_a$} 
\end{cases}
\label{estimator1}
.\end{equation}

\begin{enumerate}
 \setcounter{enumi}{3}
 \item Use $\lambda_a$ to assign a new kernel index $n_a$ according to:
\end{enumerate}

\begin{equation}
n_a=n_{0}+\Delta n\cdot f(\xi_a),\qquad\mathrm{with}\qquad \xi_a=\frac{(\lambda_a-1)}{\lambda_c}~\geq 0\,,
\label{indices}
\end{equation}
  
\noindent where $\Delta n$ is the maximum allowed jump from $n_0$, $\lambda_c\simeq 1$ is a scaling parameter, and $f(\xi)$:

\begin{equation}
f(\xi)=1-\frac{2}{\exp(\xi)+\exp(-\xi)}
\label{ffunction}
,\end{equation}

\begin{enumerate}
 \setcounter{enumi}{4}
 \item Solve Eq.~(\ref{indices}) jointly with mass conservation:
\end{enumerate}

\begin{equation}
\rho_a=\frac{m_a}{X_a}\sum_b X_b~W_{ab}(h_a,n_a)= \frac{C_a}{h_a^d}
\label{NRg1}
,\end{equation}

\noindent where $C_a= \rho_{a,0} h_{a,0}^d$ is a constant set at the beginning of the simulation jointly with $n_0$, $\Delta n$, and $\lambda_c$. After this process, we obtain the values of $h_a$, $n_a$, $\rho_a$, and $\Omega_a$. Obviously, setting $\Delta n=0$ turns the equalization off and the evolution is computed with $n_a=n_0$. Typical values for the simulations presented in the following sections are $n_0=5$, $0\le\Delta n\le 5$, and $\lambda_c=0.5$.

\section{The hydrodynamics code SPHYNX}
\label{sec:sphynx}
SPHYNX gathers all the advances on the SPH technique that have been presented in Sections~\ref{sec:generalities} and \ref{sec:preconditioning} and represents our flagship code for astrophysical applications. In the following, we present the most relevant details on its implementation and itemize the mathematical expressions of the SPH equations implemented in the Cartesian version of the code. The precise form of the equations as well as the notation follows the discussion made in the preceding Sections. They are similar to those presented in \cite{ros15b}.
\subsection{Work-flow and formulation}
\label{sec.formulation}
First of all, we specify the initial number of neighbors $n_b$ and kernel index variables: $n_0$ (baseline value), $\Delta n$ (maximum allowed index jump), and $\lambda_c$ (scaling parameter). This is followed by the choice of the volume estimator $X_a$. Then, the preconditioning stage starts with the self-consistent calculation of the volume element $V_a$, smoothing-length $h_a$, kernel index $n_a$, density $\rho_a$, and grad-h/n term $\Omega_a$ (see Sect.~\ref{updating n}).
\newline

\begin{itemize}

 \item[$\bullet$] {\bf Volume element}

\begin{equation}
V_a=\frac{X_a}{k_{a}}
\label{volEq}
,\end{equation}

with, 

\begin{equation}
\kappa_{a}=\sum_b X_b~W_{ab}(h_a,n_a)
\label{kappaEq}
.\end{equation}

For the volume elements we use $X_a=(m_a/\rho_a)^p$. From the experiments presented below, we saw that taking $p\simeq 1$ leads to the best results in most cases, but it has the disadvantage of overshooting density interpolations in contact discontinuities when $p > 0.7$. We overcome this problem by using a SPH-averaged value of the VE estimator, making it robust and allowing us to safely rise the exponent to $p=1$ (see Sect.~\ref{squaretest} for more details.)\\\\

\item[$\bullet$] {\bf Density equation}

\begin{equation}
\rho_a=\frac{m_a}{V_a}
\label{densityEq}
,\end{equation}

\noindent which we calculate jointly with $\Omega_a$ including both contributions: the \textit{grad-h} and \textit{grad-n} terms \citep{garciasenz2014},

\begin{equation}
\begin{aligned}
\Omega_a= 1-&\left[\left(\sum_b m_b\frac{\partial W_{ab}(h_a,n_a)}{\partial h_a}\right)~\frac{\partial h_a}{\partial\rho_a}\right.\\
+&~\left.\left(\sum_b m_b\frac{\partial W_{ab}(h_a,n_a)}{\partial n_a}\right)~\frac{\partial n_a}{\partial\rho_b}\right]\,.
\label{omega}
\end{aligned}
\end{equation}

Once the preconditioning is completed, the hydrodynamic equations are evaluated.\\

\item[$\bullet$] {\bf Momentum equation}

\begin{equation}
\begin{split}
\ddot x_{i,a}=-\frac{X_a}{m_a}\sum_{b=1}^{n_b}X_b\left(\frac{P_a}{\Omega_a\kappa_a^2}\mathcal A_{i,ab}(h_a,n_a)+\frac{P_b}{\Omega_b\kappa_b^2}\mathcal A_{i,ab}(h_b,n_b)\right)+
a_{i,a}^{AV}\,,
\label{momentumEq}
\end{split}
\end{equation}

\item[$\bullet$] {\bf Energy equation}

\begin{equation}
\begin{split}
\left(\frac{du}{dt}\right)_a=\frac{P_aX_a}{m_a\Omega_a~\kappa_a^2}\sum_{b=1}^{n_b}\sum_{i=1}^d (v_{i,a}-v_{i,b})\left(X_b~\mathcal{A}_{i,ab}(h_a,n_a)\right)+
\left(\frac{du}{dt}\right)_a^{AV}\,,
\label{EnergyEq}
\end{split}
\end{equation}

\end{itemize}

\noindent where $\mathcal{A}_{i,ab}$ is given by Eq.~(\ref{lagr7}), but with the $\tau_{ij,a}$ coefficients computed using the generalized volume elements, Eq.~(\ref{volEq}),

\begin{equation}
\tau_{ij,a}=\sum_b \frac{X_b}{\kappa_b}(x_{i,b}-x_{i,a})(x_{j,b}-x_{j,a})W_{ab}(h_a,n_a)\,; i,j=1,d\,.
\label{tauijsph1}
\end{equation}

The terms with superscript AV refer to the artificial viscosity acceleration and energy contributions. As detailed in the following subsection, we have slightly changed the implementation of these terms in order to make them fully compatible with the generalized volume elements.

\subsection{The artificial viscosity}
\label{AV}
Regarding the inclusion of physics, SPHYNX incorporates, by default, an artificial viscosity algorithm to handle shocks, as well as routines for the calculation of the equation of state (EOS) and  gravitational force. Heat transport by conductive and diffusive means is also included as a basic code unit. More specific routines dealing with nuclear or chemical reactions, ionization or more complex transport schemes can be modularly added to the code (see, e.g., the application of SPHYNX to simulate core collapse Supernova, including a spectral neutrino treatment, in \cite{perego16}). In the following, we explain with some detail the AV algorithm.
    
First of all, we take the viscosity $\Pi_{ab}$ as in \cite{mon97}:

\begin{equation}
\Pi_{ab}=
\begin{cases}
-\frac{\alpha}{2}\frac{v_{ab}^{sig}~w_{ab}}{\rho_{ab}} & \text{for~~$\mathbf{r}_{ab}\cdot\mathbf{v}_{ab} < 0$}\\
 0 & \text{otherwise}
\end{cases}
\label{avis}
,\end{equation}

\noindent where $v_{ab}^{sig}=c_a+c_b-3w_{ab}$ is an estimate of the signal velocity between particles $a,b$ and $w_{ab}= \mathbf{v}_{ab}\cdot\mathbf{r}_{ab}/\vert\mathbf{r}_{ab}\vert$. It is common to take $\rho_{ab}^{-1}=2~(\rho_a+\rho_b)^{-1}$ but in that case the volume element is implicitly assumed to be $m_a/\rho_{ab}$. Thus, the VE of particle $a$ is not an unequivocally defined magnitude because it also depends on the density of the neighbor particle $b$. A more compatible option between AV and VE, which can be generalized to other VE, is obtained using $\rho_{ab}^{-1}=0.5~(\rho_a^{-1}+\rho_b^{-1})$, which leads to the following viscous acceleration, 

\begin{equation}
a_{i,a}^{AV}= -\frac{1}{2}\sum_b m_b~\Pi'_{ab}\left\{\frac{f_a}{\rho_a}A_{i,ab}(h_a,n_a)+ \frac{f_b}{\rho_b}~A_{i,ab}(h_b,n_b)\right\} 
\label{accvis_1}
,\end{equation}

\noindent where 

\begin{equation}
\Pi'_{ab}=
\begin{cases}
-\frac{\alpha}{2}~v_{ab}^{sig}~w_{ab} & \text{for~~$\mathbf{r}_{ab}\cdot\mathbf{v}_{ab} < 0$} \\
 0 & \text{otherwise}
\end{cases}
\label{avis_2}
,\end{equation}

\noindent and $f_a, f_b$ are the Balsara limiters \citep{bal95}:

\begin{equation}
f_a=\frac{\vert\nabla\cdot\mathbf{v}\vert}{\vert\nabla\cdot\mathbf{v}\vert+\vert\nabla\times\mathbf{v}\vert+10^{-4}~c_a/h_a}
\label{balsara}
.\end{equation}

Expression (\ref{accvis_1}) can be written as  

\begin{equation}
\begin{aligned}
a_{i,a}^{AV}= -\frac{1}{2m_a}\sum_b &\left\{V_a~m_b~\Pi'_{ab}~f_a~A_{i,ab}(h_a,n_a)\right.\\ 
+&~\left.V_b~m_a~\Pi'_{ab}~f_b~A_{i,ab}(h_b,n_b)\right\}\,,
\label{accvis_2}
\end{aligned}
\end{equation} 
 
\noindent with $V_a=m_a/\rho_a$, $V_b=m_b/\rho_b$. Unlike $\Pi_{ab}$, the magnitude $\Pi'_{ab}$ is not divided by the density and, as a consequence, the viscous force, $m_a~a_a^{vis}$, is symmetric with respect to any pair of particles, $a,b$. We note that the volume elements are now unequivocally defined when the mass of the particles is the same. The Balsara limiters work more efficiently if they are averaged as, 

\begin{equation}
f_{ab}=2\frac{f_a~f_b}{f_a+f_b}
\label{fab}
,\end{equation}

\noindent
because it gives a lower AV than the arithmetic average in regions with strong shear flows. By default SPHYNX sets these limiters to:

\begin{equation}
f_a=f_b= max(0.05, f_{ab})
\label{fab2}
,\end{equation}

\noindent
which retains a residual viscosity to damp the smallest numerical fluctuations. Equation~(\ref{fab2}) works remarkably well in regions where strong shocks and instabilities cohabit, such as in the Triple Point test described below. In the case of strong shocks with no shear, Eq.(\ref{fab}) could be replaced by the standard arithmetic mean if necessary. 

Following \cite{spr10}, we used a constant $\alpha=4/3$ so that $\Pi_{ab}$ remains close to the classical SPH artificial viscosity introduced by \cite{mon83}. 
The contribution of the AV to the energy equation (Eq.~\ref{EnergyEq}) is then,

\begin{equation}
\left(\frac{du}{dt}\right)^{AV}_a=\frac{1}{2}\sum_{b=1}^{n_b}\sum_{i=1}^d (v_{i,a}-v_{i,b})~a_{i,a}^{AV}
\label{EnergyAV}
.\end{equation}

Additional information on the physics included in SPHYNX, such as gravity, heat transport or alternative energy equations is deferred to Appendix~\ref{App:energyequation}.

\section{Two-dimensional tests}

In this section we present the outcome of applying SPHYNX to several two-dimensional (2D) tests that have traditionally been problematic for the SPH technique. We show that the {\it sinc} kernels help against the appearance of pairing instability, that the new VE provide a better treatment of discontinuities than the often-used $m/\rho$ value, and that, when applied jointly with IAD$_0$, there is an overall improvement in the results of the shock treatment in the Sedov and wall-shock tests, and also in the development of subsonic instabilities; in particular, the challenging cases of Kelvin-Helmholtz with a density jump of 8 and Rayleigh-Taylor with very weak gravitational field ($g=-0.1$). We also found a convergence on the Gresho-Chan test closer to those of Eulerian codes. Finally, we prove that our implementation suppresses the tensile instability, enabling mixing in tests like the wind-cloud interaction and the triple-point shock test.

Our results below suggest that improving volume conservation produces better results than using volume elements that do not ensure volume equipartition. At first glance, it also renders equalization useless because the advantages of using variable adaptive kernel indices are apparently obscured by the new VE. Nevertheless, equalization and volume elements work on different bases. As demonstrated in~\cite{garciasenz2014}, the use of a variable kernel index improves the interpolation of non-linear functions, even when they are estimated using integrals instead of finite summations. It is therefore expected that the equalization will be useful when {\sl both} the number of particles, $N$, and neighbors, $n_b$, are high enough so that the error in interpolations is dominated by non-linear effects rather than by the evaluation of integrals as finite summations (see also Sect.~\ref{sedov3d}). For a moderate amount of neighbors the use of the new VE given by Eq.~(\ref{estimatorXrho}) makes equalization unnecessary. Thus, unless explicitly stated, many hydrodynamic tests presented in this work have been carried out taking $\Delta n=0$.
\label{sec:2Dtests}
\subsection{Pairing instability}
\label{pairingtest}
Here we show a numerical experiment, similar to that described in \cite{zhu15}, used to study the relationship between the kernel index and pairing instability. We set a sample of particles in a regular 2D lattice leading to a homogeneous density distribution. Then, we seed a random small fluctuation of the internal energy, taking the system out of mechanical equilibrium. Afterwards, we add a frictional dissipative force proportional to the velocity in the momentum equation so that the system returns to equilibrium after a few sound crossing times. The final equilibrium distribution is sensitive to the number of neighbors and to the exponent of the $sinc$ function. This is summarized in Fig.~(\ref{figure2}), which depicts the distribution of the minimum inter-particle distance ($q$), normalized to its initial value ($q_0$), as a function of pairs $(n,n_b)$ at several elapsed times. As we can see, a careful choice of the exponent $n$ keeps the distribution of normalized particle distances closer to 1, making particle clustering difficult, thus avoiding the pairing-instability even at large $n_b$. For example, $n=3$ keeps $q/q_0$ close to 1 when the number of neighbors is small (top-left panel of Fig.~\ref{figure2}), even at large times, while it completely fails when $n_b$ increases (top-right panel). Nevertheless, increasing the exponent $n$ of the $sinc$ kernel suppresses again the pairing instability. This numerical experiment points to $n= 3$, $5$, $7$ and $n\ge 10$ as indicative values to handle with $n_b\simeq 15$, $25$, $50$, and $100$ neighbors, respectively (in 2D). For equivalent numbers in 3D, for example $n_b\simeq 60-120$, currently used in SPH calculations, a $sinc$ kernel with $n\simeq 4-6$ will be sufficient to suppress clustering in many applications (see also Fig.~(3) in \cite{garciasenz2014}). We provide a comparison between $S_n$ and the Wendland $C_3$ in the bottom-right panel of Fig.~(\ref{figure2}). As we can see, the $C_3$~kernel works better than $S_{10}$ but the difference is small. 

The previous test was done using the standard gradient estimator. Fig.~(\ref{figure3}) depicts the results for the same numerical experiment, but carried out with \emph{IAD}$_0$. The comparison between Figs.~(\ref{figure2}) (standard gradient estimator) and (\ref{figure3}) (\emph{IAD}$_0$) suggests that the integral scheme also helps with the pairing problem, as it keeps the particle distribution closer to $q/q_0=1$.

\subsection{Static square of test}
\label{squaretest}
We consider a system of two fluids with different density but identical pressure:

\begin{equation}
\rho=
\begin{cases}
4 & 0.25\le x\le 0.75\quad\mathrm{and}\quad 0.25\le y\le 0.75\\
1 & \mathrm{otherwise}
\end{cases}
\label{twofluid}
.\end{equation}

The system is isobaric with $P=2.5$ and the EOS is that of a perfect gas with $\gamma= 5/3$. 

To carry out the simulations we use Eqs.~(\ref{momentumEq}) and (\ref{EnergyEq}) from Section~\ref{sec:sphynx}, with constant kernel index $n=5$, $n_b=40$ neighbors, and a variety of volume elements. We tried two different initial settings: using particles with the same mass (and uneven spaced grid), as well as particles with unequal mass spread on a uniform lattice. The outcome of the simulations at times $t=0$, 0.4, 1, and 2 is depicted in Fig.~(\ref{volumeelements1}). For equal-mass particles and $X_a= (m_a/\rho_a)^p$ (with $p=0.9$), the code was able to keep the hydrostatic equilibrium during several crossing times (first row). The evolution calculated using $X_a=1$, $X_a=m_a$, and $X_a=P_a^k$ $(k=0.5)$ (second row) led to the same wrong results in much shorter times. Rows 3 and 4 in Fig.~(\ref{volumeelements1}) summarize the evolution using a homogeneous lattice of particles with different mass. Hydrostatic equilibrium is very well preserved using $X_a=1$, $X_a= P_a^{0.5}$, and $X_a= (m_a/\rho_a)^{0.9}$ (third row), while $X_a=m_a$ (fourth row) does not give a satisfactory result.

In Fig.~(\ref{volumeelements2}) we show the magnitudes $\sum_b V_b W_{ab}$ (left panel) and $\langle\Delta\mathbf{r}\rangle = \sum_b V_b~(\mathbf{r}_b - \mathbf{r}_a) W_{ab}$ (right panel) along a 1D cut taken around $y=0.5$, from $x=0.1$ to $x=0.5$ (center of the system) in the hydrostatic system described above after 20 integration steps. As can be seen in the left panel, choosing $X_a=(m_a/\rho_a)^p$ $(p=1)$ gives the best results, leading to an almost perfect volume normalization, while the case with $p=0$ (equivalent to $X_a=1$) shows a large oscillation around the contact discontinuity. The case with $p=0.9$ is also quite satisfactory although, a small oscillation is still present. The right panel in Fig.~(\ref{volumeelements2}) suggests that an adequate choice of the volume element may improve even further the accuracy of the \emph{IAD}$_0$ method to estimate gradients because, as we can see, a value of $p\simeq 1$ helps to keep $\langle\Delta\mathbf{r}\rangle\simeq 0$, which is the necessary condition for \emph{IAD}$_0$ to exactly reproduce the gradient of linear functions.

Unfortunately, taking $p=1$ has the unwanted side effect that density tends to overshoot close to the contact discontinuity. The overshooting in density drops the value of the estimator $X_a=m_a/\rho_a$, so that in the next integration step the density undergoes a further overshooting. Occasionally, the feedback between $\rho_a$ and $X_a$ might rise the density to unrealistic values after several time steps. This problem can be avoided by reducing the value of the exponent $p$ but in that case the adequate optimal value becomes problem dependent. For example, $p=0.9$ works well in the hydrostatic test depicted in Fig.~(\ref{volumeelements1}) but it produces worse results for the Sedov test described later in Sect.~\ref{sec:2Dtests}. For that test $p\simeq 0.6-0.7$ was a better choice. A way to circumvent the overshooting problem is to consider 

\begin{equation}
X_a=\left(\langle m/\rho\rangle_a\right)^p
\label{xrhobis}
,\end{equation}

\noindent
where $\langle m/\rho\rangle_a$ is the SPH average of the standard volume element. Although less efficient than $X=(m/\rho)^p$, this simple recipe enhances the robustness of the scheme so that the exponent of the estimator can be safely raised to $p\simeq 1$, independently of the problem to be simulated. This procedure is especially well suited to handle contact discontinuities hosting large density contrasts, and it has been checked in the blob and Evrard tests described in Sections~\ref{wind-cloud} and \ref{evrard}. A more quantitative discussion on the convergence rate of the estimator $X_a$ as a function of $p$ and the density contrast is given in Appendix~\ref{App:convergence}.

\subsection{Shocks}
\label{shocks}
\subsubsection{Point explosion}
\label{sedov2D}
The study of the propagation of a point-like explosion in more than one dimension is of great usefulness for checking hydrodynamics codes. We have carried out several simulations with SPHYNX to explore the impact of different combinations of volume element and level of equalization on handling shock waves. The explosion was triggered by depositing a considerable amount of energy inside a Gaussian surface of characteristic width $\sigma=0.05$ located at the center of a uniform lattice of size $[0,1]\times [0,1]$. The initial density profile is $\rho(\mathbf{r},0)=1$ and the total injected energy $E=1$. All simulations used $N=256\times 256$ particles and the initial number of neighbors was set to $n_b=50$. The different parameters used in the simulations are summarized in Table~\ref{table3}.

Figure~(\ref{sedov_1}) shows the results of the Sedov test for several combinations of volume elements $V_a$ and equalization $\Delta n$ at $t=0.1$. Models $S_1$, $S_2$, $S_3$, and $S_4$ were calculated using $X_a=(\frac{m_a}{\rho_a})^p$ and different values for the exponent $p$. All models are quite satisfactory but model $S_2$, calculated with $p=0.7$ and $\Delta n=0$, leads to the best results. The density peak is closer to the analytic value and the pressure profile close to the origin behaves well. Model $S_4$, calculated using $X_a=(\langle\frac{m_a}{\rho_a}\rangle)^p$~with $p=1$, also provides good results while models $S_1$, $S_3$, with the choice $p=0$ (standard VE) lead to lower peak values and oscillating pressure tails that depart from the analytic result.

The lower panels of Fig.~(\ref{sedov_1}) depict the profile of the moduli $\vert\langle\Delta\mathbf{r}\rangle\vert_a=\vert\sum_b~V_b~(\mathbf{r}_b-\mathbf{r}_a) W(h_a,n_a)\vert$ (left panel) and that of the normalization condition $\sum_b V_b W(h_a,n_a)$ (right panel). These results are encouraging because they strongly suggest that the new volume elements not only enhance volume conservation (lower right panel of Fig.~\ref{sedov_1}) but also make gradient estimation more accurate, because Eq.~(\ref{approxI}) is better fulfilled, ensuring that the \emph{IAD}$_0$ approximation is, in fact, close to the full \emph{IAD} in terms of accuracy (lower left panel of Fig.~\ref{sedov_1}). Again models $S_2$ and $S_4$ lead to the best results whereas $S_1$ and $S_3$ show worse volume conservation and lower fulfillment of Eq.~(\ref{approxI}).

It has been suggested that other choices of VE could be of interest for handling specific problems. In particular, it has been claimed that the pressure-based estimator $X_a=P_a^k$ ($k<1$) could be well suited to handling contact discontinuities \citep{hop13}. We carried out one simulation using that estimator, with the recommended value $k=0.05$ \citep{ros15}. We found that the results with the pressure-based estimator were not as good as those of models $S_2$ and $S_4$ of Table~\ref{table3}. In Fig.~(\ref{sedov_2}) we compare the results of using the $X_a=P_a^k$ estimator with those of $S_1$ and $S_2$. In the right panel it is clear that the volume normalization is not so well preserved as in model $S_2$, obtaining values similar to those of using standard VE (model $S_1$)}. Moreover, it leads to a lower peak density value and a spurious precursor shock is clearly seen ahead of the shock in the left panel. Although small, such precursor shock is a numerical artifact which was also reported by other authors \citep{ros15}. For that reason, hereafter we have limited the choice of volume elements to those given by Eq.~(\ref{estimatorXrho})\footnote{We note that the standard $V_a=m_a/\rho_a$ is a particular case of Eq.~(\ref{estimatorXrho}) when $p=0$ and all particles have the same mass. Therefore, we can use the same expression to explore the impact of using both the standard and the new VE.}.

\begin{table*}
\centering
\begin{tabular}{@{}lcrrrrrr@{}}
\hline
Model & Dimensions &$N$ & $n_b^0 $& $\Delta n$& $p$&$\rho_{peak}$ \\
\hline
\hline
$S_1$&2D&$256^2$&$50$&$0$&$0$&$3.20$ \\
$S_2$&2D&$256^2$&$50$&$0$&$0.7$&$3.48$ \\
$S_3$&2D&$256^2$&$50$&$5$&$0$&$3.13$\\
$S_4$&2D&$256^2$&$50$&$0$&$1$&$3.44$ \\
$S_5$&3D&$40^3$&$110$&$0$&$0$&$2.01$ \\
$S_6$&3D&$40^3$&$110$&$0$&$0.7$&$2.33$ \\
$S_7$&3D&$40^3$&$110$&$5$&$0$&$2.06$\\
$S_8$&3D&$40^3$&$110$&$0$&$1$&$2.32$ \\
$S_9$&3D&$82^3$&$220$&$0$&$0$&$2.45$ \\
$S_{10}$&3D&$82^3$&$220$&$0$&$0.7$&$2.88$ \\
$S_{11}$&3D&$82^3$&$220$&$5$&$0$&$2.46$\\
$S_{12}$&3D&$82^3$&$220$&$5$&$0.7$&$2.94$ \\
\hline
\end{tabular}
\caption{Settings of the calculated models in the Sedov test. Symbols are, $N$, total number of particles, $n_b^0$, initial number of neighbors, $\Delta n$, linked to variable kernel indices calculated according to Eq.~(\ref{indices}), $p$, exponent of the volume estimator $X_a=(\frac{m_a}{\rho_a})^p$ in Eq.~(\ref{estimatorXrho}) (except models $S_4$~and $S_8$~which uses $X_a=(\langle\frac{m_a}{\rho_a}\rangle)^p)$, and $\rho_{peak}$ the maximum value of density.}
\label{table3}
\end{table*}

\subsubsection{The wall-shock test}
\label{wallshock}
Along with the point explosion (Sedov test), the so called wall-shock (or Noh) test is also used to check the performance of multidimensional hydrodynamics codes. Unlike the Sedov test, the Noh experiment is an implosion towards a geometrical center. The interaction of the converging inflow of gas with the stagnated material at the central volume provokes the formation of a shock-wave moving outwards. The Noh problem has an analytic solution \citep{noh87} to compare with. It is, however, not an easy test owing to the large jump in density across the shock front, $16$ in 2D, $64$ in 3D, for $\gamma=5/3$. SPH codes traditionally had difficulties in handling this test because: a) many particles are needed to correctly reproduce the shock region (especially in 3D); and b) close to the center the internal energy shows a pronounced spike and, consequently, the density abruptly drops to keep the pressure constant. In fact, both problems are connected to the use of the AV and it is not clear to what extent a change in the SPH formalism may alleviate these shortcomings \citep{brookshaw03}.

A sample of $256^2$ equal-mass particles was distributed in an \textit{bcc} lattice with circular perimeter. The radius of the initial configuration is $R=1$ and the density is homogeneous with $\rho=1$ (except at the outer edge of the system). A radial profile of $v_r=-1$ was imparted to all particles at $t=0$, so that the system implodes. The evolution was tracked with SPHYNX until $t=0.5$, when a clear steady shock moving outwards is formed. The results, calculated with $n=5$, $p=0$, and $p=0.7$ for the VE and $n_b=50$, have been compared to both, the analytic predictions and to the output of GADGET-2 for the same initial model.

Table~\ref{table4} and Figs.~(\ref{noh_1}) and (\ref{noh_2}) summarize initial parameters of each simulated model and the main results of the simulations. The density profiles at $t=0.5$ for the three calculated models, $WHS_1$, $WHS_2$, $WHS_G$ are shown on the leftmost panel of Fig.~(\ref{noh_1}). A sharp shock, moving towards the right, is clearly seen at $r\simeq 0.17$. The density jump across the shock front is $\Delta\rho\simeq 16$, hence close to the analytic value. Close to the center of the configuration there is a pronounced dip in density caused, as expected, by the use of the AV. The SPHYNX calculation with $p=0.7$ (model $WHS_2$ with green dots) gives slightly better results at the shock position than $WHS_1$ and $WHS_G$ because the density profile is steeper and the maximum density is closer to the analytic value of 16. We note that all three calculations do show density oscillations at the shock front and that the GADGET-2 simulation is more blurred. The color-maps of density for models $WHS_G$ and $WHS_2$ are also shown in the central and right panels of Fig.~(\ref{noh_1}). As we can see, the shock region is sharp and well defined in the SPHYNX calculation whereas the simulation with GADGET-2 shows stronger oscillations close to the shock location and an inhomogeneous particle distribution close to the origin.

Additionally, model $WHS_2$ has both a better volume normalization than $WHS_1$ (i.e., closer to 1), as shown in Fig.~(\ref{noh_2}) (left panel) and a better behavior of $\vert\Delta \mathbf{r}\vert$ (i.e., closer to 0) at the shock location (right panel). 

\begin{table*}
\centering
\begin{tabular}{@{}llrrrrrr@{}}
\hline
Model &name & N&$\rho_0$& $v_r$&$p$& $(\Delta\rho/\rho)_{max}$ \\
\hline
\hline
$WHS_1$&SPHYNX&$256^2$&$1$&$-1$&$0$&$15.4$ \\
$WHS_2$&SPHYNX&$256^2$&$1$&$-1$&$0.7$&$15.7$ \\
$WHS_G$&GADGET-2&$256^2$&$1$&$-1$&$-$&$15.4$ \\
\hline
\end{tabular}
\caption{List of the calculated models in the wall-shock tests. Symbols are, $N$, total number of particles, $\rho_0$, initial density, $v_r$, initial radial velocity, $p$, exponent of the volume estimator $X_a=(\frac{m_a}{\rho_a})^p$ in Eq.~(\ref{estimatorXrho}), and $(\Delta\rho/\rho)_{max}$ the maximum density jump at $t=0.5$.}
\label{table4}
\end{table*}

\subsection{Fluid instabilities}
\label{instabilities}
Instabilities play a central role in hydrodynamics because they are usually connected to shear mixing and turbulence. In the cosmos, instabilities are especially important because the large Reynolds numbers involved in astrophysical processes make these systems prone to turbulence. Particle- and grid-based codes have been applied with different levels of success to simulate the evolution of hydrodynamical instabilities such as Kelvin-Helmholtz (KH) or Rayleigh-Taylor (RT) instabilities. Here we re-visit them with SPHYNX.

\subsubsection{Kelvin-Helmholtz}
\label{KH}
Grid-based codes are, for the most part, almost free of numerical viscosity, leading to a good match between simulations and the analytic predictions during the linear phase of growth of the KH instability \citep{jun10}. On the other hand, SPH codes are Galilean invariant and do not suffer from numerical diffusion, but intrinsically have more numerical viscosity. It has been pointed out that SPH may not be appropriate for handling fluid instabilities across contact discontinuities with large density jumps \citep{age07}. Although there have been proposals to enhance the modeling of these systems \citep{pri08}, the SPH calculation of shear flows with large density contrast is still an open question. 

We ran two sets of models simulating the growth of the KH instability around the boundary layer separating two flows with moderate and large density contrasts, respectively. Each set was in turn calculated using two values for the volume elements namely $p=0$ and $p=0.7$ in Eq.~(\ref{estimatorXrho}). The initial setting was the same as in \cite{mcn12}, where three stratified fluid layers inside a 2D box of size $[0,1]\times [0,1]$ were considered. The fluid layers span for $y \le y_1$, $y_1\le y \le y_2$ and $y \ge y_2$ with densities $\rho_1$, $\rho_2$, and $\rho_3$, respectively. The adopted values for $y_1$, $y_2$, $\rho_1$, $\rho_2$, and $\rho_3$, as well as the number of particles used in the simulation, are shown in Table~(\ref{table5}). Periodic boundary conditions were implemented at all sides of the box.

\begin{table*}
\centering
\begin{tabular}{@{}llrrrrrrr@{}}
\hline
Model &N&$m_a$ &$y_1$& $y_2$& $\rho_1$& $\rho_2$&$\rho_3$&$X_a$ \\
\hline
\hline
$KH_1$&$256^2$&$C$&$0.25$&$0.75$&$1$&$2$&$1$&$1$\\
$KH_2$&$256^2$&$C$&$0.25$&$0.75$&$1$&$2$&$1$&$(\frac{m_a}{\rho_a})^{0.7}$ \\
$KH_3$&$256^2$&$V$&$0.25$&$0.75$&$1$&$8$&$1$&$1$\\
$KH_4$&$256^2$&$V$&$0.25$&$0.75$&$1$&$8$&$1$&$(\frac{m_a}{\rho_a})^{0.7}$ \\
\hline
\end{tabular}
\caption{Settings of the calculated models in the 2D Kelvin-Helmholtz tests. Symbols are $N$ total number of particles; $m_a$ the nature of the mass-particles -constant (C) or variable mass (V); $y_1, y_2$ the location of the contact layers; $\rho_1, \rho_2, \rho_3$ the densities at the fluid layers. The last column shows the estimator $X_a$ used to compute the volume elements.}
\label{table5}
\end{table*}

A velocity $v_x=0.5$ is given to the central strip, whereas the rest of the box moves in the opposite direction with $v_x=-0.5$. Prior to the calculations, the density and $v_x$ distributions were smoothed following the method described in \cite{mcn12} (their Eqs.~1 and 3). Thus, the growth-rate of the instability obtained with SPHYNX can be compared to the templates obtained by McNally and collaborators using the PENCIL Code\footnote{Available at \url{pencil-code.nordita.org}}, a state-of-the-art hydrodynamics code of Eulerian type \citep{brandenburg2002}. The pressure is set to $P=2.5$ everywhere, with $\gamma=5/3$. The fluid layer is in almost vertical equilibrium except for a small seeded perturbation in $v_y$ given by,

\begin{equation}
v_y(x,y)= w_0 \sin (4\pi x)
\label{kh1}
,\end{equation}

\noindent 
with $w_0=0.01$. The velocity perturbation has therefore a wavelength $\lambda=0.5$ so that the box hosts two complete waveforms. 

Models $KH_1$ and $KH_2$ in Table~(\ref{table5}) were calculated assuming a density contrast of two, whereas an initial jump of 8 was imposed to models  $KH_3$ and $KH_4$. To build the initial models, a sample of $256\times 256$ particles were first spread in a squared lattice and then stretched in the vertical direction so that the ensuing density profile adapts to that of Eq.~(1) in \cite{mcn12}. This setting naturally leads to a density contrast of two for equal-mass particles and provides the necessary smoothness around the separation layers between the fluids. To build the high-density contrast models, the initial lattice was stretched until the density ratio was $2\sqrt{2}$. The mass of the particles was then made proportional to the density of the stretched grid, so that a smooth profile with a density contrast of 8 was achieved.

The Balsara limiter $f_a$ given by Eqs.~(\ref{balsara}, \ref{fab}, and \ref{fab2}) was applied to all models to reduce the shear viscosity. For these initial conditions, previous simulations with the traditional formulation of SPH predict the growth of the KH instability only for the low-density contrast case \citep{age07,jun10}. When the density contrast rises well above $2$ it is necessary either to add an artificial heat flux to keep isobaricity \citep{pri08} or to redefine the volume elements \citep{sai13,hop13}. In particular, it was shown that the formulation which uses volume elements linked to pressure is able to reproduce the KH instability across high density jumps. Also, codes which directly use a smoothed pressure in the SPH equations, such as the PSPH formulation described in the Appendix of \cite{hop15}, can also handle with high density contrasts. Here we show that the \emph{IAD} approach combined with new volume elements, which preserve kernel normalization, is also able to successfully simulate the KH instability with high density jumps.

Figure \ref{kh_1} shows a color-map of density of the evolution of models $KH_1$ and $KH_2$ at times $t=1.5$ and $2.5$. We see that there is a growth of the instability,  qualitatively similar to the results obtained using other novel formulations of the technique such as the PSPH scheme with $\simeq 200$ neighbors (see for example  Fig.~(19) in \cite{hop15}). The results of SPHYNX are also very similar to those obtained with the recent CKRSPH formulation (see for example the upper-row of figure 17 in \citealt{fro17}). Model $KH_2$, calculated using the new volume elements, appears to evolve slightly faster than $KH_1$, computed with the standard choice, $VE=m/\rho$, showing more evolved structures in the non-linear stage. Unlike the PSPH scheme, our method does not estimate the pressure by kernel smoothing nor does it incorporate any artificial flux of heat to smooth the pressure.

Figure \ref{kh_2} depicts the mode-amplitude evolution of $v_y$ calculated taking the Fourier transform (FT) of the $v_y$ field. The FT was calculated with the expressions given in \cite{mcn12} so that a comparison with the amplitude growth calculated with the PENCIL code (continuum-red line) and the NDSPMHD code, described in  \cite{pri12}, (dashed lines) is straightforward. The results are encouraging as SPHYNX reproduces the KH growth closer to the reference simulation of the PENCIL code than the NDSPMHD code, the latter even having a factor 2 higher resolution than our simulations with SPHYNX (see also Fig.~7 in \cite{mcn12} and comments therein). Our results are close to those obtained with the PSPH scheme with the same resolution. In Fig.~\ref{kh_2} we also show the L$_1$ errors of each model with respect to the reference values, calculated as: $L_1=\frac{1}{N}\sum_{i=1}^N\vert A^r_i-A^s_i\vert$, where $A^r$ and $A^s$ stand for the reference and simulated amplitudes, respectively. These results are an indication of the importance of using an accurate gradient evaluation, this being the most relevant difference between SPHYNX and NDSPMHD.

Model $KH_2$, calculated with $p=0.7$, exhibits  better volume normalization than $KH_1$, computed with $p=0$, as shown in Fig.~(\ref{kh_3}). That figure depicts the value of the normalization condition $\sum_b V_b W_{ab}$ as a function of the density of the particles. For both runs, most of the particles cluster around the expected value of 1, but the summation falls below one in the low-density region and exceeds it in the high-density region. Nevertheless, the dispersion around the correct value is considerably  lower in the case with $p=0.7$, showing its greater capability to achieve equipartition in disordered particle distributions.

Figure~\ref{kh_4} summarizes the simulations with an initial density ratio $\rho_2/\rho_1=8$. The upper and lower panels were calculated with $p=0$ and $p=0.7$ in Eq.~(\ref{estimatorXrho}), respectively. The snapshots correspond to times $t=1.5$ and $2.5$. First of all, we see that despite the large jump in density, the instability is able to develop in both cases. 
These results, therefore, suggest that the use of the \emph{IAD} scheme to calculate gradients improves by itself the quality of the simulations of hydrodynamic instabilities. Although the instability evolves slightly faster when $p=0.7$, and shows more structure, it is probably affected by the noise introduced by the VE estimator $(m/\rho)^p$ when the mass of the particles is not constant. As in the precedent tests, the model calculated with $p=0.7$ has the best kernel normalization properties (not shown in the figures).

\subsubsection{The Rayleigh-Taylor instability}
\label{RT}

\begin{table*}
\centering
\begin{tabular}{@{}llrrrrr@{}}
\hline
Model &N&$\rho_u$& $\rho_d$&$w_0$&g&p \\
\hline
\hline
$RT_1$&$256^2$&$2$&$1$&$0.0025$&$-0.5$&$0$\\
$RT_2$&$256^2$&$2$&$1$&$0.0025$&$-0.5$&$0.7$ \\
$RT_3$&$256^2$&$2$&$1$&$0.0025$&$-0.1$&$0$\\
$RT_4$&$256^2$&$2$&$1$&$0.0025$&$-0.1$&$0.7$\\
$RT_5$&$256^2$&$2$&$1$&$0.0025$&$-0.1$&$1.0$\\
\hline
\end{tabular}
\caption{Settings of the calculated models in the 2D Rayleigh-Taylor tests. Symbols are $N$ total number of particles; $\rho_u, \rho_d$ are the densities at the upper and lower regions; $g$ is the gravitational acceleration and $p$ is the exponent of the volume estimator $X_a=(\frac{m_a}{\rho_a})^p$ in Eq.(\ref{estimatorXrho}). Model $RT_5$ uses the averaged version of $X_a$ (Eq.~\ref{xrhobis}).}
\label{table6}
\end{table*}

In \cite{garciasenz2012} it was shown that the \emph{IAD}$_0$ scheme is able to simulate the gross features of the growth of the RT instability for initial perturbations as low as $\Delta v_y=0.01$ in the velocity field. In this section we present clear proof that SPHYNX is able to cope with the growth of the RT instability also when the gravitational force is small.
           
Our initial model is similar to that described in \cite{spr10a}, where the numerical experiment takes place in a box sizing $[0.5\times 1.5]$. The lower and upper halves of the box are filled with equal-mass particles. The particle distribution is then arranged in the vertical direction so that $\rho_u=2$ (upper region) and $\rho_d=1$ (lower region). 

The contact discontinuity around the interface was smoothed using:

\begin{equation}
\rho=\frac{\rho_d+\rho_u\exp(\frac{y-y_0}{\Delta y})}{1+\exp(\frac{y-y_0}{\Delta y})}
\label{rt1}
,\end{equation}

\noindent with $\Delta y=0.0083$ and $y_0=0.75$. The integration of the hydrostatic equilibrium equation $\frac{1}{\rho}\frac{dP}{dy}=-g$, where $g$ is the gravitational acceleration, with an ideal EOS with $\gamma=1.4$ gives the pressure profile in the vertical direction, 

\begin{equation}
\begin{split}
P(y)=P_0+\rho_d g~\Delta y\left[\frac{y-y_0}{\Delta y}+
\log\frac{2}{1+\exp(\frac{y-y_0}{\Delta y})}\right]+\\
\rho_u g~\Delta y\log\left(\frac{1+\exp(\frac{y-y_0}{\Delta y})}{2}\right)\,,
\end{split}
\label{rt2}
\end{equation}

\noindent with $P_0=2.5$. We used $256^2$ particles and an initial number of neighbor particles $n_b=50$. The boundary conditions are periodic at the left and right sides of the box but reflective on the bottom and the top. The main features of the models used in this test are summarized in Table~\ref{table6}. 

The fluid was initially at rest everywhere, except for a small, single-mode perturbation applied to the vertical velocity field,

\begin{equation}
v_y(x,y)=w_0 \left[1-\cos(4\pi x)\right]\left[1-\cos(4\pi y/3)\right]\,,
\label{rt3}
\end{equation}

\noindent where $w_0=0.0025$. In Fig.~(\ref{rt_1}) we show the density color-map of the evolution of the instability when $g=-0.5$ is adopted. As we can see, the instability grows at a good rate and it enters into the non-linear regime at $t> 3.5$. The contact discontinuity between both fluids is well marked, but not totally sharp, owing to both the smooth initial conditions and to the spread of the physical magnitudes over the characteristic smoothing length $h$. During the non-linear stage, the calculation with $p=0.7$ (model $RT_2$ in Table~\ref{table5}) displays more structure than that of model $RT_2$ calculated with $p=0$. At $t=7.5$ the vertical extension of the unstable region is nearly the same but model $RT_2$ shows a richer structure inside the mushroom-like instability. 

The impact caused by the choice of VE is much more marked when a smaller gravitational force, $g=-0.1$, is adopted. This was the case reported in \cite{spr10} concerning the simulations of the RT instability with the AREPO code. As far as we know, there are no satisfactory calculations on this same scenario using the SPH technique. The combination of a small initial perturbation plus a tiny gravitational force makes the problem difficult because the $E_0$ errors and the tensile instability conspire to totally suppress the growth of the instability \citep{val12}. The upper panels in Fig.~(\ref{rt_2}) depict the evolution of the instability as simulated by SPHYNX taking standard VE (i.e., $p=0$ in Eq.~13). As we can see, the instability is able to grow, although the growth-rate is slow and there is a clear lack of structure during the non-linear stage. The lower row of panels in the same figure depict the evolution when VE are calculated with $p=0.7$. In this case there is a clear boost in the development of the instability as it grows faster and shows a richer structure in the non-linear regime. 

Quantitative numbers on this test can be extracted from the evolution of the  points located on the tip/bottom of the bubbles/spikes. Additionally, the numerically inferred terminal velocity of the vertex of the bubble can be compared with the analytical results by \cite{gon02}. The results are summarized in the Figure \ref{rt_3}, where the left panel indicates the bubble/spike evolution of the farthest point achieved by the lighter and heavier fluids (bubble and spike respectively), and the right panel indicates the evolution of the velocity of the bubble. As we can see, the evolution of the bubble/spike sample is faster when the VE are calculated with $p=0.7$ and $p=1$ (in this last case the averaged estimator, Eq.\ref{xrhobis}, was used). We note that at $t\simeq 15$, the tip/bottom of the bubble/spike are close to colliding with the box limits. Consequently the rising velocity of the bubble velocity was computed in the interval $0\le t \le 15$. As we can see, the terminal velocity of the bubble matches well the analytical estimation by \cite{gon02},

\begin{equation}
v_{term}=\sqrt{\frac{2~A_t}{(1+A_t)}\frac{g}{C_gk}}
\label{velbubble}
,\end{equation}

\noindent
where $A_t=(\rho_2-\rho_1)/(\rho_2+\rho_1)=1/3$ is the Atwood number, $k=4\pi$ is the wave-number of the applied single perturbation and $C_g=3$ in two-dimensions. The terminal velocity calculated using Eq.~\ref{velbubble} (solid pink horizontal line in Fig.\ref{rt_3}) is close to those obtained using SPHYNX.

Although our results are not as good as those obtained with the code AREPO for the same initial conditions (see Fig.~35 in \cite{spr10} for a comparison), they are encouraging, as they suggest that the simulation of the RT instability in a tiny gravitational field is also affordable for SPH codes when an E0-suppressing technique like \emph{IAD}$_0$ is used jointly with equipartition-preserving volume elements.

\subsection{The Gresho-Chan vortex}
\label{gresho}
The simulation of a stationary vortex that is in stable equilibrium is a very demanding test for any hydrodynamics code. This is a fundamental test for the accuracy of the numerical scheme, the preservation of symmetry and the conservation of angular momentum. The Gresho-Chan vortex has been especially problematic for SPH, which, in spite of having excellent angular momentum conservation, shows very poor convergence \citep{spr10} due to numerical deviations from the initial conditions that degenerate the stability of the vortex within short timescales. This is specially relevant in the simulation of self-gravitating disks, where the centrifugal force and pressure gradient should be (up to a certain extent) balanced. Therefore, numerical noise can trigger deviations from this equilibrium configuration that lead, for example, to an artificial fragmentation of the disk.

Several recent works \citep{deh12,zhu15,ros15} have managed to keep the vortex in steady state, with low dispersion, during more than one orbit-cycle. In his work, \cite{ros15} proves that the use of high-order kernels is crucial to obtain accurate results, and in combination with \emph{IAD}$_0$ and an improved artificial viscosity with a noise dissipation trigger, leads to a roughly linear convergence rate.

In this test, we use the common initial conditions for the azimuthal velocity and pressure profile: 

\begin{equation}
  v_\phi(r)=v_0
\begin{cases}
    \psi & \text{for $\psi<=1$}\\
    2-\psi & \text{for $1<\psi<=2$}\\
    0 & \text{for $\psi>2$,}\\
\end{cases}
\end{equation}

\begin{equation}
  P(r)= P_0+4v_0^2
\begin{cases}
    \psi^2/8 & \text{for $\psi<=1$}\\
    \left(\psi^2/8-\psi +\ln \psi +1\right) & \text{for $1<\psi<=2$}\\
    \left(\ln 2 -1/2\right) & \text{for $\psi>2$,}\\
\end{cases}
\end{equation} 

\noindent
where $\psi=r/R_1$, $R_1=0.2$, $v_0=1$, $P_0=5$, for an ideal gas with $\gamma = 5/3$. This setting corresponds to a low Mach number test with $\mathcal M=1/\sqrt{\gamma P_0}\simeq 0.35$. We use $N=256^2$ particles evenly distributed in a square lattice of size $[0,1]\times [0,1]$ so that the density is uniform with value $\rho=1$. The artificial viscosity parameter in Eq.~(\ref{avis}) is set to the current value $\alpha=4/3$~and the Balsara coefficients estimated with Eqs.(\ref{balsara}, \ref{fab}, and \ref{fab2}). We note that because the vortex is in steady conditions, the divergence of the velocity should vanish everywhere and $f_a\simeq 0$. As in other simulations, however, we allow a low level of AV to control the numerical noise and to reduce the dispersion of the particles. For this test, the particular choice of the volume element is not relevant because the particle distribution is homogeneous due to the constant density, hence the kernel is properly normalized at all times, independently of the exponent. Therefore, we take the nominal value of $p=0.7$ in Eq.~(\ref{estimatorXrho}) in all calculated models.

We carried out several simulations of the vortex evolution for different choices of the kernel index $n$ and initial number of neighbors\footnote{In a square homogeneous lattice the equivalent number of neighbors in 1D, 2D, and 3D is related by $n_b(2D)= n^2_b(1D)\frac{\pi}{4}$, $n_b(3D)= n^3_b(1D)\frac{\pi}{6}$, and $n_b(3D)=n_b^{\frac{3}{2}}(2D)\frac{4}{3\sqrt\pi}$}, $n_b(2D)$. We also calculated models with the Wendland $C_6$ kernel which, according to \cite{deh12} and \cite{ros15}, was especially suited to handle this test. The quality of the simulations and convergence rate were checked with the parameter $L_1=\frac{1}{N}\sum_{i=1}^N \vert\mathbf{v}_i^t-\mathbf{v}_i^s\vert$ where $\mathbf{v}^t$ and $\mathbf{v}^s$ stand for the theoretical and the simulated value of velocity, respectively, and the summation runs over all particles.

In Fig.~(\ref{gresho_1}), we show the profile of the tangential velocity at time $t=1$ (roughly the elapsed time needed by the vortex to complete one turn) for three kernel choices and different number of neighbors ($n_b$). Using a low $n_b$ ($\simeq 30$, in 2D) the best results were obtained with the $sinc$ ($n=5$) whereas the choice $n=6.315$ or the Wendland $C_6$ leads to qualitatively similar results. The situation is reverted when $n_b=50$, where the $C_6$ interpolator gives a slender velocity profile and the lowest $L_1$ value. However, the $sinc$ $(n=6.315)$ also shows low dispersion profiles and $L_1$ values. Increasing the number of neighbors to $n_b=70$ ($\simeq 500$ in 3D) reduces the dispersion, as expected. The $C_6$ provides the lowest $L_1$ value here but closely followed by the $sinc~(n=7)$. 

Figure~(\ref{gresho_2}) summarizes the functional dependence of the estimator $L_1$ with respect to the index $n$ (left panel) and number of neighbors $n_b$ (right panel). Looking at the left panel we see that for $n_b\simeq 20$ the $sinc~(n=4)$ (similar to the $M_5$  and $C_2$ kernels) gives the best results. Increasing the number of neighbors moves the minimum value of $L_1$ to the right (higher $n$ values). At $n_b\simeq 30$, the index $n=5$ becomes the optimal choice. Setting the initial number of neighbors to $70$ and $80$ demands larger kernel indices ($n\simeq 7-8$) at the position of minimum $L_1$. The right panel of Fig.~(\ref{gresho_2}) explores the convergence rate of the Gresho-Chan vortex test with respect to the number of neighbors at a {\sl constant} total particle number $N=256^2$. When $n_b\le 30$ it seems that the low-order interpolators with  $n\le 5$ give the lowest $L_1$ values. Nevertheless, above $n_b\ge 40$ it is necessary to raise the index of the $sinc$~kernel, or use the Wendland $C_6$, to achieve convergence. The $C_6$ gives the lowest $L_1$ values when $n_b\ge 40$. The curve referred as $nmix$ (dotted black line) is a fit which takes advantage of the optimal $n$ index at each $n_b$. Both, the $nmix$ profile and the $C_6$ curve can be used to estimate the convergence  rate in this test. These profiles show a rapid initial decline: $L_1(n_b\le 50)\propto n_b^{-1.25}$, followed by a more gentle reduction of the convergence rate $ L_1(n_b\ge 50)\propto n_b^{-0.6}$.               

Figure~(\ref{gresho_3}) shows the convergence rate of $L_1$ at $t=1$ when the one-dimensional (1D) equivalent number of particles, $N_{1D}$, is varied in the range $50\le N_{1D}\le 400$. In this study we have only considered the $sinc$ kernels, but with several exponents. The convergence rate is explored using the following $(n,n_b)$ pairs: $(5,30)$, $(6.315,50)$, $(8,80)$ and $(10,100)$. As can be seen, the high-order $sincs$ with $n=8$, $n=10$ achieve a linear convergence of $L_1$, provided that a sufficient number of neighbors is taken. Moreover, the convergence rate $L_1\propto N_{1D}^{-1}$ is larger than those reported in \cite{spr10} ($L_1\propto N_{1D}^{-0.7}$)~or in \cite{deh12} ($L_1\propto N_{1d}^{-0.6}$) and closer to the value obtained with Eulerian codes, $L_1\propto N_{1D}^{-1.4}$ \citep{spr10a}; see also \cite{val16} where the ability of the \emph{IAD}$_0$ scheme to reduce sampling errors is proved in a similar subsonic test, but with Mach number as low as $\mathcal M=0.02$.

\subsection{Interaction of a supersonic wind with a cold cloud of gas}
\label{wind-cloud}
  
Popularly known as the `blob' test \citep{age07}, this problem has challenged SPH codes for a long time. The basic setting of this test gathers many pieces of physics, such as strong shocks and mixing due to the KH and RT instabilities in a multi-phase medium with a large density contrast. The initial configuration consists in a dense spherical cloud of cold gas embedded in a hotter ambient medium. The cloud is initially at rest while the ambient background (the wind) moves supersonically. The wind-cloud interaction generates a bow shock and, in short time, the cloud is fragmented and mixed with the background owing to the combined effect of ablation and KH and RT instabilities. 
 
Modern grid-based codes are able to handle this scenario and all of them agree in the gross features of the evolution. Nevertheless, simulations using the SPH technique have historically had difficulty disrupting the cloud because of the poor development of the hydrodynamic instabilities that drive the erosion of the cloud \citep{age07}. As a matter of fact, the challenge introduced by this problem has led to interesting advances of the SPH technique in recent years, such as the pressure, PSPH, formulation of the SPH equations by \cite{sai13} and \cite{hop13}, and the conservative reproducing kernel method, CRKSPH by \cite{fro17}. As shown in these works, it is feasible to suppress the tensile-instability acting at the contact discontinuity between the two fluids, leading to a  better agreement between SPH and grid-based codes. 

Here we show that the new volume elements given by Eq.(\ref{estimatorXrho}) also suppress the tensile-instability at the contact discontinuity. The code SPHYNX is thus able to  cope with the blob test {\sl without leaving} the  density-formulation of SPH. Our initial setting is similar to that described in \citep{age07}, although restricted to two dimensions where the dense cloud is no longer a sphere but a circle (a cylinder in 3D). The wind is simulated with  $N_w=256^2$ particles spread following a glass-like stable (previously relaxed) distribution in a box sizing $[1,1/4]$. The cloud is reproduced with $N_C=68^2$ particles, spread in a regular square lattice tailored in a circle of radius $R_C=1/40$, and centered at coordinates $(1/8,1/8)$. All particles have the same mass. The features of the wind and cloud at $t=0$ are $\rho_w=1, P_w=1, u_w=3/2, v_w=2.7$ and $\rho_c=10, P_c=1, u_c=3/20, v_c=0$, respectively, where the wind velocity $v_w$ is supersonic with Mach number $\mathcal M=2.7$. As the particles in the cloud are fully ordered we add a small random initial radial velocity, $v_r=v_r^0~\times rand[-1:1]$ with $v_r^0=0.03$.

We carried out two calculations of the blob test using SPHYNX: with the standard VE (model $W_0$) and with the new VE (model $W_1$), as well as with GADGET-2, (model $W_G$). Because of the large density contrast between the wind and the cloud the model $W_1$~was calculated with the averaged estimator $X=(\langle m/\rho\rangle)^p$~and $p=1$.

In Fig.~(\ref{wc_1}) we show several snapshots of the simulations at normalized times $(t/t_{KH}=0.5, 2, 3)$, where $t_{KH}$ is the characteristic KH time, as defined in \cite{age07}. As we can see, the fragmentation and mixing of the cloud is only attained in model $W_1$, calculated with $p=1$. Models $W_0$ and $W_G$ give worse results, but we note that even in that case SPHYNX with $p=0$ yields a slightly larger fragmentation of the cloud than GADGET-2 (see Fig.~\ref{wc_2}). Our results are also in good agreement with those obtained by \cite{fro17} using the CRKSPH method (see their Fig.~21). The large suppression of the tensile instability in the $W_1$ calculation is the characteristic that allows the fragmentation of the cloud. This is clearly visible in the detailed density snapshots shown in Fig.~(\ref{wc_3}). It is therefore evident that the long-standing tensile instability problem in SPH is bounded to kernel normalization. Having a good partition of unity considerably reduces the strength of that instability.      
    
The evolution of the surviving fraction of the cloud as a function of time is depicted in Fig.~(\ref{wc_2}). The expected outcome is that the surviving fraction of the cloud reduces to 0 in about $t/t_{KH}\sim 2.5$, as obtained by grid codes (see, e.g., Fig.~25 of \citealp{hop15}). As we can see, the $W_1$ calculation leads to almost complete destruction of the cloud in $t/t_{KH}\simeq 4$, while in the other two calculations hardly $50\%$ of the cloud is mixed. Nevertheless, unlike in the grid-based calculations, the cloud is not totally mixed with the ambient gas as $\simeq 10\%$ of it remains unmixed in the $W_1$ simulation. The difference may arise from the different dimensionality of the calculations,  because in 2D the cloud is a cylinder whereas in 3D the shape of the blob is spherical. However, a calculation of the same model in 3D with $3\cdot 10^6$ particles led to similar results. Actually, the complete dissolution of the cloud is only attainable if some heat-transport is included in the scheme \citep{hop15}.

\subsection{Shock plus vorticity: The triple-point shock test}

The similarities among the results obtained by SPHYNX and the CRKSPH method, as found in the KH and the Blob tests above, are even more evident when considering the so called triple-point shock test. This is a shock-tube-like setting where three materials with different densities and pressures are put into contact through discontinuity lines. The first region contains a high-density, high pressure material, located in the vertical band on the left of the first snapshot in Fig.~(\ref{triplepoint}) (see also the sketch by \cite{fro17} shown in their Fig.~24). That region is in contact with two low-pressure horizontal bands, the upper band being considerably less denser than the lower band.  The particular point where the three discontinuity lines intersect is the triple point. The vertical high-pressure band on the left launches a shock through the lower-pressure horizontal bands. Because of the pronounced difference in the speed of sound, the shock moves faster in the upper low-density, low-pressure region. As a consequence, a shear is induced around the line separating both low-pressure materials, and soon the KH instability develops, rolling-up the interface between these two regions. At late times, the reflected shocks at the boundaries of the box also induces the Richtmyer-Meskhov instability when they cross the interfaces between the different materials.  

Our main aim is to simulate the triple-point shock scenario and to compare the results obtained with SPHYNX to those given by the CRKSPH method. A comparison among the performances of CRKSPH, the standard SPH formulation and the ReALE (Reconnecting arbitrary Lagrangian Eulerian) method by \cite{lou10} is provided in the work by \cite{fro17} and we refer the reader to that paper. Their results show that for the triple-point shock scenario the CRKSPH scheme gives the best results.  

The setting of the initial model, particle number in each region and EOS is exactly the same as described in \cite{fro17}. The evolution of the system with SPHYNX and two choices for the VE, $p=0$ and $p=1$ in Eq.~(\ref{xrhobis}), was tracked until $t=8$. We show in Fig.~(\ref{triplepoint}) the density color-maps for both cases at times $t=1$, 3, 5, 7. As we can see, the results with SPHYNX and $p=1$ are similar to those obtained with the state-of-the-art CRKSPH scheme. The main features of the shock wave are well simulated for both, the $p=0$ and $p=1$ cases, but the amount of structure seen in the calculation with $p=1$ at $t=5$ and $t=7$ is  larger than that of the standard VE choice. It is worth noting that our recipe for implementing the artificial viscosity is relatively basic and there is still some room for improvement in this area, such as, for instance, the use of time-dependent AV coefficients \citep{cul10,ros15}.

\section{Three-dimensional tests}

In this section we present three test cases where SPHYNX is applied to three-dimensional (3D) scenarios. We show here that the combined use of the new VE, the $sinc$ kernels, and \emph{IAD}$_0$ leads to better results when trying to obtain an homogeneous density field, decreasing discretization errors. In the following, we re-do the Sedov test, now in 3D, where the new VE stand out by better handling shock-waves and the equalization enhances the outcome in simulations with high number of neighbors. Finally, the collapse of an isothermal cloud tests the implementation of SPHYNX coupled with gravity evaluation. The results show that the new VE enhance the kernel normalization, and consequently the overall interpolations, leading to a better shock treatment when compared with standard implementations.

\label{sec:3Dtests}
\subsection{Approaching a uniform scalar field}

The initial models used in many SPH simulations are often built so that the density and/or the pressure are as homogeneous as possible. These systems should ideally remain in equilibrium during many sound-crossing times. Although very common, arranging particles in ordered lattices is not the best option because the particles move off the lattice and introduce noise. In this section we discuss the ability of SPHYNX to build a uniform density/pressure field, which is obtained after relaxing a pseudo-ordered initial distribution of particles. Several features of the resulting glass-like structure are discussed as a function of the VE choice. To obtain a homogeneous density profile we follow a similar procedure such as that explained in \cite{ros15}.  

We first spread $N=128,000$~particles in an ordered 3D squared \emph{bcc} lattice. Afterwards, their position is displaced at random in each direction, with maximum amplitude $40\%$ of the size of the smoothing length. Then, we allow the system to evolve keeping internal energy constant, at $u_0=1$, to regain equilibrium. During the relaxation, the velocities are set to zero but the particles are displaced with the simple recipe:  $\Delta\mathbf{r} = 10^{-3}~h~(\mathbf{a}_c/{\mathrm a_c}$)~where $h$ is the smoothing-length and $\mathbf{a}_c$~stands for the acceleration vector. The value of $h$~was the same for all particles, so that the number of neighbors is $n_b \simeq 100$~with low dispersion. During the evolution, the $L_1$~error estimators were monitored for several magnitudes, such as density, partition of unity, and $\langle\Delta \mathbf{r}\rangle$. These are defined as: 

\begin{align}
L_1(\rho)&= \frac{1}{N}\sum_{a=1}^N \vert(1-\rho_a)\vert, \\
L_1(\textrm{Vol})&=\frac{1}{N}\sum_{a=1}^N \vert(1-\sum_b V_b W_{ab})\vert, \\
L_1(\langle\Delta\mathbf{r}\rangle)&=\frac{1}{N}\sum_{a=1}^N\vert(0 -\langle\Delta\mathbf{r}\rangle_a)\vert.
\end{align}

We note that as $P=(\gamma-1)u_0\rho$, with $\gamma=5/3$, the error estimator $L_1(\rho)$ also provides the dispersion in pressure.

The $L_1$~values after 800 iterations are summarized in Table \ref{table7}. All settings of the VE lead to stable glass-like structures with small departures from homogeneity ($L_1 (\rho)\leq 10^{-3}$). Model D, calculated with $X_a=(\langle m_a/\rho_a \rangle)^p;~ (p=1)$, displays the lowest $L_1 (\rho)$~values. As expected, the calculation with  $X_a =(m_a/\rho_a)^{0.7}$~leads to the best partition of unity but, paradoxically, is the most inhomogeneous of the sample. Model C, calculated with a negative exponent $p= -0.7$~is very similar to model A computed with the standard VE choice (p=0.0). Because the internal energy was kept constant during the relaxation, a negative $p$~value in Eq.~(\ref{estimator}) is equivalent to taking a VE linked to pressure. We therefore conclude that SPHYNX is able to handle homogeneous systems in pressure equilibrium for a wide range of the parameter $p$. For these systems, the best choice of VE is that given by Eq.~(\ref{xrhobis}) with $p\simeq 1$.

\begin{table*}
\centering
\begin{tabular}{@{}ccccc@{}}
\hline
Model &Exponent $p$& \multicolumn{1}{c}{$L_1~(\rho$)} & \multicolumn{1}{c}{$L_1~(Vol)$} & \multicolumn{1}{c}{$L_1~(\langle\Delta\mathbf{r}\rangle) $} \\
\hline
\hline
A &0.0& $5.4~10^{-4}$ & $3.9~10^{-4}$ & $1.4~10^{-5}$  \\
B& 0.7& $1.2~10^{-3}$  & $3.5~10^{-4}$ & $1.5~10^{-5}$  \\
C&-0.7& $5.3~10^{-4}$  & $3.8~10^{-4}$ & $1.4~10^{-5}$  \\
D&1.0& $4.5~10^{-4}$  & $3.9~10^{-4}$ & $1.4~10^{-5}$  \\
\hline
\end{tabular}
\caption{Values of the parameter $L_1$~for different magnitudes. Models A, B, and C were calculated with $X_a=(m_a/\rho_a)^p$, whereas model D uses $X_a=(\langle m_a/\rho_a\rangle)^p$.} 
\label{table7}
\end{table*}

\subsection{Point explosion}
\label{sedov3d}
The description of an explosion in the 3D space is usually a challenge for any hydrodynamics code. The lower resolution (in comparison with 2D calculations) makes it difficult to capture the density, pressure, and velocity at the peak of the blast and also to correctly reproduce the post-tail values of these variables. We have investigated the ability of our code to cope with a point-like explosion in 3D and compared the results with those obtained using GADGET-2.

We have run two sets of models labeled as $\{S_5, S_6, S_7, S_8\}$ and $\{S_9, S_{10}, S_{11}, S_{12}\}$ in Table~\ref{table3}. The first set consists of low-resolution calculations whereas the second set of simulations has not only more resolution, but also uses more neighbors to make the interpolations.

Figure~(\ref{sedov_3}) summarizes the results for the set of low-resolution models. As in the 2D simulations, models $S_6$~and $S_8$ in Table~\ref{table3}, with the new volume elements, lead to the best results. Both models give a higher density peak and a better post-shock evolution of density and pressure. They also exhibit an enhanced volume normalization and $|\Delta\mathbf{r}|$ has a better behavior across the wave front. The model calculated with $p=0$ and $\Delta n=5$ ($S_7$) is the third-best case, followed by model $S_5$ (standard VE setting $p=0, \Delta n=0$). Figure~(\ref{sedov_3}) also includes the density and pressure profiles (light blue lines) calculated using GADGET-2 for the same initial setting. Its density profile shows a similar post-shock tail evolution as model $S_6$, but lower peak value and the post-tail pressure profile diverges considerably from the analytical curve. 

Figure~(\ref{sedov_4}) is the same as Fig.~(\ref{sedov_3}) but for models $\{S_9, S_{10}, S_{11}, S_{12}\}$ calculated with a larger number of particles and taking two times more neighbors in the summations. Therefore, these models are characterized by both a higher resolution and a less noisy estimation of gradients. According to our calculations, model $S_{12}$ computed using $p=0.7$, $\Delta n=5$ is the one closer to the analytic expectations. The density peak of the blast was the highest and at the correct position, while the values of density and pressure close to the origin behave well. Further, model $S_{12}$ also has the best volume normalization and minimum $|\Delta\mathbf{r}|$ at the position of the shock front. The profile of the $sinc$ kernel index $n$ of model $S_{12}$ is shown in Fig.~(\ref{sedov_5}). We can see that the highest value of the exponent ($n\simeq 9.8$) is reached at $x=0.3$ behind the density peak. These results are closely followed by model $S_{10}$ ($p=0.7$ and no equalization), and, in decreasing quality, by $S_{11}$ ($p=0$ with equalization), and $S_{9}$ ($p=0$ and no equalization). 

The results above suggest that: 1) the use of the new volume elements is beneficial when handling shock-waves, 2) when using these VE, the equalization algorithm (i.e., considering $n(\mathbf{r},t)$, variable kernel exponents) is only able to enhance the results when the number of neighbors is high (as in the cases depicted in Fig.~\ref{sedov_4}).

\subsection{Collapse of an isothermal cloud}
\label{evrard}

A relevant test in Astrophysics is the numerical study of the gravitational collapse of a gaseous configuration, also known as the Evrard test \citep{evr88}. This test includes gravity and has been used many times to check hydrodynamics codes (see for example \cite{spr10a} and references therein).  

We used here the initial configuration described in \cite{spr10a} consisting of a sphere of gas with unit radius and mass. The density profile at $t=0$ is 
\begin{equation}
\rho(r)=
\begin{cases}
 M/(2\pi R^2 r) &\text{for $r \le R$}\\
 0 &\text{otherwise}
\end{cases}
\label{evrardprofile}
,\end{equation}

\noindent where $R=1$ and $M=1$. The initial configuration is assumed isothermal with $u_0=0.05$ per unit mass, obeying an ideal EOS with $\gamma=5/3$. All particles are initially at rest. For this choice the internal energy is much lower than the gravitational energy (assuming G=1) and the system collapses.

To build the initial model we have conveniently stretched a uniform grid with $N=40^3$ particles so that the profile given by Eq.~(\ref{evrardprofile}) is obtained. This procedure reproduces  well the 1D density profile, except near the surface layers. We have tracked the collapse of the sphere for two different values of the parameter $p$ in both Eq.~(\ref{estimator}) and in its smoothed counterpart, Eq.~(\ref{xrhobis}). The results at time $t=0.8$ are compared in Fig.~(\ref{evrardfig1}) not only with an accurate 1D-PPM calculation \citep{ste93}, but also with the results obtained using GADGET-2 for the same initial model. 

The evolution of models calculated using Eq.~(\ref{estimator}) with $p=0.7$, and with $p=1$ in Eq.~(\ref{xrhobis}) were rather similar, so we describe the collapse of the configuration only for the latter case. The upper panels of Fig.~(\ref{evrardfig1}) depict the density and radial velocity profiles at $t=0.8$. We see only minor differences among the three calculated models. The density profiles are almost indistinguishable. Nevertheless, a close inspection of the region around the shock front (see the zoomed region in Fig.~\ref{evrardfig1}) reveals that the model calculated with $p=1$ in Eq.~(\ref{xrhobis}) has a slightly better match to the exact solution than $p=0$ and that obtained with GADGET-2. The profile of the radial velocity is similar in all three calculations. The evolutions calculated with GADGET-2 and with SPHYNX with $p=0$ were almost identical and some minor differences may be due to the details in the implementation of the AV and gravity.

The new volume elements provide a better kernel normalization at the center and front locations (bottom-right panel in Fig.~\ref{evrardfig1}). Nevertheless, unlike in the point explosion tests, we have not seen a clear enhancement of $\vert\Delta\mathbf{r}\vert$ (Eq.~\ref{approxI}) when $p\simeq 1$ is used to compute the volume elements, although, overall, the profile is smoother (bottom-left panel).

\section{Discussion and conclusions}
\label{sec:conclusions}
In this paper, we present a new density-based SPH code, named SPHYNX, and test it in a series of traditionally problematic simulations for SPH codes in 2D and 3D. In particular, we have been able to perform a Rayleigh-Taylor simulation in a weak gravitational field $g=0.1$. Additionally, the shock-blob interaction test proved that SPHYNX can efficiently suppress the tensile instability that prevents the rise of hydrodynamical instabilities and mixing in many scenarios simulated with SPH. Additionally, the outcome of other tests, such as the hydrostatic square, Kelvin-Helmholtz instability, Gresho-Chan vortex, Sedov explosion, Noh wall-shock, Evrard collapse and Triple point-shock, prove that our implementation provides results competitive with other state-of-the-art calculations. For these problems, SPHYNX produces better results than many of the extant density-based SPH codes, being qualitatively similar to those obtained with the recently developed CRKSPH scheme  \citep{fro17}. But, unlike the CRKSPH method, our approach ensures the angular momentum conservation from the onset.

To achieve this, SPHYNX benefits from recent advances in the field and gathers together the latest methodologies to perform numerical simulations of astrophysical scenarios via the smoothed particle hydrodynamics technique. These methodologies include, as a novelty, a new generalized volume element estimator and a consistent update of the smoothing length and the sharpness of the interpolating kernel along with the particle density. Additionally, it counts with an integral approach to calculate gradients and a pairing-resistant family of interpolators. These features are summarized and discussed in the following.

The choice of non-standard volume elements to approximate the Euler integral equations as finite summations has a significant impact on the simulations. Following the works by \cite{sai13} and \cite{hop13}, who generalized the VE so that they are not necessarily the trivial $m/\rho$ choice, we postulate a new volume element which enhances the normalization of the kernel. As discussed in Sect.~\ref{choice_volel}, the VE assigned to a particle is $V_a= X_a/\sum_b X_b W_{ab}$, where $X_a = (m_a/\rho_a)^p$ is the weighting estimator of the kernel and $0\le p\le 1$ is a parameter chosen by the user. The value $p=0$ reduces the VE to $1/\sum_b W_{ab}$, which is the standard VE when the mass of the particles is the same. For $p=1$, we have $V_a = (m_a/\rho_a)/\sum_b (m_b/\rho_b) W_{ab}$ which is simply the re-normalized traditional volume element. As expected, a better kernel normalization (between a factor 2 and a factor 5) is obtained when these VE are used. A negative feature of the proposed VE is their tendency to overshoot the density estimation in the presence of sharp gradients when $p\simeq 1$. Actually, that is the fundamental reason for not taking $p= 1$ in the estimator $X_a= (m_a/\rho_a)^p$. The optimal value of $p$ depends on the particular problem at hand, but the range $0\le p\le 0.7$, explored in this work seems to be safe. Nevertheless, a most robust implementation that allows taking $p=1$ is to consider $X_a= (\langle m_a/\rho_a\rangle)^p$, where $\langle .\rangle$ is the SPH average of the magnitude. Although this last procedure requires the computation of the averages $\langle m_a/\rho_a\rangle$, it is the recommended default choice because of its robustness and ability to keep track of strong shocks and instabilities in the presence of sharp density gradients.

Another important feature is the dynamical choice of the interpolating kernel function. A large body of calculations carried out with SPH in the past made use of the $M_4$ cubic-spline function to perform interpolations. The $M_4$ polynomial has, however, a serious drawback: it is prone to the pairing-instability when the number of neighbors increases (e.g., exceeding $n_b\simeq 60$ in 3D calculations which uses the $M_4$~kernel). This is clearly a limitation, because in practical applications it is advisable to take as many neighbors as possible to reduce the $E_0$ errors in the SPH equations. A growing number of kernel candidates has been proposed during the last decade to alleviate this problem. For example, one option is to consider the natural extension of the $M_n$ family to higher polynomial degrees, such as the quartic ($M_5$) or the quintic ($M_6$) kernels. More recently, a different family of interpolators has been proposed based on the Wendland functions, as discussed in \cite{deh12}, which shows a strong endurance against the pairing-instability. A third family of interpolators, called the $sinc$ (harmonic-like) kernels, was introduced by \cite{cabezon2008}, which are also implemented in SPHYNX. As mentioned in Sect.~\ref{Sec.sinc}, the definition of the $sinc$ kernels is directly linked to that of the Dirac-$\delta$ function. Unlike the $M_n$ family, which is discrete in the index $n\in \mathbf{Z (+)}$, the $sinc$ kernels do form a continuous family, which depends on a leading exponent $n\in \Re (+)$. Actually, the $M_n$ family could be considered as a subset of the $sinc$ family \citep{garciasenz2014}. Using the $sinc$ family of kernels endows the SPH technique with a flexible engine, as the shape of the kernel can be dynamically changed, in a continuous way, during run-time. This feature can be used, for example, to suppress the pairing instability (see Sect.~\ref{pairingtest}) or to equalize the resolution behind a shock-wave (as shown in Fig.\ref{sedov_5}).

Additionally, SPHYNX estimates gradients by an integral approximation (\emph{IAD}$_0$) which is more accurate than the traditional procedure based on the analytic derivative of the kernel function, and reduces the $E0$ errors caused by the particle sampling of the fluid. We fully confirm in this work the importance of this new approach, especially for handling hydrodynamic instabilities, in agreement with previous publications \citep{garciasenz2012,cabezon2012,ros15,ros15b,val16}.

SPHYNX has been validated with several standard tests in two and three dimensions, ranging from strong shocks and subsonic fluid instabilities in boxes, to larger systems where the gravitational force takes over. From the analysis of these test cases we summarize the following conclusions. 

The use of the Integral Approach to calculate gradients along with the traditional volume elements, $V_a= m_a/\rho_a$ ($p=0$ in Eq.~\ref{estimatorXrho}) and a $sinc$ kernel with $n=5$ improves the simulation of hydrodynamics instabilities subjected to small initial perturbations with respect the standard SPH. The quantitative amplitude growth-rate of the KH instability is closer to the correct growth-rate (as computed with state-of-the-art Eulerian codes) than current density-based  SPH codes (with smaller L$_1$ errors by a factor $1.5-4$), being similar to the results of the modern PSPH formulation. It is also able to reproduce the KH instability in stratified fluids with high density contrasts ($\rho_2/\rho_1\simeq 8$). In the case of the RT instability, the scheme is also able to cope with small perturbations ($w_0=0.0025$) and tiny gravity values ($g=-0.1$), although in the latter case the non-linear evolution scarcely shows structure. In shocks, the results are similar to those provided by the standard method in identical conditions.

When the new VE are switched-on there is, in general, an increase of the quality of the simulations. We have monitored the volume normalization condition $\sum_b V_b W_{ab}=1$ in all calculated models and, without exception, it is better fulfilled (usually in the range of 10-20\% closer to unity) with the new volume elements. This change has an impact on the overall evolution of the simulation, considerably improving the results of the simulations. A paradigmatic case is the RT instability, where the use of the VE leads to an increase of the growth-rate of the instability and to a  richer evolution in the linear stage, even for the low-gravity simulation. In shock-waves, the front of the blast becomes steeper and the density peak is 10-25\% higher, even in 3D. Regarding the Sedov test, the post-shock evolution of density and pressure is 5-10\% closer to the analytic expectations. It is also worth noting that the VE also improve the condition $\vert\Delta\mathbf{r}\vert=0$ which, according to Eq.~(\ref{approxI}), is a necessary condition to exactly compute the gradient of linear functions when the $IAD_0$ scheme is used.

During the course of the simulations we did not see any sign of pairing instability, even when working with $\simeq 50$ neighbors in the 2D tests. In any case, to avoid the instability it is enough to raise the exponent of the $sinc$ kernel above the adopted default value $n=5$. We stress that, unlike other recent SPH schemes, the simulations of the KH and RT instabilities were carried out without including any artificial flux of heat or any other procedure to smooth the pressure.       

Among the several improvements left for future work, we plan to improve the calculation of gravity by including a better treatment of the gravitational softening on short distances. The best way to do that is to include the gravity into the discretized SPH Lagrangian as described in \cite{pri07} and \cite{spr10}. Also, the implementation and validation of switches to ensure that the AV is only added in regions where there are shocks \citep{cul10}, as well as noise triggers to control the velocity in subsonic flows \citep{ros15} could be done with moderate effort. A more ambitious goal would be to directly calculate the volume elements solving implicitly the equation $\sum_b V_b W_{ab}=1$ on each particle of the system. Even though the strong coupling between particles renders any implicit calculation computationally expensive, it will probably solve the density overshooting problem seen in our explicit approach.

\section*{Acknowledgments}
The authors want to thank S.~Rosswog, M.~Liebend\"orfer, and R.~K\"appeli for useful discussions and comments. We also thank M.~Steinmetz for providing the PPM calculations of the Evrard test and C.~McNally for giving us the reference model describing the growth-rate of the KH instability. This work has been supported by the Swiss Platform for Advanced Scientific Computing (PASC) project DIAPHANE and by the European Research Council (FP7) under ERC Advanced Grant Agreement No. 321263 - FISH. (R. Cabez\'on). It has also been supported by the MINECO-FEDER Spanish projects AYA2013-42762-P, AYA2014-59084-P and by the AGAUR (D. Garc\'\i a-Senz and J. Figueira). The authors acknowledge the support of sciCORE (http://scicore.unibas.ch/) scientific computing core facility at University of Basel, where some of the calculations were performed. 
  
\bibliographystyle{aa}
\bibliography{bibliography_r}

\clearpage
\begin{appendix} 

\counterwithin{figure}{section}

\section{Kernel separability}
\label{App:separability}
\begin{figure*}
\includegraphics[angle=-90,width=\textwidth]{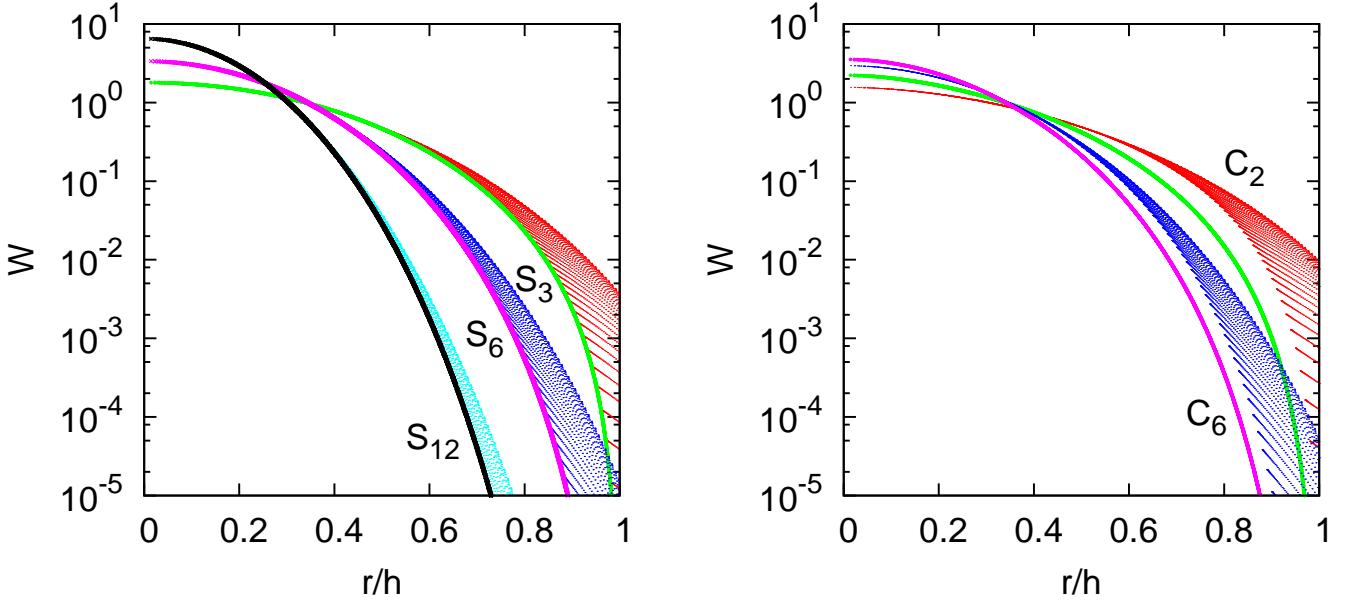}
\caption{Value of $S_n$, for indices $n=3$, $n=6$, and $n=12$ (left), and Wendland kernels $C_2$ and $C_6$ (right) in 2D (continuum thick lines). We show in dots the direct product $S_n^{1d}(x)\cdot S_n^{1d}(y)$ (left) and $C_n^{1d}(x)\cdot C_n^{1d}(y)$ (right) for the same set of indices. Increasing $n$, the $sinc$ kernels approach separability in each spatial direction, while Wendland kernels do not.}  
\label{figure1}
\end{figure*}

Spherically symmetric kernels sacrifice tensorial features in order to preserve the second-order accuracy. The notable exception is the Gaussian kernel, which is both symmetric and separable in each axis direction. But, unlike the Gaussian function, the compact-supported spherically symmetric kernels are not separable, hence neither is the $sinc$ kernels ($S_n$) used in this work. Nevertheless, the $S_n$ family approaches separability as $n\rightarrow\infty$. This property is shown in Fig.~(\ref{figure1}) (left panel) which depicts the profiles of $W_{2d}^S(\sqrt {x^2+y^2},h,n)$ (continuum lines) and that of the direct product $W_{1d}^S~(x,h,n)\cdot W_{1d}^S~(y,h,n)$ (dotted lines) for $n=3$, $n=6$, and $n=12$, in green, pink, and black, respectively. As we can see, the dispersion in the kernels product becomes narrower as the kernel index $n$ increases. As a comparison, we show the same profiles for some Wendland kernels in Fig.~(\ref{figure1}) (right panel). In this case, it is clear that Wendland kernels are not separable as the product of the unidimensional components departs from the direct multidimensional calculation. Furthermore, the result does not improve as we increase the order of the Wendland kernel. The study of the consequences that this property may have is beyond the scope of this paper, but it is undoubtedly worthy to explore in future works.

\section{A better partition of unit improves the estimation of the gradient of a linear function}
\label{App:demo}

We  consider the following integral in 1D:

\begin{equation}
I_0=\int_{-\infty}^{+\infty}~x~W(x,h)~dx
\label{integral1}
.\end{equation}

Because the integrand is an odd function, $I_0$ is trivially equal to zero. Nevertheless, approaching $I_0$ with finite summations does not necessarily ensure $I_0=0$. We know that a necessary and sufficient condition to exactly reproduce the gradient of a linear function in SPH is that $I_0=\sum_b  V_b~x_b~W_{ab}=0$, (Eq. \ref{approxI}), where $V_b$~is the volume element of particle $b$.

Integrating $I_0$ by parts,

\begin{equation}
I_0= \left[x~G(x,h)\right]_{-\infty}^{\infty}-\int_{-\infty}^{+\infty} G(x,h)~dx,
\label{integral2}
\end{equation}

\noindent
where $G(x,h)$~is the primitive integral of $W(x,h)$: $G(x,h)=\int W(x,h)~dx$.

We now consider a very simple kernel $W(x,h)$ whose primitive $G(x,h)$ can be obtained analytically:
\begin{equation}
W(x,h)=\frac{C}{h}\exp(-\vert x\vert/h).
\label{kernelexp}
\end{equation}

This kernel \citep{fulk96} is spherically symmetric but currently it is not the preferred one because it is not a flat-top kernel. The normalization constant in 1D is $C=0.5$. Nevertheless, it is an adequate kernel for this proof, as it admits an analytical primitive,

\begin{equation}
G(x,h):
\begin{cases}
G_1(x,h)=C\exp{(x/h)}~; x< 0 \\
G_2(x,h)=-C\exp{(-x/h)}~;  x\geq 0
\end{cases}
.\end{equation} 


The first term of the RHS  of Eq.~(\ref{integral2}) is zero, because $G(x,h)$ vanishes very quickly when $x/h$ increases. For spherically symmetric kernels $I_0=0$, thus the integral giving the second term should also be zero.
The integral $I_0$~can be estimated as
$I_0=I_1+I_2$~where,

\begin{equation}
I_1=-\int_{-\infty}^{0} G_1(x,h)~dx=-\int_{-\infty}^{0} C\exp{(x/h)}~dx= -h\int_{-\infty}^0 W(x,h)~dx
\label{I1}
,\end{equation}

\begin{equation}
I_2=-\int_{0}^{+\infty} G_2(x,h)~dx=\int_{0}^{+\infty} C\exp{(-x/h)}~dx=h\int_{-\infty}^0 W(x,h)~dx
\label{I2}
,\end{equation}

\noindent
with $W(x,h)$~given by Eq.~\ref{kernelexp}. The analytical calculation of these integrals gives $I_1=-Ch$, $I_2=+Ch$, and $I_0 = I_1+I_2 = 0$, which is obvious because Eq.~\ref{integral1} is an odd function. Things are different when we approach these integrals by finite summations:

\begin{equation}
\begin{split}
I_0=\sum_b V_b x_b W_{ab}(x_b,h_a) = - h_a\sum\limits_{b,~x<0} V_b W_{ab}(x_b,h_a)+ \\
h_a\sum\limits_{b,~ x\geq 0} V_b W_{ab}(x_b,h_a)
\label{integral0}
\end{split}
.\end{equation}

On the other hand, if the $V_b$ are well chosen, they should fulfill the equipartition condition:
\begin{equation}
\sum_b V_b~W_{ab}(x_b,h)=1
\label{norm}
.\end{equation}

Now we make the reasonable ansatz that improving $V_b$ is actually enhancing each term inside the summation in Eq.~\ref{norm}. Thus,

\begin{equation}
\sum\limits_{b,~x<0} V_b W_{ab}= \sum\limits_{b,~x\geq 0} V_b W_{ab}=0.5 
,\end{equation}

\noindent
and, according to Eq.~\ref{integral0}, $I_0$ vanishes (or, in any case, it becomes very small) even when it is approached using summations. We note that this proof is strictly valid only for exponential kernels. Nevertheless, the main conclusion also holds (at least qualitatively) for other centrally-peaked, spherically-symmetric interpolating functions. The positive feedback between $IAD_0$ and the VE is also supported by many of the test cases presented in this work.

\section{Partition of the unity: Convergence rate of the estimator $\mathbf X= (m/\rho)^p$}
\label{App:convergence}
\begin{figure*}
\includegraphics[angle=-90,width=\textwidth]{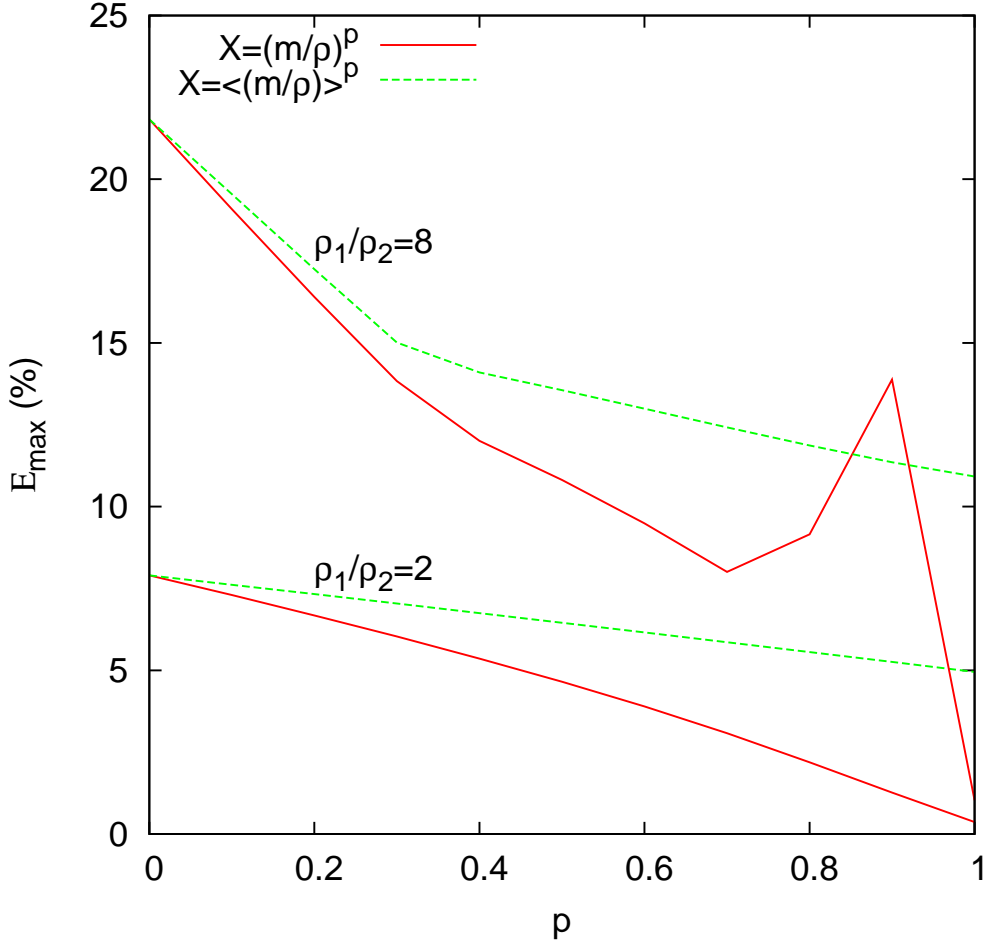}
\caption{Static shock-tube problem described in the Appendix~\ref{App:convergence}: Maximum relative error in the partition of unit as a function of the exponent $p$~used to infer the particle volume. The red solid line is for $X=(m/\rho)^p$~and the green dashed line is for $X=(\langle m/\rho\rangle)^p$.}
\label{AppendixFig1}
\end{figure*}

We discuss here a simple 2D static test which explores the effect of changing the exponent $p$ of $X_a= (m_a/\rho_a)^p$ into the normalization of the kernel. We considered a shock-tube filled with a two-density fluid separated by a contact discontinuity. The jump of the density across the frontier is estimated for several values of $0\le p\le 1$. The SPH particles were arranged in an ordered grid according to the density value at both sides of the discontinuity. The contact between both fluids was not smoothed. We have considered two density ratios: $\rho_1/\rho_2=2$ and $\rho_1/\rho_2=8$, and calculated the magnitude $\sum_b V_b W_{ab}$ for different values of $p$. The results are depicted in Figure(\ref{AppendixFig1}) which shows the maximum relative error in the kernel normalization $\vert\sum_b V_b W_{ab}-1\vert$ as a function of $p$, the density contrast and the particular procedure to estimate $X_a$, either $X_a=(m_a/\rho_a)^p$ (continuum red lines) or $X_a=(\langle m_a/\rho_a\rangle)^p$ (green dashed lines), where $\langle .\rangle$ is the SPH average. The smoothing length was adjusted so that there were $n_b=50$ neighbors contributing to the summations. 

For not too large density ratios, raising the exponent $p$ leads to a linear improvement in kernel normalization. The convergence rate is slower when $X_a=(\langle m_a/\rho_a\rangle)^p$, as expected. Things are different for large density ratios where the convergence rate follows a parabolic line with a local minimum around $p\simeq 0.7$~ (uppermost red line in Figure \ref{AppendixFig1}). Above that value it seems that any increase of $p$ makes the convergence worse. The reason is that, in static configurations with sharp contact discontinuities, the density under/over-shoots at the two sides of the discontinuity. Taking both a large density ratio and a high $p$~value induces a catastrophic feedback between the overshooting and $X_a$. Such behavior is due to the explicit nature of the implementation of the volume elements in SPH and would probably be overcome using an implicit approach. A simple recipe to circumvent the problem is to take $X_a=(\langle m_a/\rho_a\rangle)^p$, which leads to a slower, albeit safer, convergence rate (uppermost green line in Figure \ref{AppendixFig1}).

\section{Included physics and alternative formulations of the energy equation.}
\label{App:energyequation}

Most astrophysical applications need to calculate gravity, and to that extent SPHYNX incorporates an octal-tree structure \citep{her89} with several levels of particle clustering. When the particle inter-distance $r_{ab}$ is shorter than $h_a+h_b$ we apply a simple smoothing to the gravitational force,

\begin{equation}
\mathbf{g}_a= -G\frac{m_b}{(h_a+h_b)^3}\mathbf{r}_{ab}
\label{grav1}
.\end{equation}  

This softening usually gives satisfactory results but it is not fully compatible with the Euler-Lagrange derivation of the SPH equations \citep{pri07}. 

SPHYNX also incorporates a thermal conductive transport equation compatible with the IAD formulation. That equation was described and checked in \cite{cabezon2012} and it is reproduced here for completeness,

\begin{equation}
\frac{du_a}{dt}=\sum_{b=1}^{n_b}\frac{m_b}{\rho_a~\rho_b}\frac{(\kappa_a+\kappa_b)(T_b-T_a)}{r_{ab}^2}\sum_{i=1}^d (x_{i,a}-x_{i,b})~\widetilde{\mathcal A}_{i,ab}\,, 
\label{conduciad}
\end{equation}
\noindent where $d$ is the dimension of the space and the tilde symbol means the arithmetic average of the magnitude.

In high density plasmas with finite temperature, which characterize compact objects such as white dwarfs and neutron stars, it is often preferable to directly compute the temperature instead of the internal energy. SPHYNX can switch the energy equation, Eq.~(\ref{EnergyEq}), to the temperature equation, 

\begin{align}
C_{v_a}\left(\frac{dT}{dt}\right)_a=&\sum_{b=1}^{n_b}\sum_{i=1}^d (v_{i,a}-v_{i,b})\left(\frac{T_aX_aX_b}{m_a\Omega_a~\kappa_a^2}~\left(\frac{\partial P}{\partial T}\right)\mathcal{A}_{i,ab}(h_a,n_a)\right) \nonumber\\
+&\frac{1}{2}\sum_{b=1}^{n_b}\sum_{i=1}^d (v_{i,a}-v_{i,b})~a_{i,a}^{AV}\,,
\label{TempEq}
\end{align}

\noindent
where $C_v$ is the specific heat. This formulation of the temperature equation also leads to an almost perfect energy conservation. The internal energy can be obtained through the EOS when necessary. Eq.~(\ref{TempEq}) was recently used to calculate explosion models of Type Ia supernova with the SPHYNX code \citep{gar16}.

Also, the energy equation may be substituted without much effort by an entropy equation which ensures perfect entropy conservation in pure adiabatic fluxes \citep{spr02}\footnote{Nevertheless, the energy equation (\ref{EnergyEq}) also conserves entropy if the smoothing length is self-consistently calculated with density using Eq.~(\ref{NRg1}) }. An isentropic flow is characterized by 

\begin{equation}
u_a(\rho_a)= A_a\frac{\rho_a^{\gamma-1}}{\gamma-1}
\label{entropy1}
,\end{equation}

\noindent
where $A_a$ is a constant determined by the initial conditions of the flux. The entropy equation was incorporated as default in the GADGET-2 code\citep{spr05}, bringing excellent results for ideal EOS. However, it has to be adapted to more complex and realistic EOS where the adiabatic index $\gamma$ can be time-dependent and not so straightforward to know.     

\end{appendix}

\clearpage

\begin{figure*}
\includegraphics[width=\textwidth]{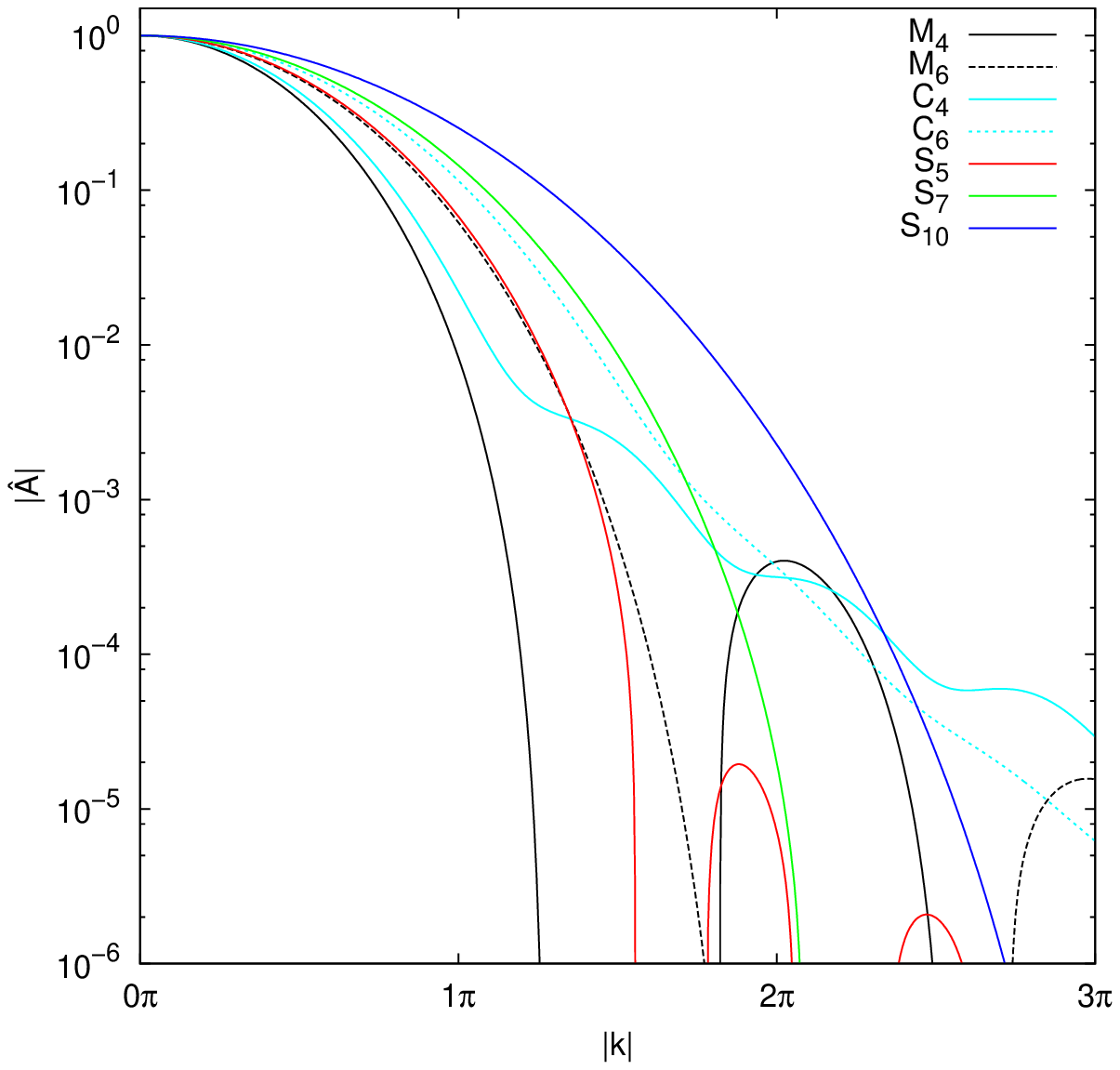}
\caption{Fourier transform for the spline kernel $B_4$ and $B_6$ (black lines), the Wendland kernels $C_4$ and $C_6$ (light blue lines), and the $sinc$ kernels $S_5$, $S_7$, and $S_{10}$. Only the positive part of the transform is shown.}   
\label{figure1.1}
\end{figure*}

\clearpage
\begin{figure*}
\includegraphics[angle=-90,width=\textwidth]{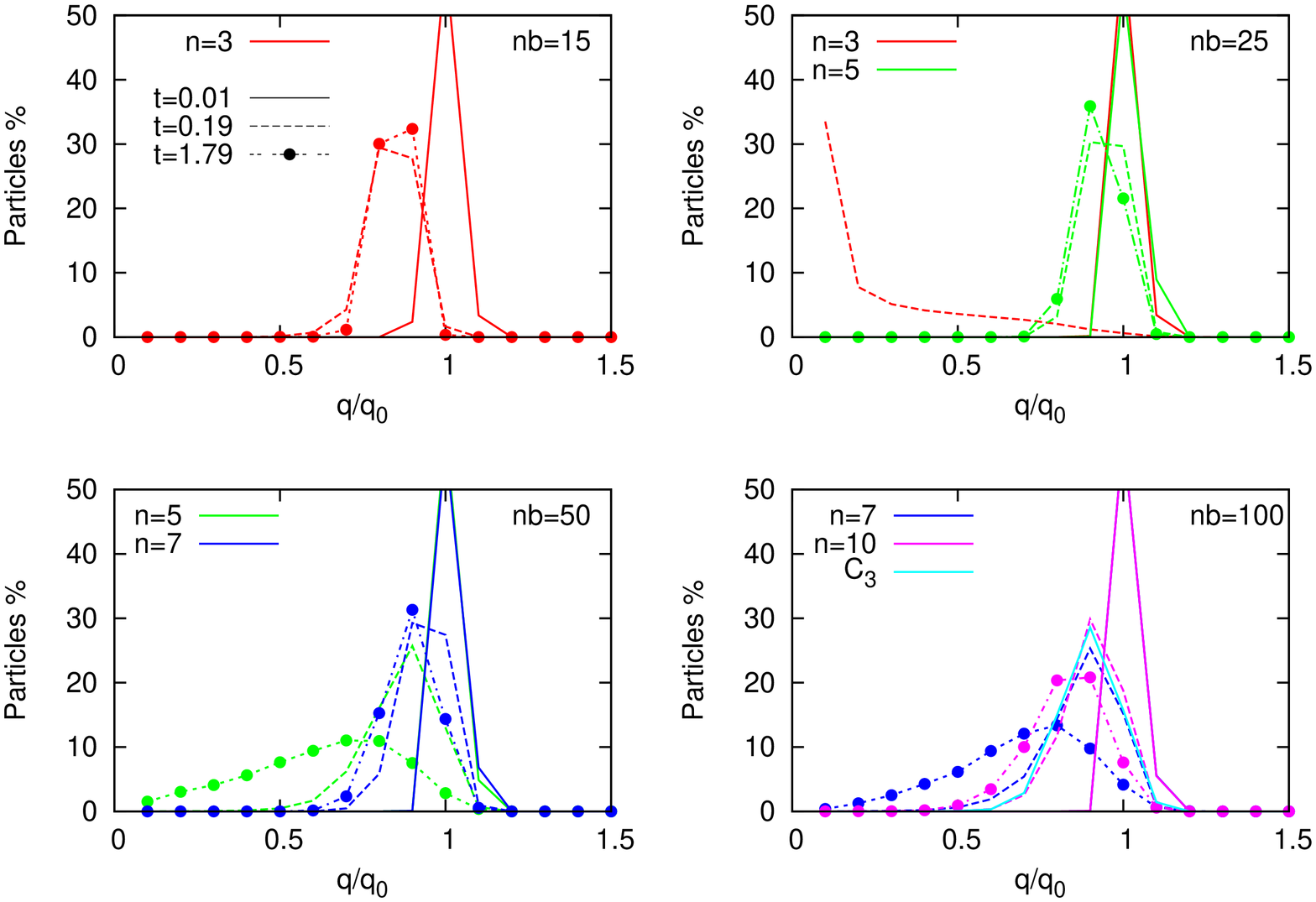}
\caption{Impact of changing the $sinc$ kernel index $n$ and number of neighbors $n_b$ on the pairing-instability during the relaxation towards equilibrium of a perturbed homogeneous system. In the X-axis the minimum interparticle distance normalized to its initial value is represented. The Y-axis shows the corresponding percentage of particles at different times. We show the results for $n=3$ (red), $n=5$ (green), $n=7$ (blue) and $n=10$ (pink). Continuum lines are for $t=0.01$, dashed lines for $t=0.19$ and big dots for $t=1.79$. The continuum line in light blue for $n_b=100$ was calculated using the Wendland $C_3$ interpolator and shows the results at $t=0.19$.}
\label{figure2}
\end{figure*}

\clearpage
\begin{figure*}
\includegraphics[angle=-90,width=\textwidth]{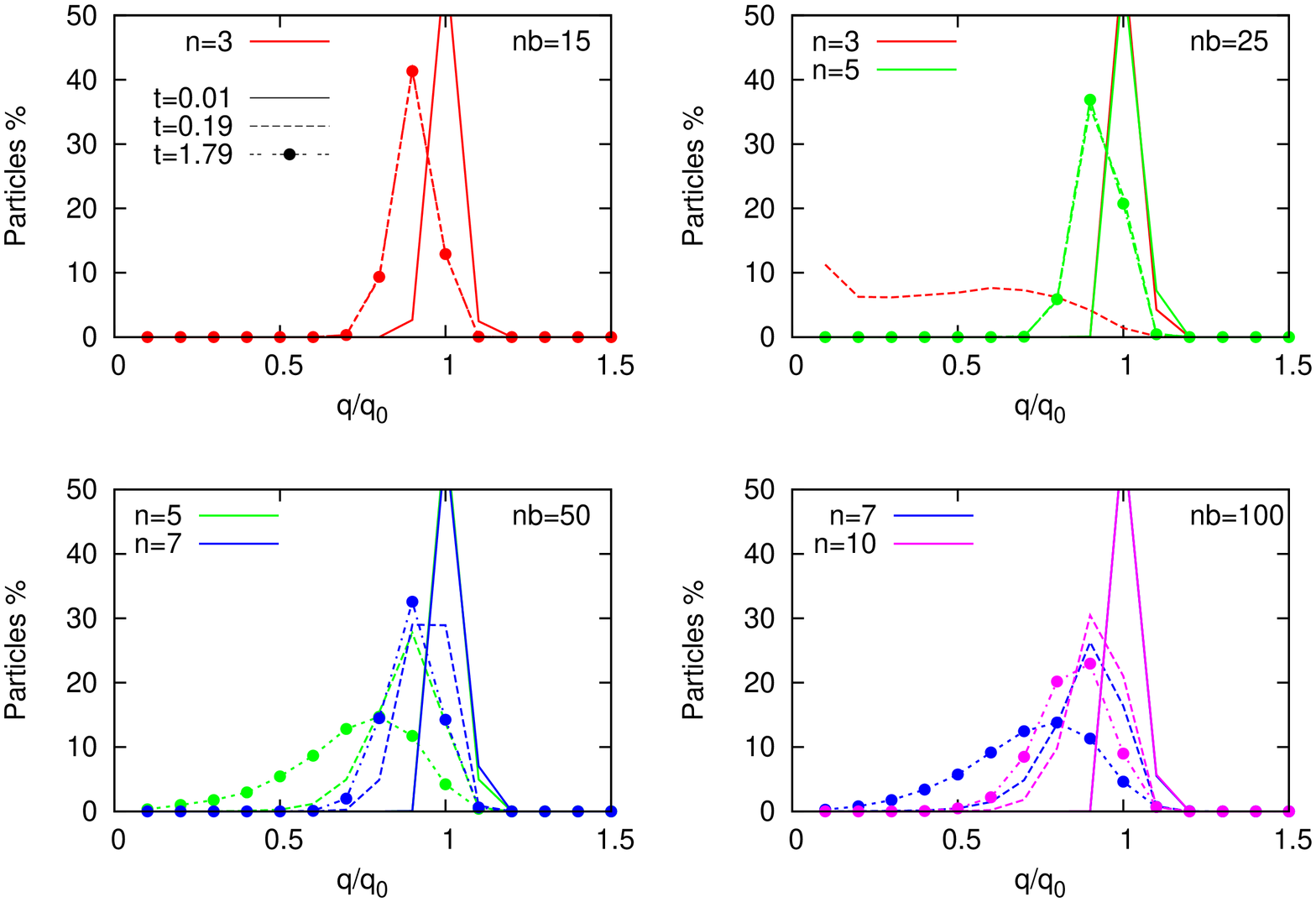}
\caption{Same as Fig.~(\ref{figure2}) but using \emph{IAD}$_0$ instead of the standard gradient estimator. In this case, particle distributions peak at higher $q/q_0$ ratios compared to those in Fig.~(\ref{figure2}), showing more resistance to particle clumping.}
\label{figure3}
\end{figure*}

\clearpage
\begin{figure*}
\includegraphics[width=\textwidth]{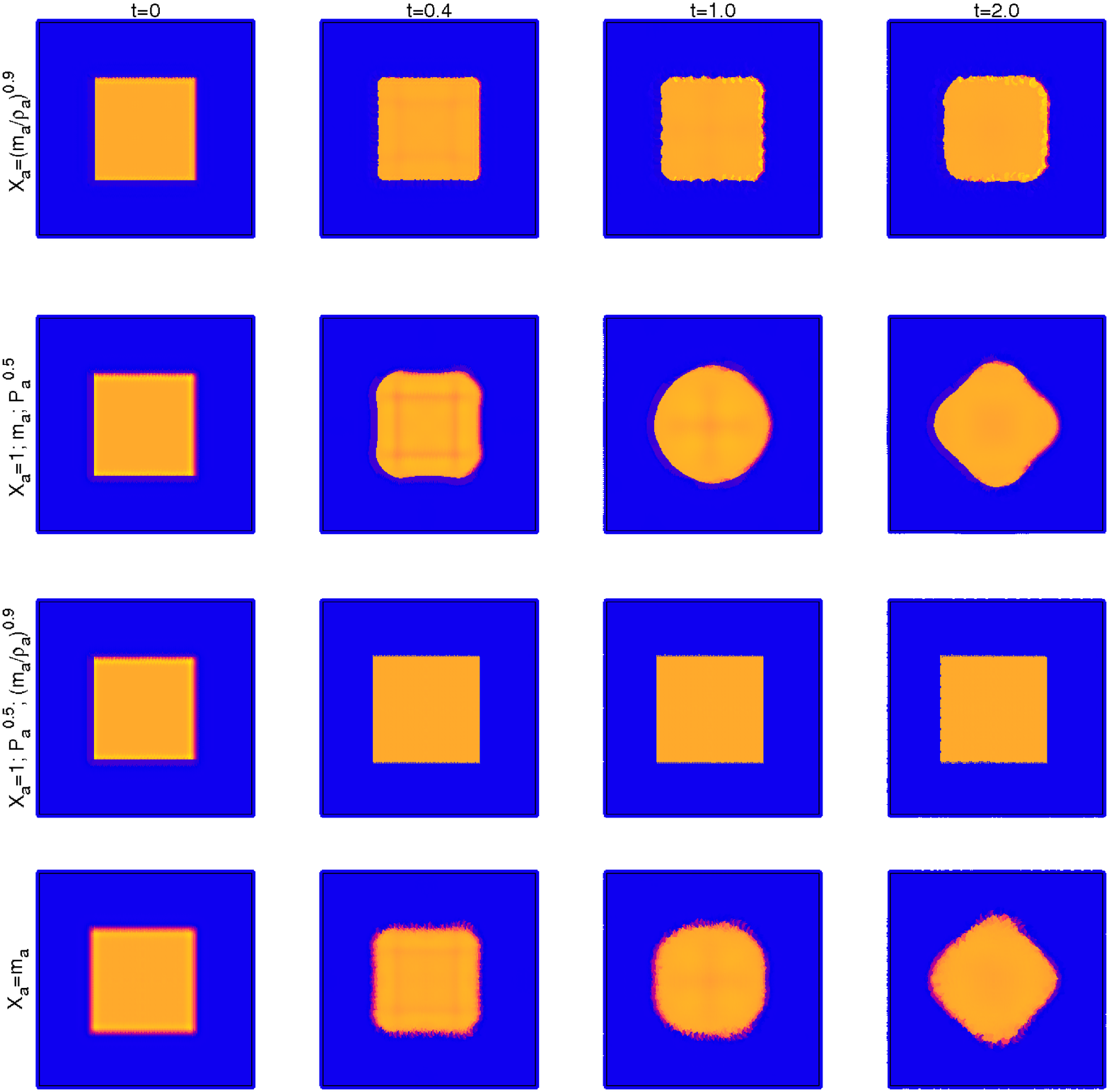}
\caption{The unmixed isobaric two-fluid test at times t=0, t=0.4, t=1 and t=2 (columns from left to right) calculated with different volume elements. The initial model is that of two nested squares with density contrast of four (the one with higher density being the inner yellow square). First and second rows use particles with identical mass but unevenly spaced grid, while the opposite applies for the two last rows. The first row depicts the evolution with the volume estimator $X_a=(\frac{m_a}{\rho_a})^p$ ($p=0.9$). The second row is the same, but using either $X_a=1$, $m_a$, and $P_a^k$ $(k=0.5)$ (all three giving similar results). The evolution shown in the third row was calculated using either $X_a=1, P_a^{0.5}$, and $(\frac{m_a}{\rho_a})^{0.9}$, while the outcome using $X_a=m_a$ is shown in the fourth row.}
\label{volumeelements1}
\end{figure*}

\clearpage
\begin{figure*}
\includegraphics[angle=-90,width=\textwidth]{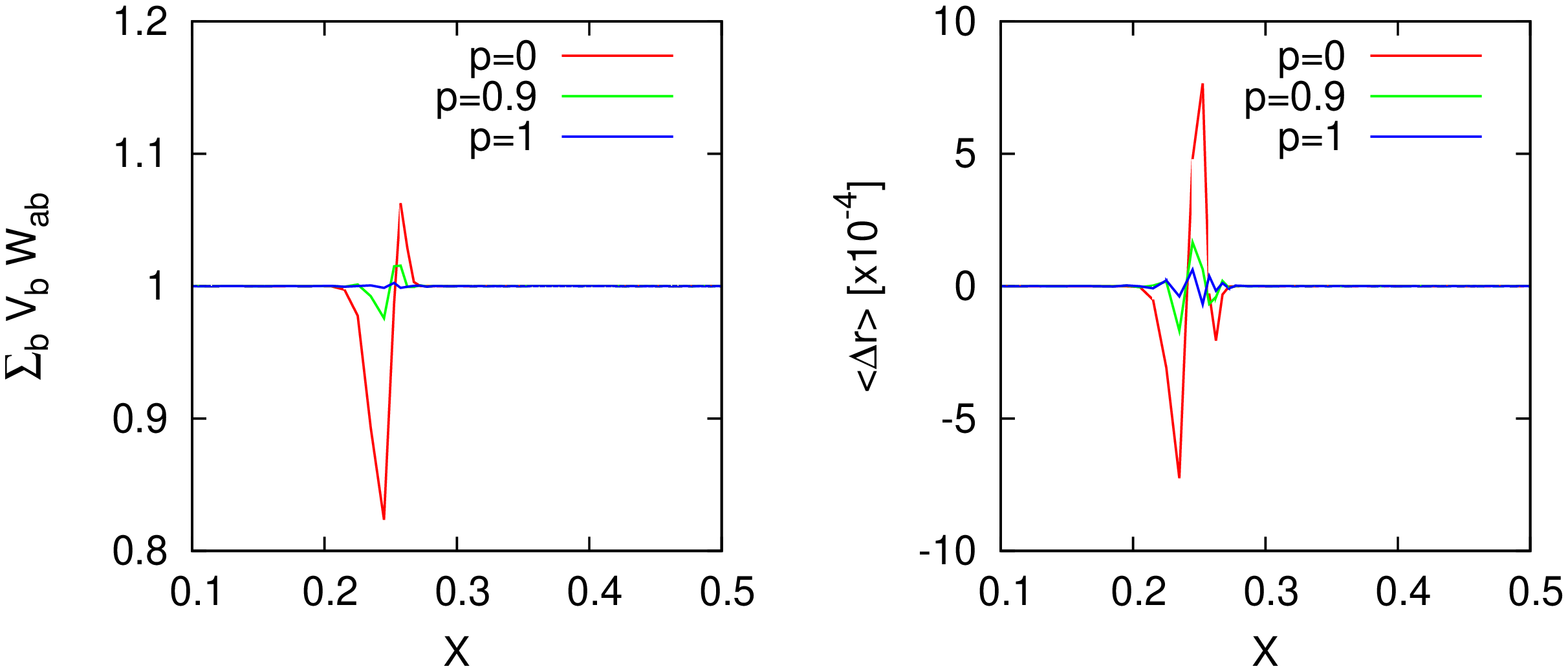}
\caption{Impact of $X_a=(m_a/\rho_a)^p$ in both the volume normalization, $\sum_b V_b W_{ab}=1$, (left) and  $\langle\Delta\mathbf{r}\rangle =0$ condition (right) for different values of the exponent. The figure depicts the profile of these magnitudes along a 1D cut taken around $y=0.5$ in the hydrostatic system shown in Fig.~(\ref{volumeelements1}). We notice how the jump at the contact discontinuity is largely reduced in both conditions when $p\simeq 1$.}
\label{volumeelements2}
\end{figure*}

\clearpage
\begin{figure*}
\includegraphics[angle=-90,width=\textwidth]{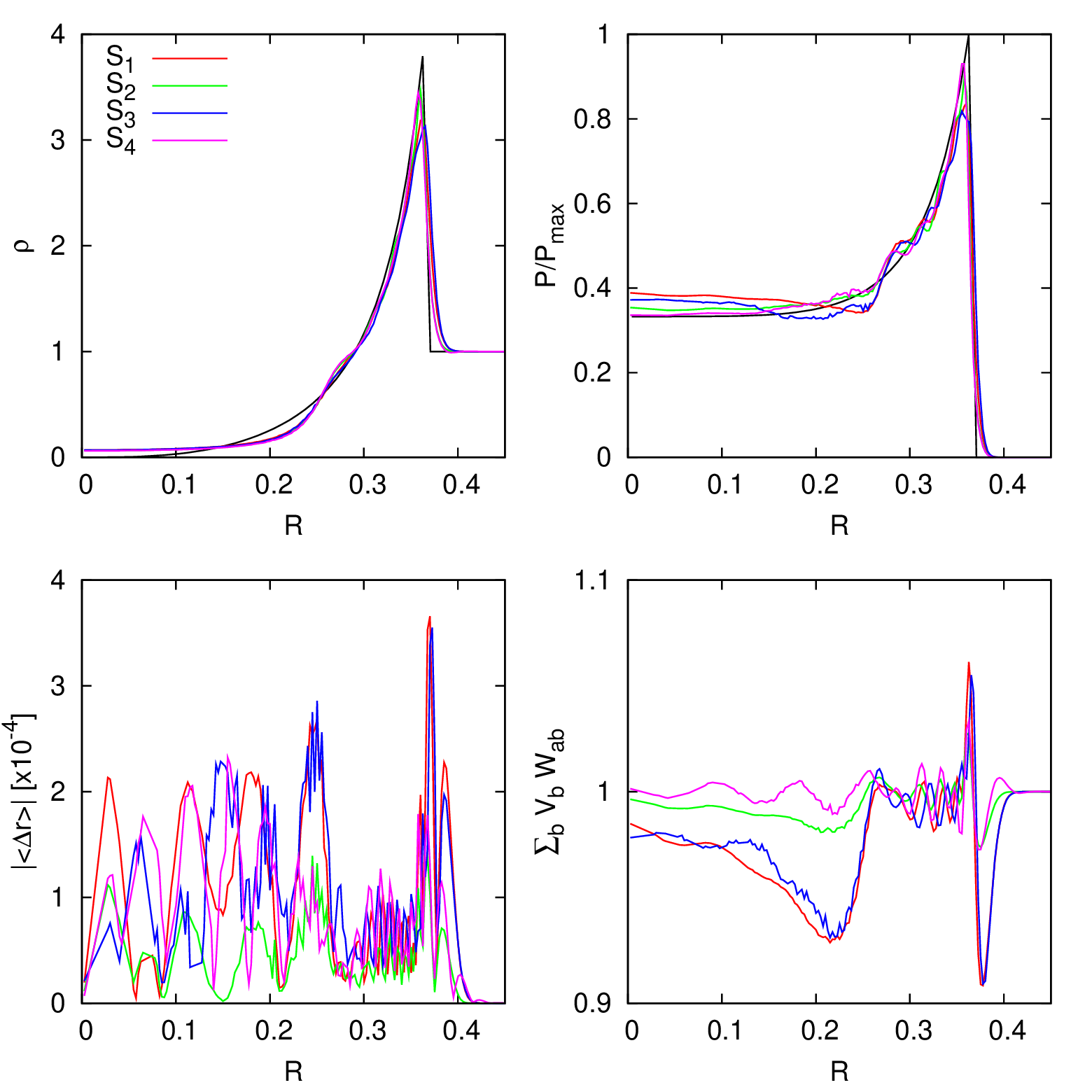}
\caption{Results for the 2D Sedov test at $t=0.1$. $S_1$ corresponds to $p=0$ (standard VE) and $\Delta n=0$ (no equalization). $S_2$ shows the effect of the VE with $p=0.7$~in $X_a=(\frac{m_a}{\rho_a})^p$~ and no equalization, while $S_3$ corresponds to $p=0$ and equalization with $\Delta n=5$. The profile $S_4$ shows the outcome when $p=1.0$~in $X_a=(\langle\frac{m_a}{\rho_a}\rangle)^p$~ and $\Delta n=0$. We show here the radial profiles of density (upper-left), normalized pressure (upper-right), and the fulfillment of the conditions $|\langle\Delta\mathbf{r}\rangle|=0$ (bottom-left) and $\sum_b V_b W_{ab}=1$ (bottom-right).}
\label{sedov_1}
\end{figure*}

\clearpage
\begin{figure*}
\includegraphics[angle=-90,width=\textwidth]{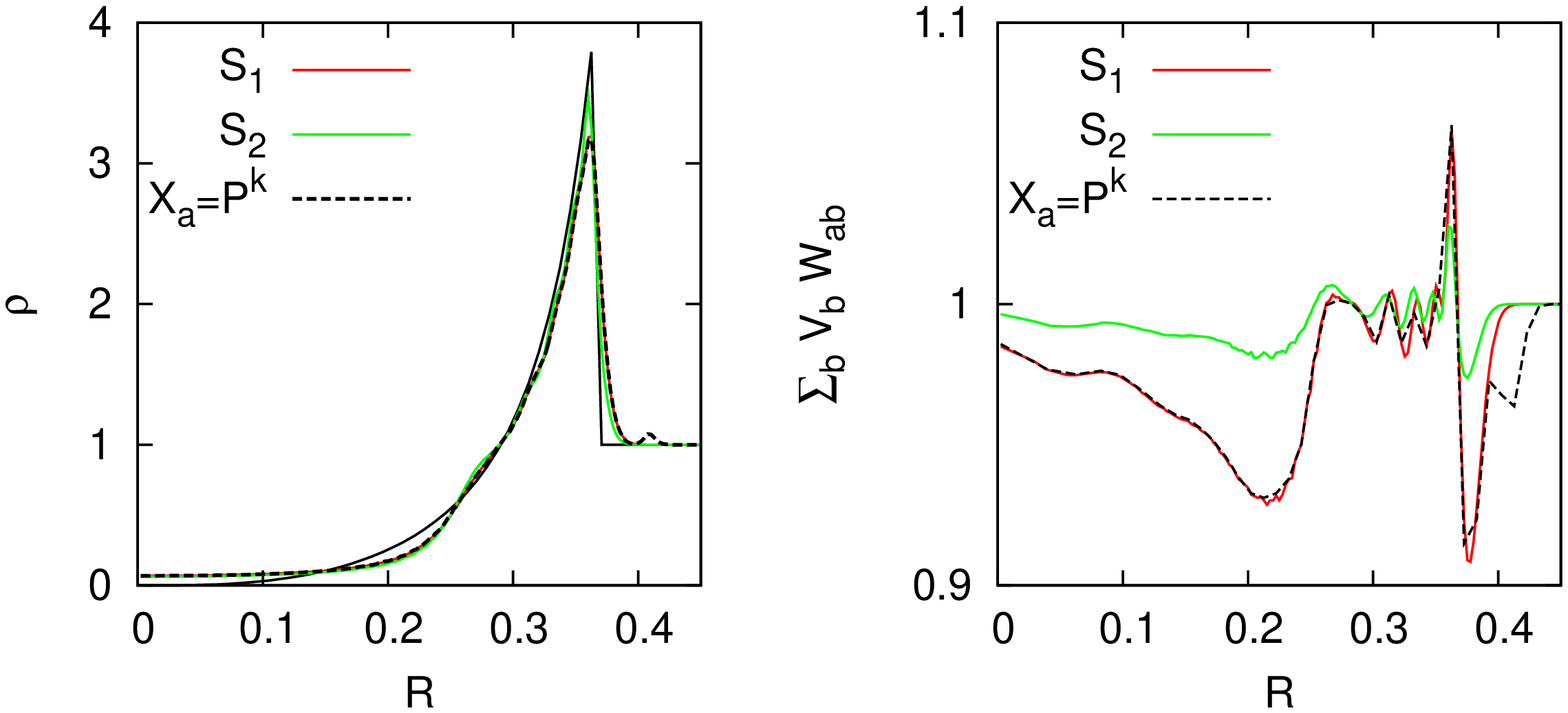}
\caption{Results for the 2D Sedov test at $t=0.1$ taking $X_a=P^k$, where P is the pressure and $k=0.05$ (black dashed line) compared to models $S_1$ (red) and $S_2$ (green). We show here the radial profiles of the density (left panel) and the equipartition condition (right panel). Solid black line is the analytic solution for the density profile. We note on the left panel that there is a small precursor bump in density for the calculation with $X_a=P^k$ which, for models $S_1$ and $S_2$, is absent.}
\label{sedov_2}
\end{figure*}

\clearpage
\begin{figure*}
\includegraphics[angle=-90,width=\textwidth]{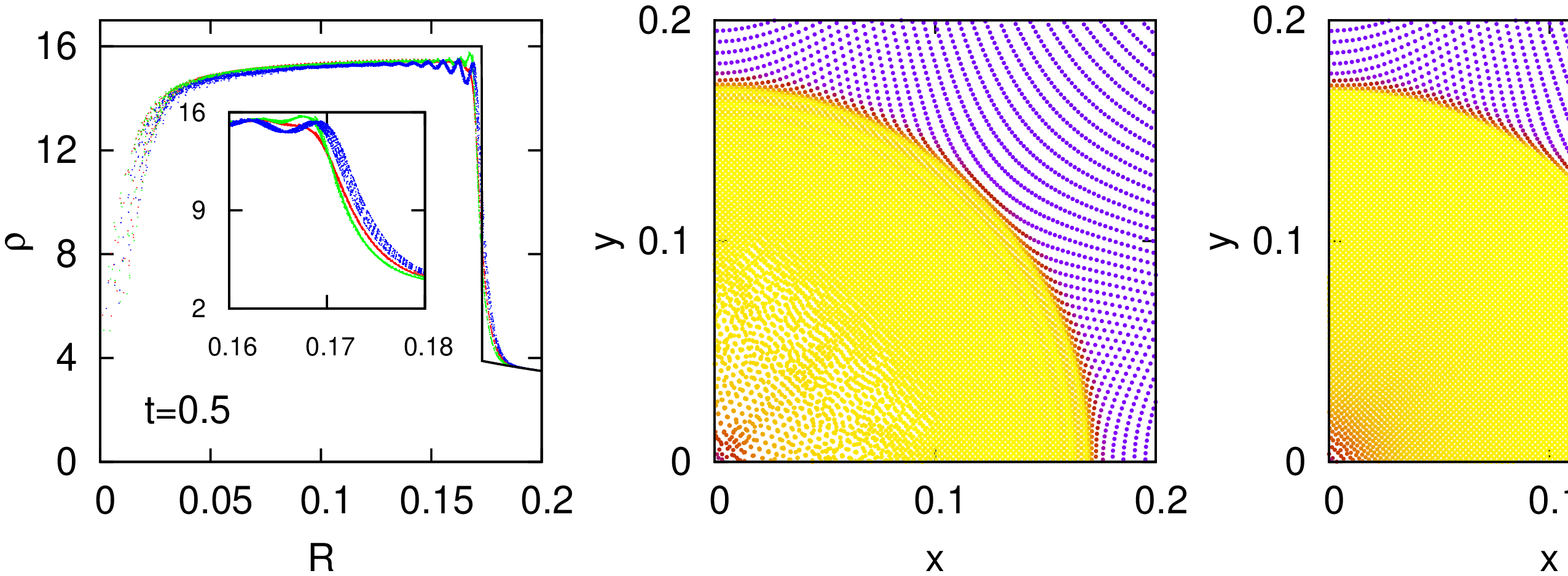}
\caption{Two-dimensional wall-shock test at time t=0.5. Left: density profiles for models described in Table~\ref{table4}: $WHS_1$ (red), $WHS_2$ (green), and $WHS_G$ (blue) in dots. The analytic result is the solid black line. The central panel depicts the particle distribution with density as color at $t=0.5$ as obtained with GADGET-2 whereas the last panel is that of model $WHS_2$. We note how $WHS_2$ achieves lower particle disorder in the central region, and smaller density oscillations at the shock position than GADGET-2.} 
\label{noh_1}
\end{figure*}

\clearpage
\begin{figure*}
\includegraphics[angle=-90,width=\textwidth]{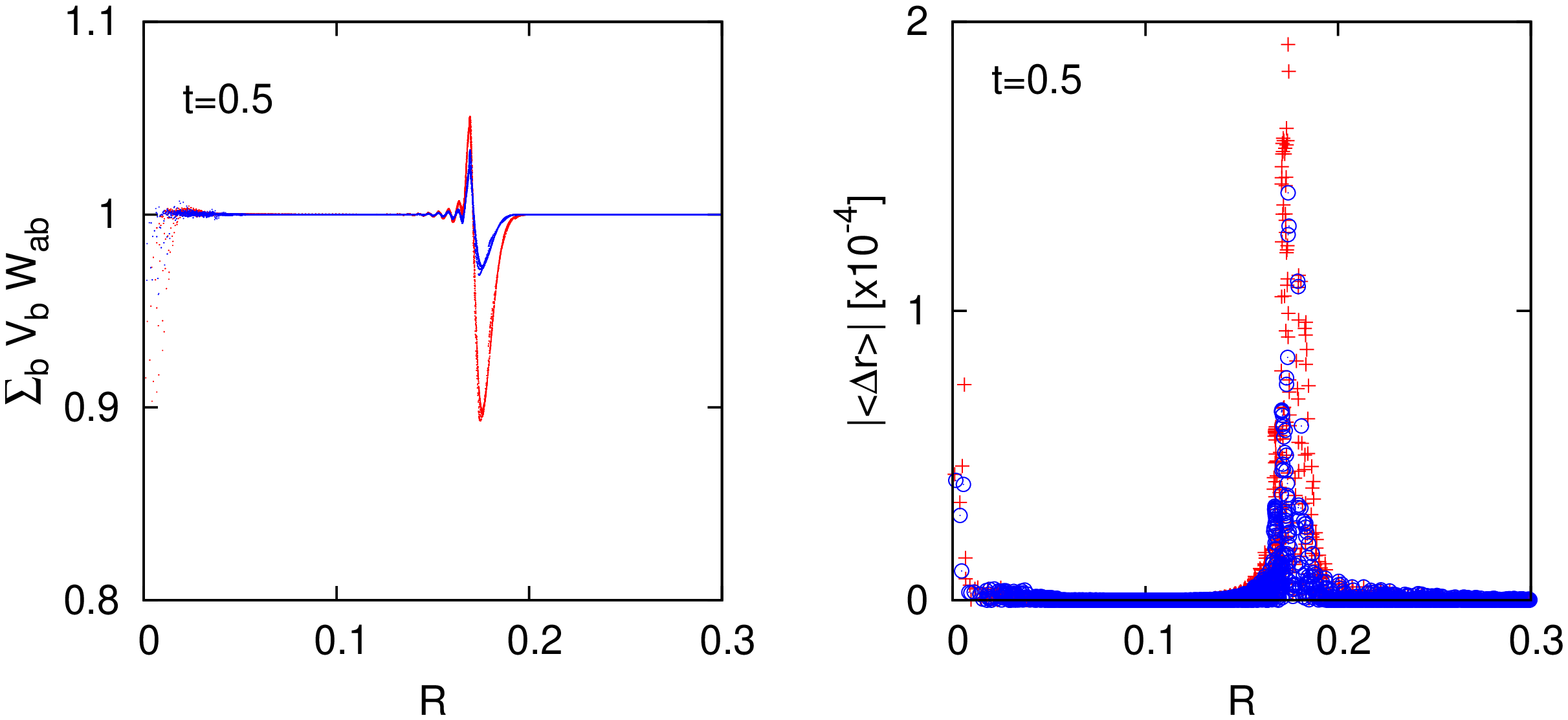}
\caption{Two-dimensional wall-shock test. Left: Volume normalization at t=0.5 for model $WHS_1$ (red) and $WHS_2$ (blue). Right: same but for the condition, $\vert\langle\Delta\mathbf{r}\rangle\vert=0$, among mass points within the kernel range of $2h$. $WHS_2$ (with new VE) fulfills both conditions better than $WHS_1$ (with the standard VE).}
\label{noh_2}
\end{figure*}

\clearpage
\begin{figure*}
\includegraphics[width=\textwidth]{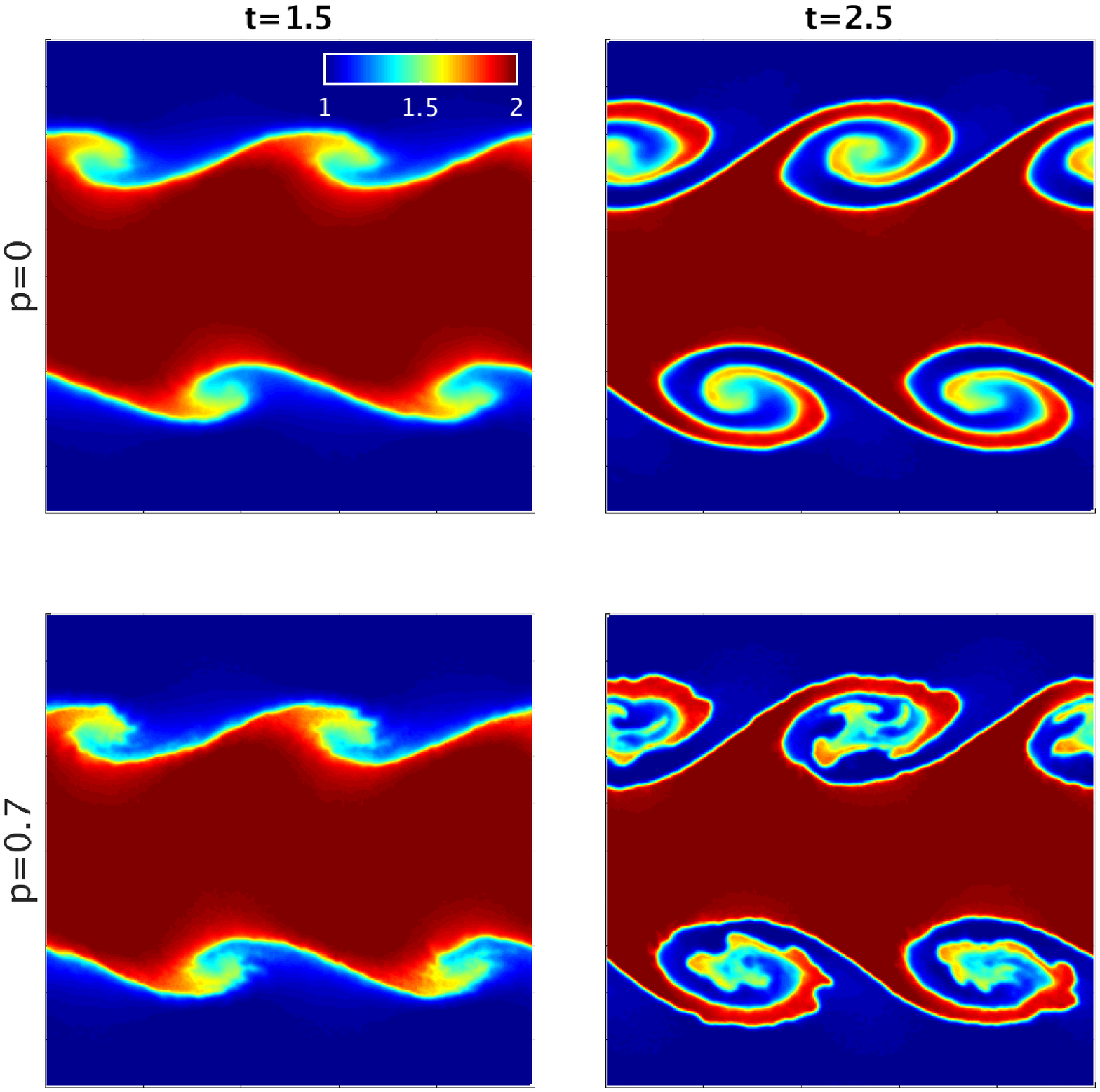}
\caption{Density color-map depicting the evolution of the Kelvin-Helmholtz instability for models $KH_1$ ($p=0$, upper panels) and $KH_2$ ($p=0.7$, lower panels) of Table~\ref{table5}. Each snapshot gives the density color-map at times $t=1.5$ (left) and $t=2.5$ (right).}
\label{kh_1}
\end{figure*}

\clearpage
\begin{figure*}
\includegraphics[angle=-90,width=\textwidth]{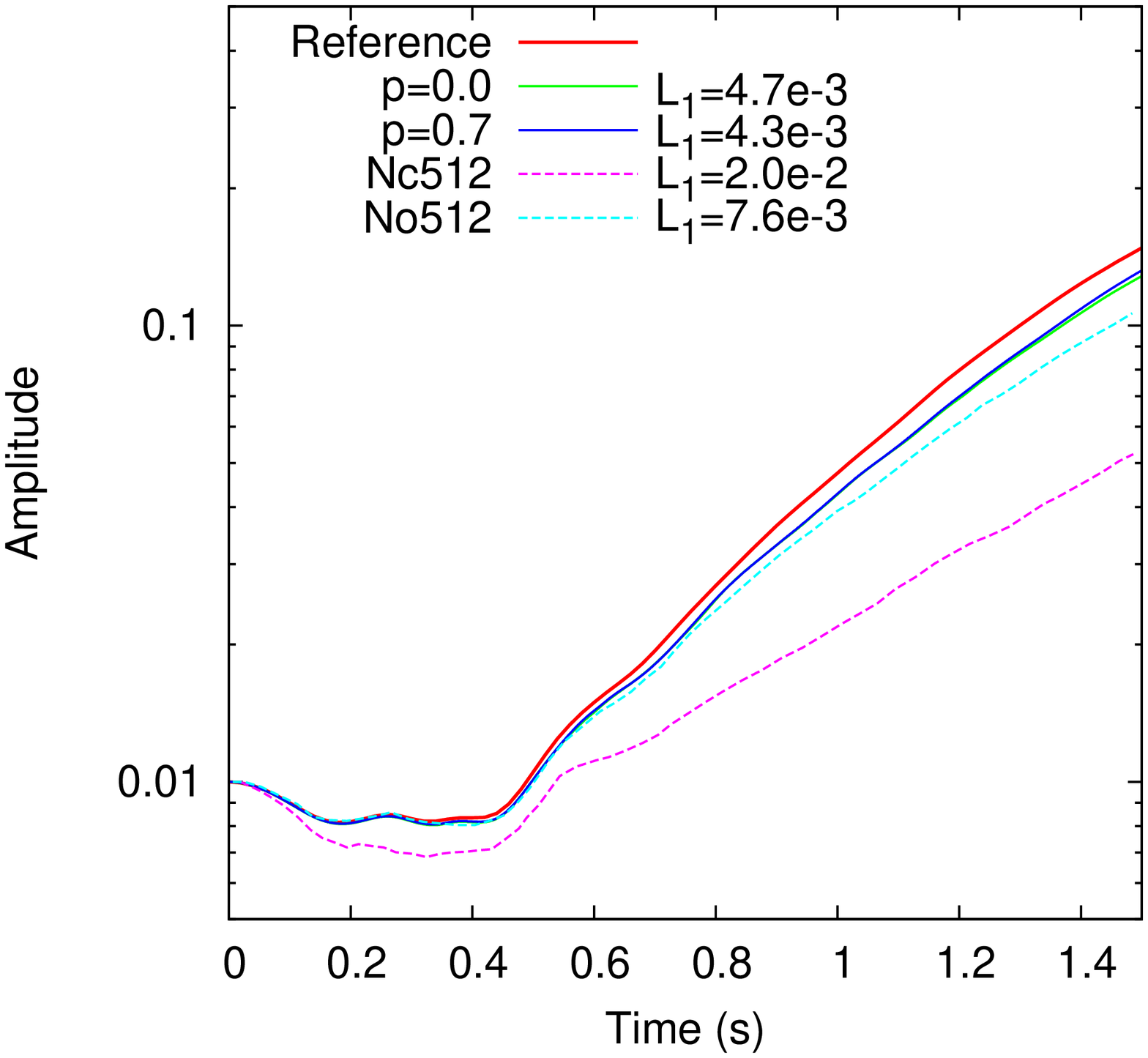}
\caption{Amplitude mode growth of models  $KH_1$ ($p=0$) and $KH_2$ ($p=0.7$) of Table~\ref{table5}. The solid red line is the reference solution \citep{mcn12} obtained with the PENCIL code \citep{brandenburg2002}, while the solid green and blue lines were obtained with SPHYNX for two different values of the $p$ parameter. The dashed lines correspond to calculations performed with the NDSPMHD code \citep{pri12} with a resolution of $512^2$ and kernel $B_4$ (pink) and $B_6$ (light blue). See \cite{mcn12} for further details. We also show the L$_1$ errors with respect to the reference values for each calculation.}
\label{kh_2}
\end{figure*}

\clearpage
\begin{figure*}
\includegraphics[angle=-90,width=\textwidth]{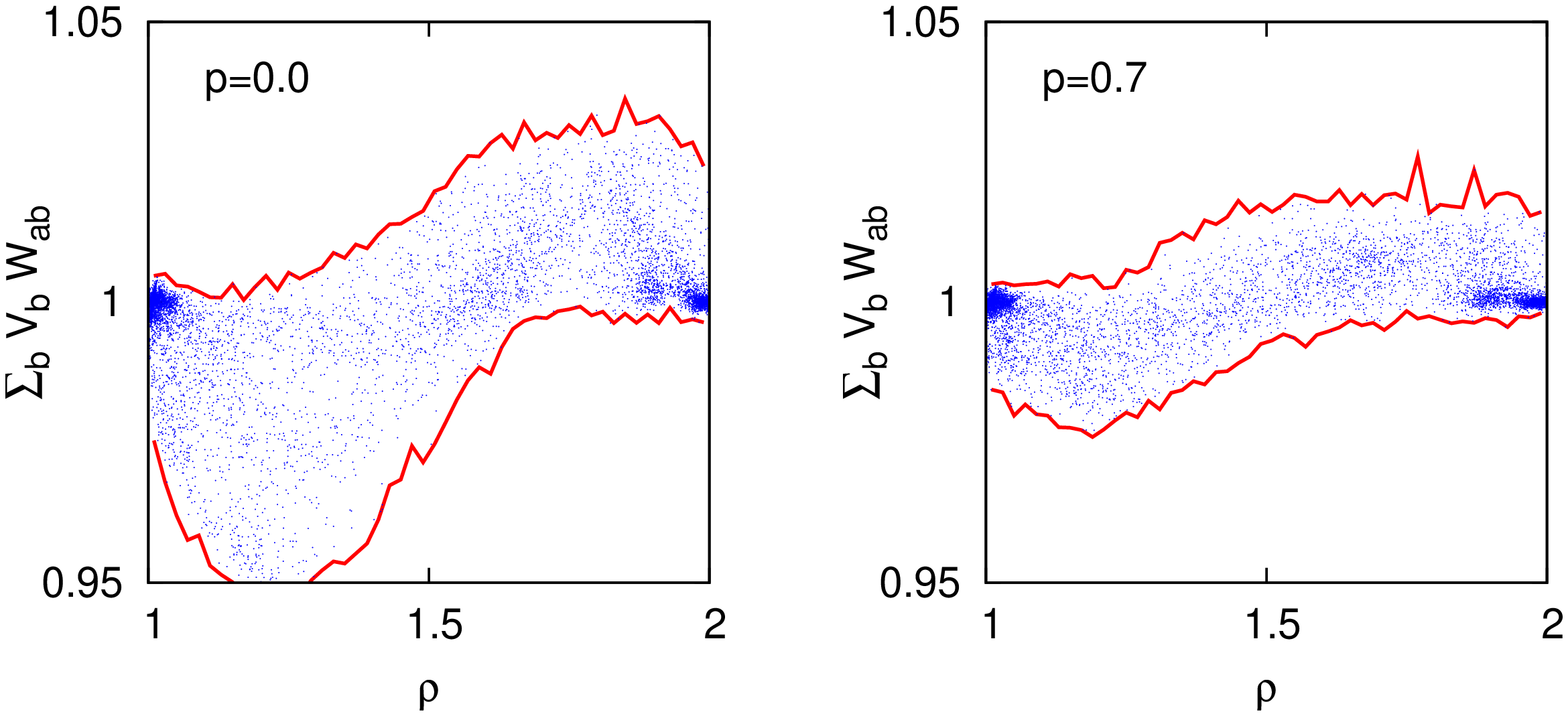}
\caption{Kernel normalization of models  $KH_1$ (left) and $KH_2$ (right) of Table~\ref{table5} at t=2.5. The dispersion of $\sum_b V_b~W_{ab}$ around unity is shown for two options of the exponent of the volume estimator, $X_a=(m_a/\rho_a)^p$, p=0 (left) and p=0.7 (right). Solid red lines are maximum and minimum values to help as a visual aid.} 
\label{kh_3}
\end{figure*}

\clearpage
\begin{figure*}
\includegraphics[angle=0,width=\textwidth]{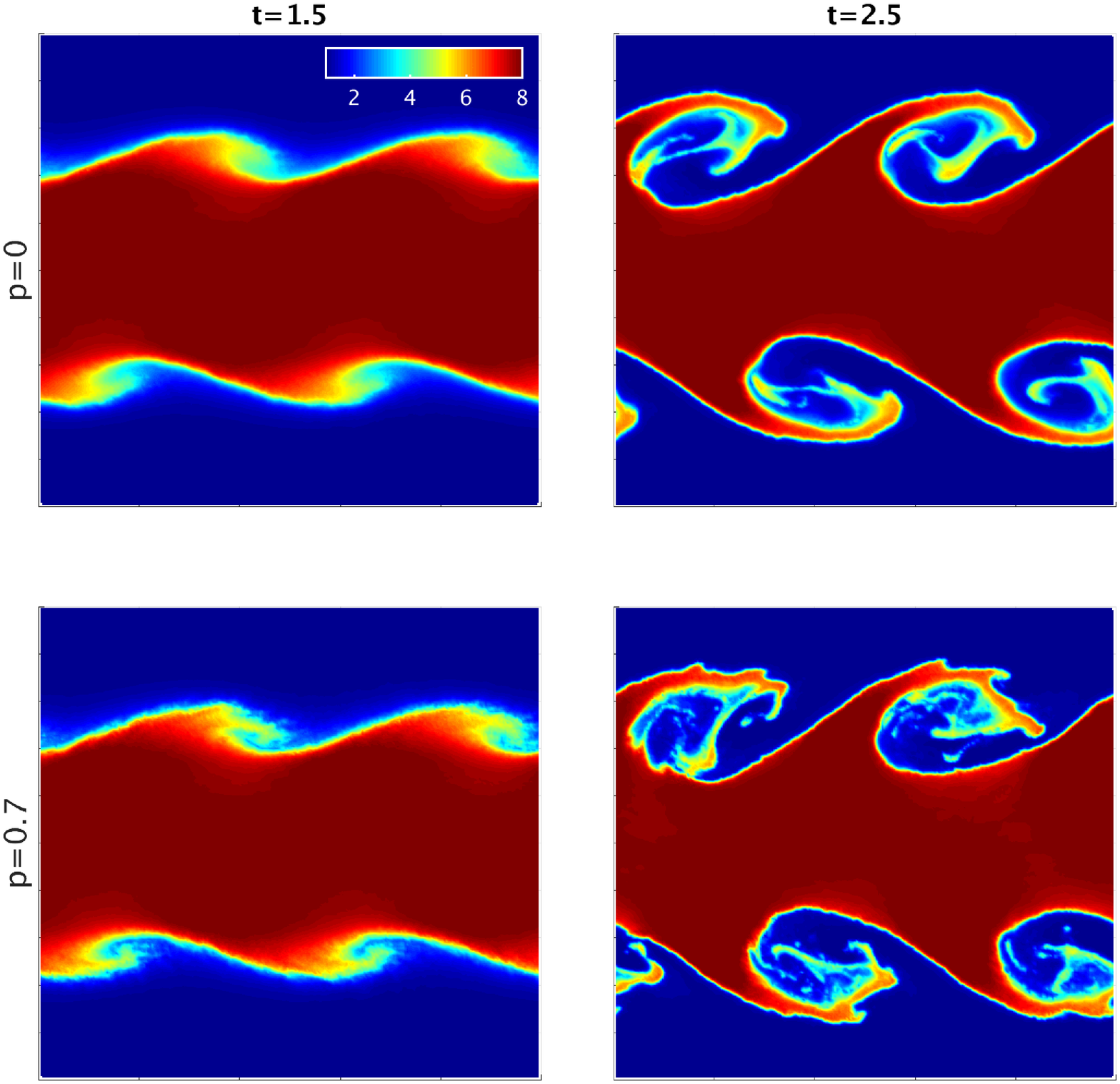}
\caption{Density color-map depicting the evolution of the Kelvin-Helmholtz instability for models $KH_3$ ($p=0$, first row) and $KH_4$ ($p=0.7$, second row) of Table~\ref{table5}, corresponding to a density jump of a factor 8, at times $t=1.5$, 2.5. We notice that, despite the much larger jump in density, the KH instability is still able to develop in both cases.}
\label{kh_4}
\end{figure*}

\clearpage
\begin{figure*}
\includegraphics[width=\textwidth]{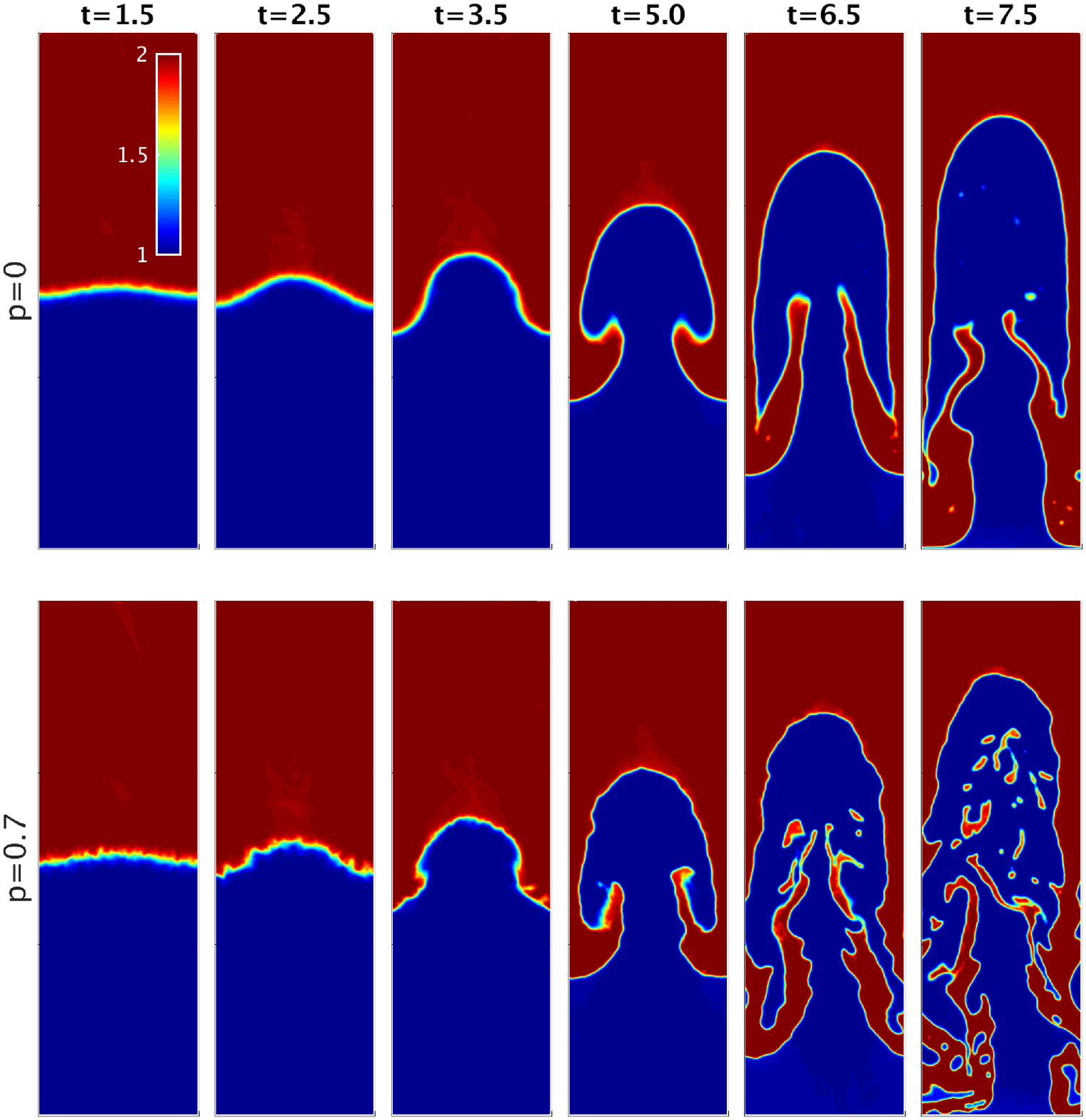}
\caption{Density color-map depicting the evolution of the Rayleigh-Taylor instability for models  $RT_1$ ($p=0$, first row), 
and $RT_2$ ($p=0.7$, second row) of Table~\ref{table6} at times $t=1.5$, 2.5, 3.5, 5.0, 6.5, and 7.5, respectively. The estimator $X_a=(\frac{m_a}{\rho_a})^p$ was used to compute the volume elements.} 
\label{rt_1}
\end{figure*}

\clearpage
\begin{figure*}
\includegraphics[width=\textwidth]{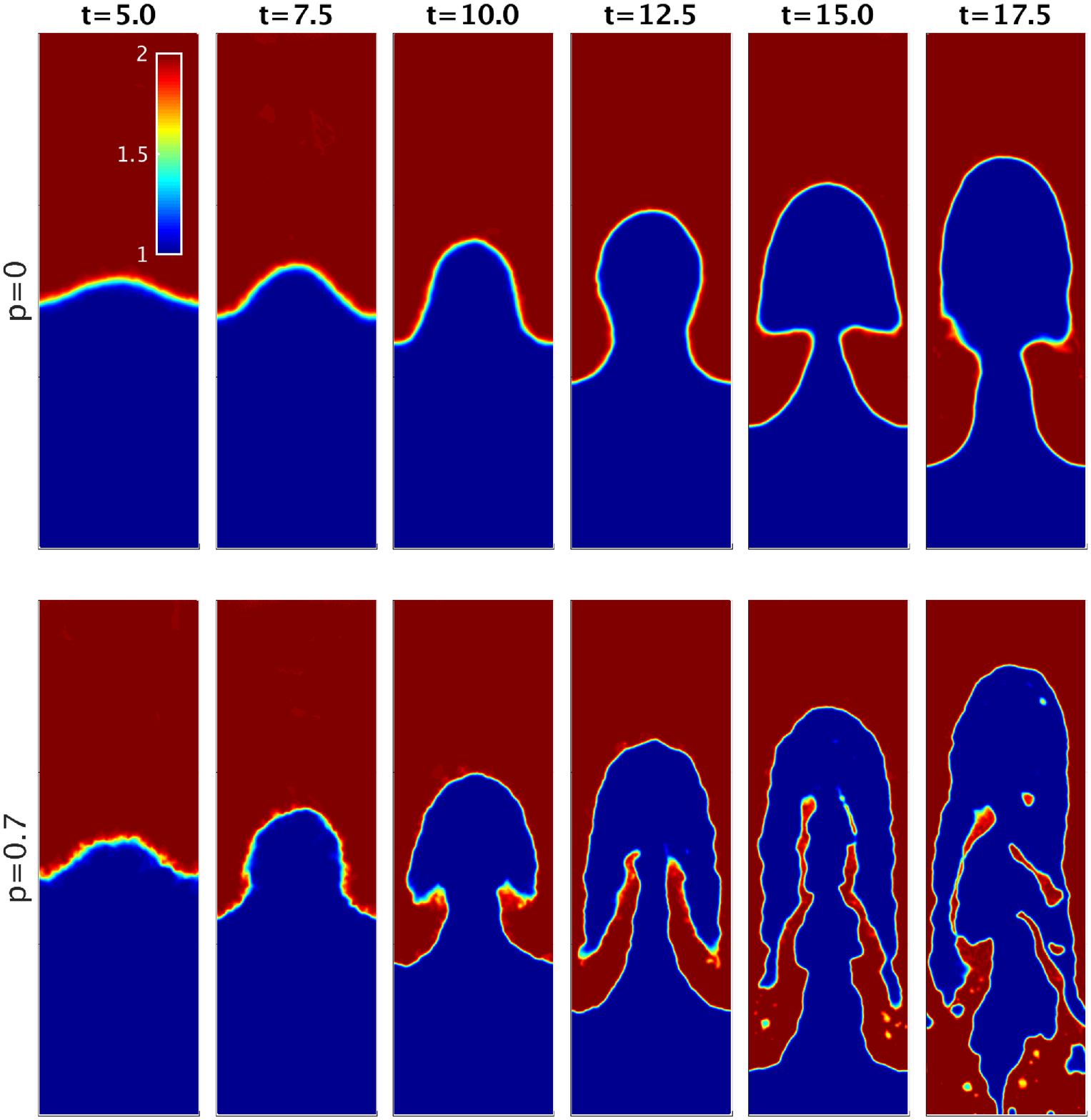}
\caption{Same as Fig.~(\ref{rt_1}) but for models $RT_3$ ($p=0$, first row), and $RT_4$ ($p=0.7$, second row) of Table~\ref{table6} (with a very reduced gravity field of $g=-0.1$). Times are $t=5.0$, 7.5, 10.0, 12.5, 15.0, and 17.5, respectively.}  
\label{rt_2}
\end{figure*}

\clearpage
\begin{figure*}
\includegraphics[angle=-90,width=\textwidth]{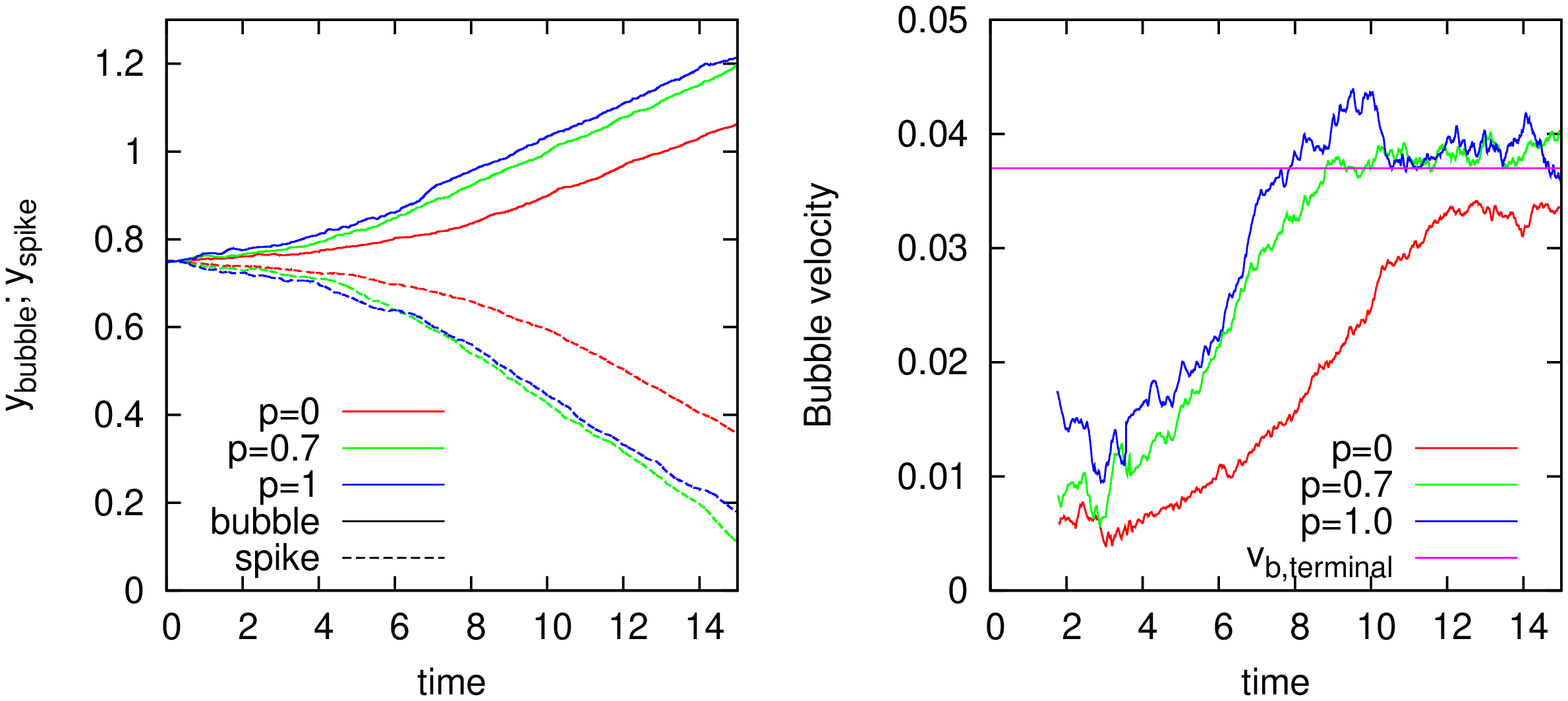}
\caption{Rayleigh-Taylor test in 2D for models $RT_3$ ($p=0$, red lines), $RT_4$ ($p=0.7$, green lines) , and $RT_5$ ($p=1$, blue lines). In the left panel we show the time evolution of the tips of the bubble (solid lines) and the spikes (dashed). From these it is clear that the inclusion of the new VE elements enhances the growth of the instability in the linear stage. Once the non-linear stage is reached all models grow at similar rates. It is worth noting here that simulations without $IAD_0$ and standard VE (not shown) failed to develop the RT instability. In the panel on the right, we show the time evolution of the bubble velocity. It can be seen how a plateau is reached by all three models, which is closer to the expected terminal velocity (pink solid line) for the models using the new volume elements.}  
\label{rt_3}
\end{figure*}

\clearpage
\begin{figure*}
\includegraphics[angle=-90,width=\textwidth]{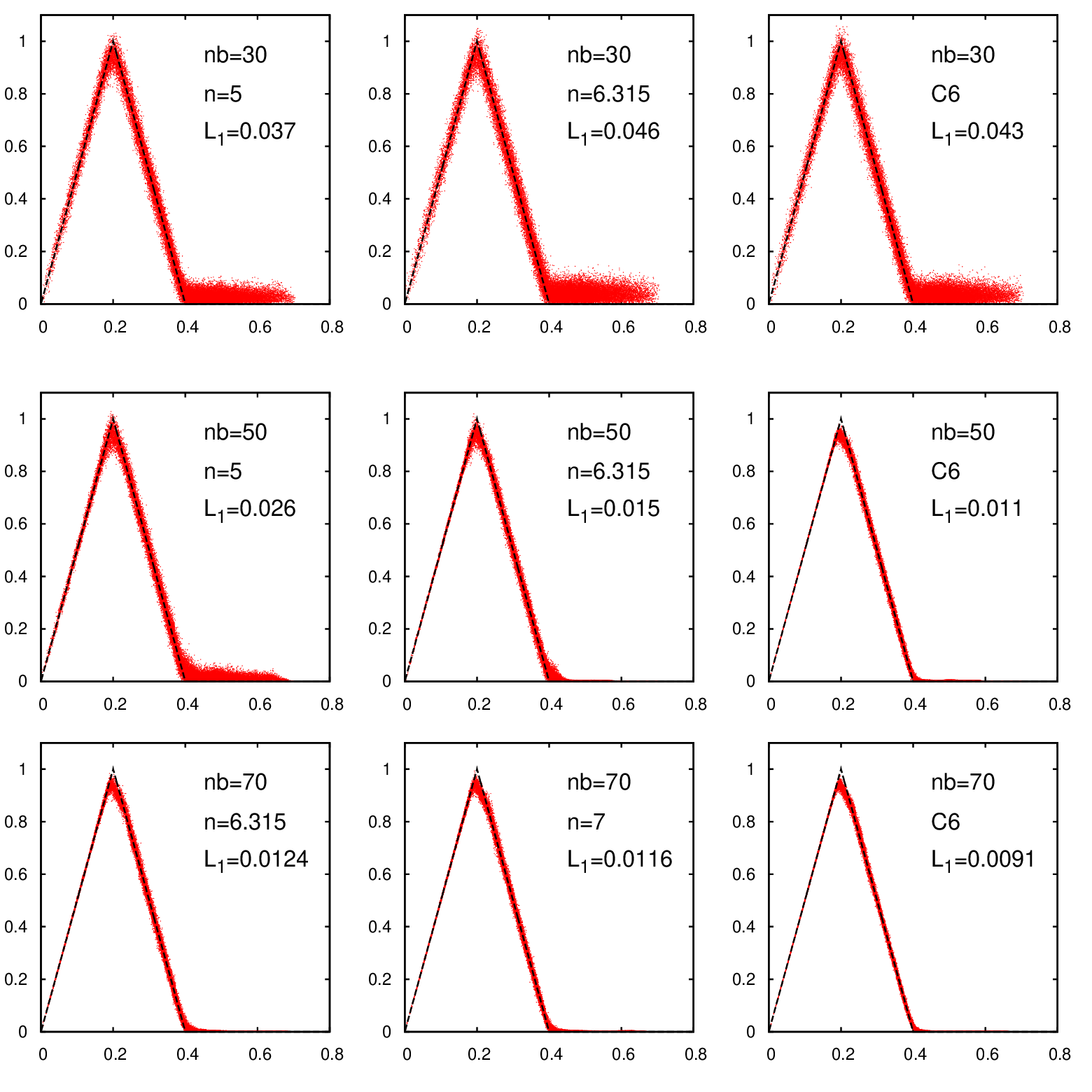}
\caption{Gresho-Chan vortex test in 2D. Radial profiles of the tangential velocity for several kernel choices and different number of neighbors, at time t=1. The interpolators are the $sinc$ $(n=5,~6.315,~7)$ and the Wendland $C_6$. The theoretical velocity profile is also shown (black-dashed line).}
\label{gresho_1}
\end{figure*}

\clearpage
\begin{figure*}
\includegraphics[angle=-90,width=\textwidth]{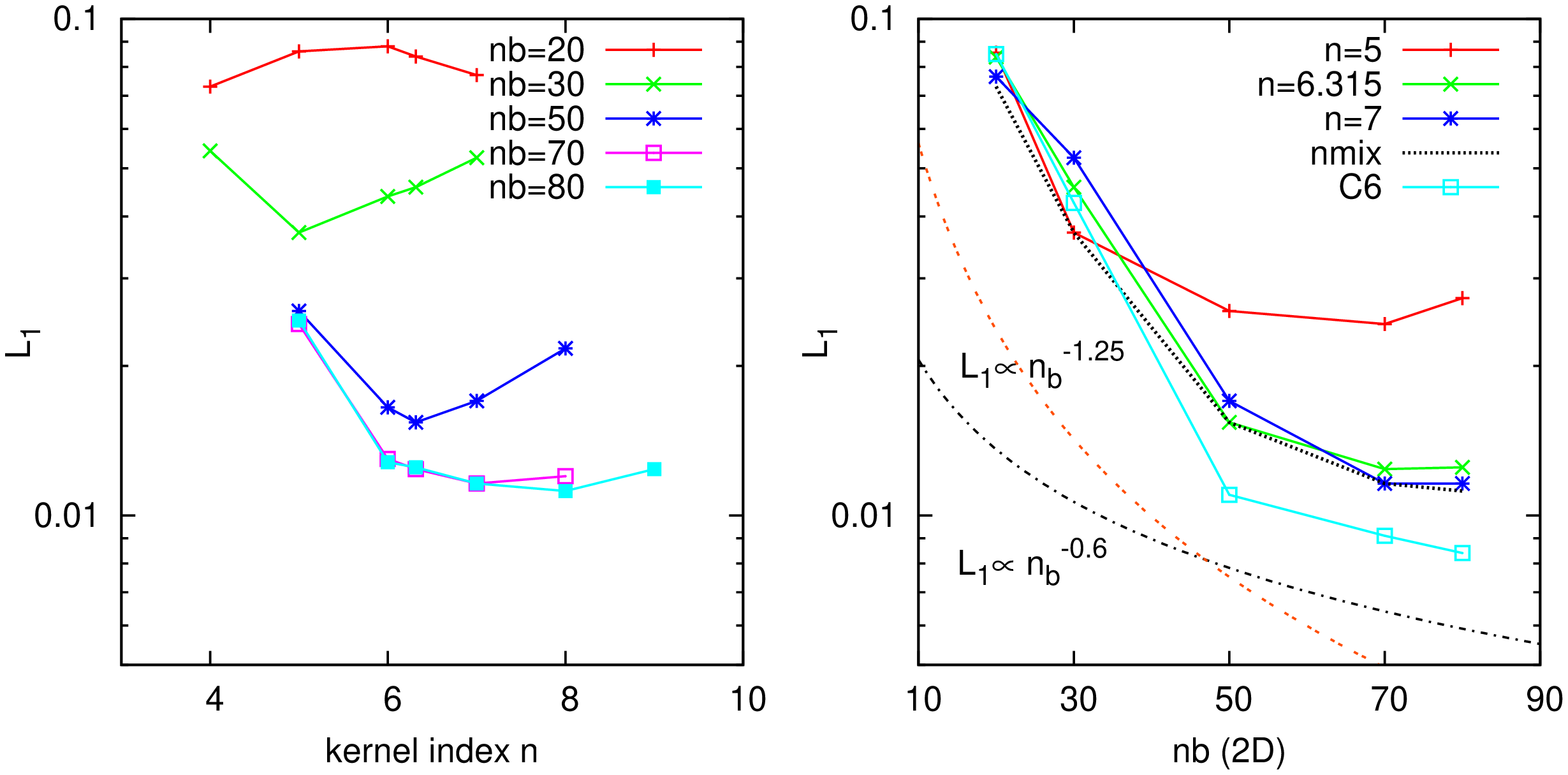}
\caption{Gresho-Chan vortex test in 2D. Left: Value of the magnitude $L_1$, at time $t=1$, calculated with the $sinc$ kernels as a function of the kernel index $n$ and number of neighbors $n_b$ (with constant total number of particles $N=256^2$). Right: Convergence rate of  $sincs$ and Wendland $C_6$ kernels as a function of the number of neighbors.}
\label{gresho_2}
\end{figure*}

\clearpage
\begin{figure*}
\includegraphics[angle=-90,width=\textwidth]{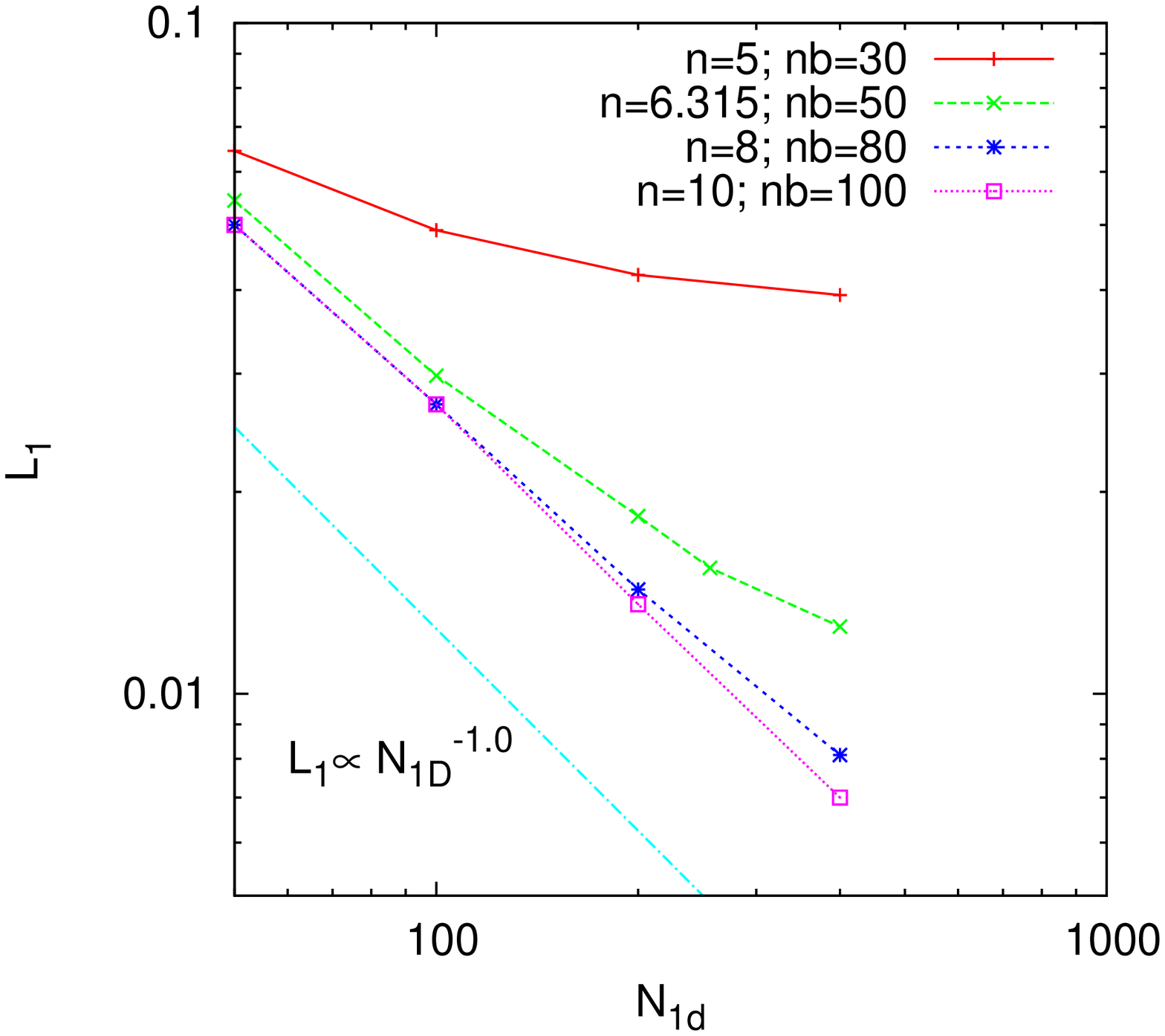}
\caption{Gresho-Chan vortex test in 2D. Convergence rate of $L_1$ as a function of the equivalent 1D number of particles, $N_{1d}$, at time $t=1$, for different $sinc$ kernels.}
\label{gresho_3}
\end{figure*}

\clearpage
\begin{figure*}
\includegraphics[width=\textwidth]{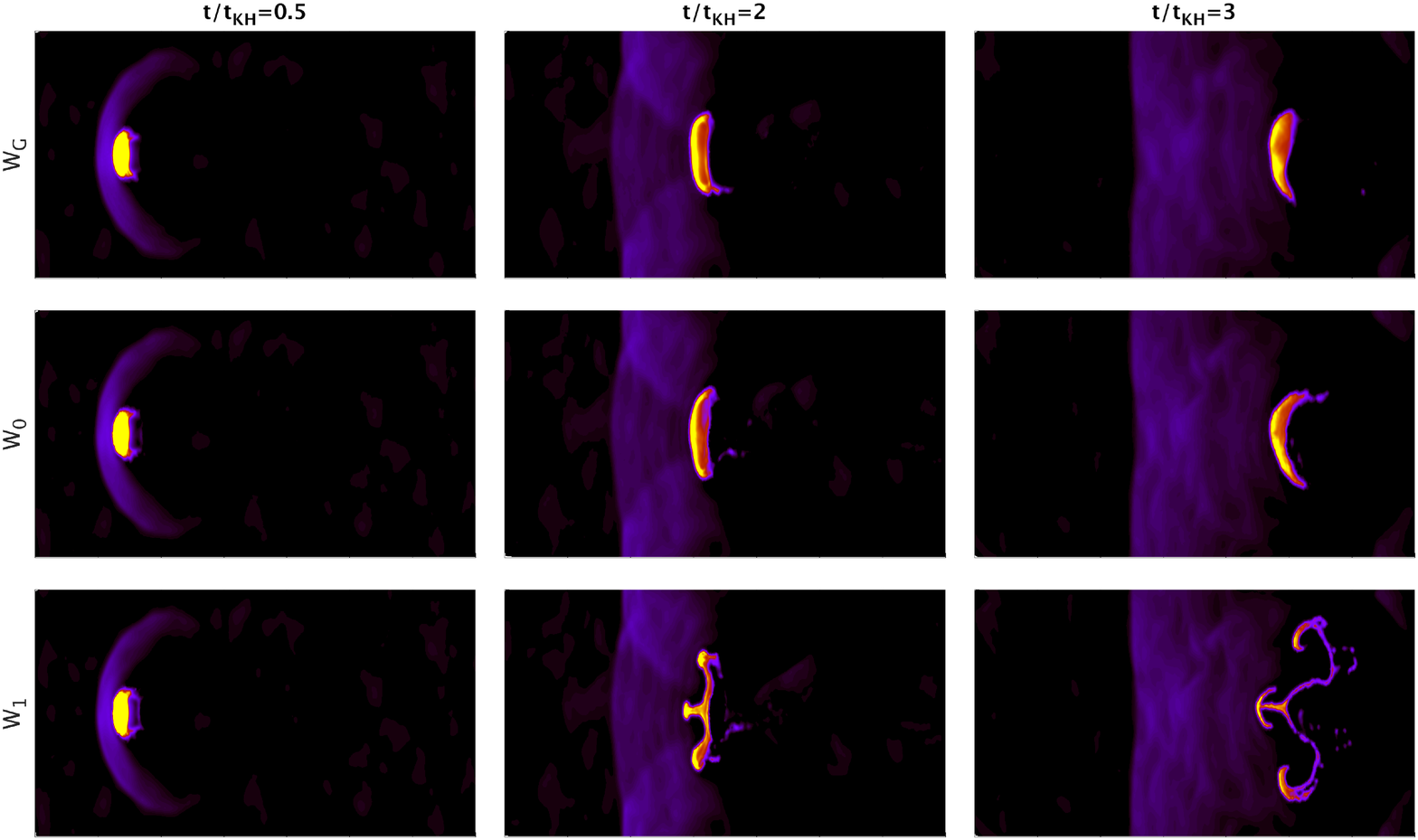}
\caption{Wind-Cloud interaction: Density color-map depicting the evolution of the collision and cloud fragmentation at three fiducial times $t/t_{KH}=0.5$, 2, and 3. Model $W_G$ refers to the calculation using GADGET-2. Models $W_0$ and $W_1$ were calculated with SPHYNX using the standard and the new VE, respectively.}
\label{wc_1}
\end{figure*}

\clearpage
\begin{figure*}
\includegraphics[angle=-90,width=\textwidth]{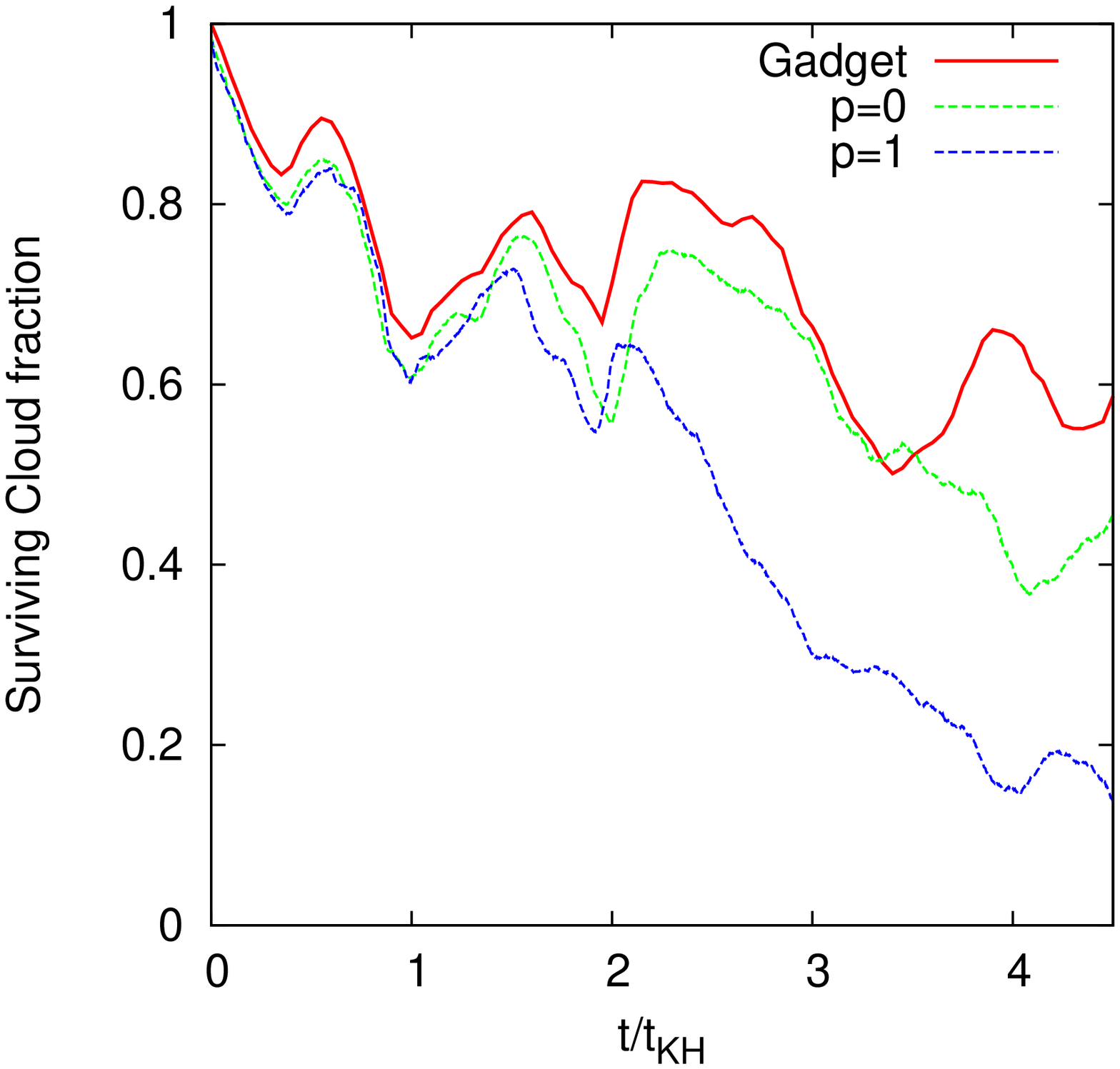}
\caption{Wind-Cloud interaction: Evolution of the surviving fraction of the cloud for the three calculated models. Even though a small remnant of the cloud is still present at $t/t_{KH}\simeq 4$ the calculation $W_1$ (SPHYNX with the new VE, that is, $p=1$) leads to the best results. The other two calculations (SPHYNX with standard VE, i.e. $p=0$, and GADGET with standard VE and without $IAD_0$) are more inefficient in diluting the cloud as between 40-60\% of it is still present even at $t/t_{KH}>4$.}
\label{wc_2}
\end{figure*}

\clearpage
\begin{figure*}
\includegraphics[angle=-90,width=\textwidth]{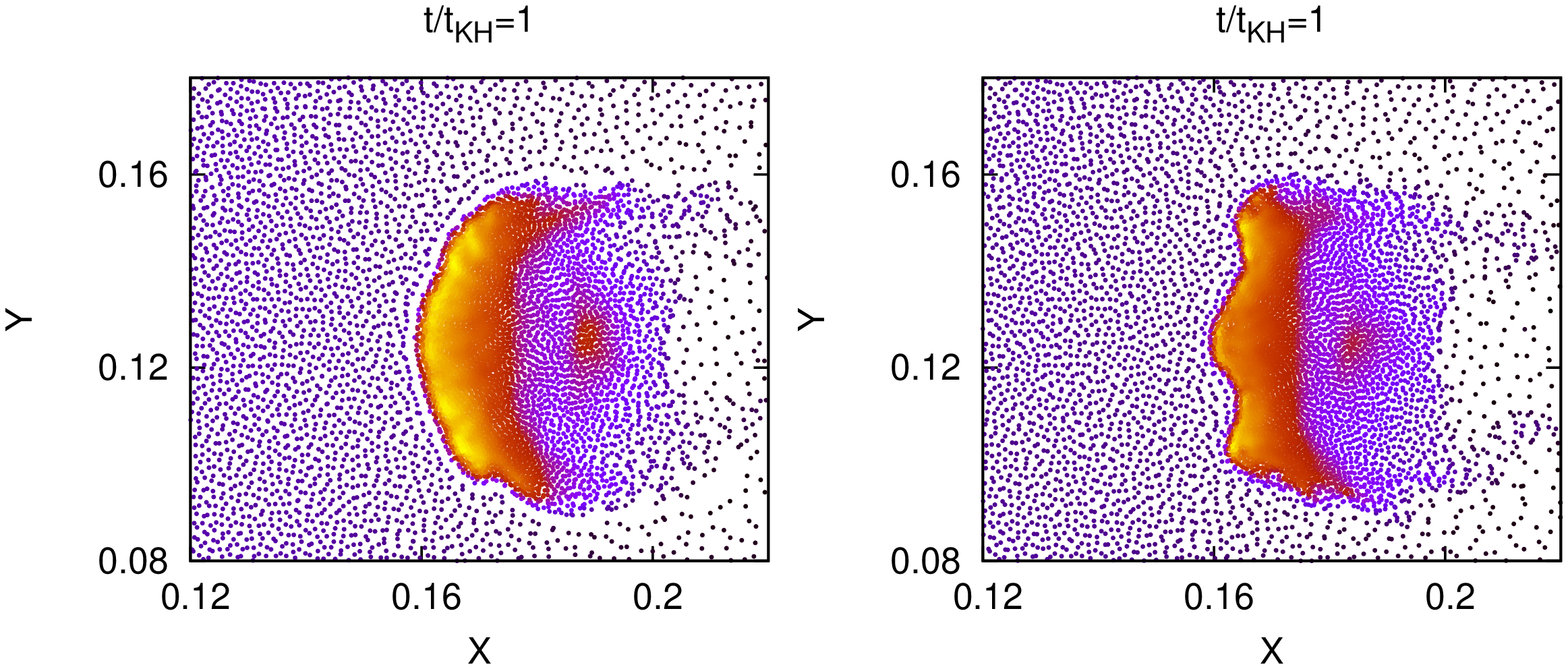}
\caption{Wind-Cloud interaction: Impact of the VE choice in the tensile instability. Near the contact discontinuity the standard VE, $m/\rho$, produces the segregation of the two fluids, preventing mixing (left). Choosing $VE_a= \left(\langle m/\rho\rangle_a\right)^p/\sum_{ab} \left(\langle m/\rho\rangle_b\right)^p W_{ab}$, where $\langle m/\rho\rangle$ is the SPH average of $m/\rho$ and $p=1$, suppresses the tensile instability leading to mixing of the fluids and the subsequent bubble fragmentation (right).}
\label{wc_3}
\end{figure*}
\clearpage

\begin{figure*}
\includegraphics[width=\textwidth]{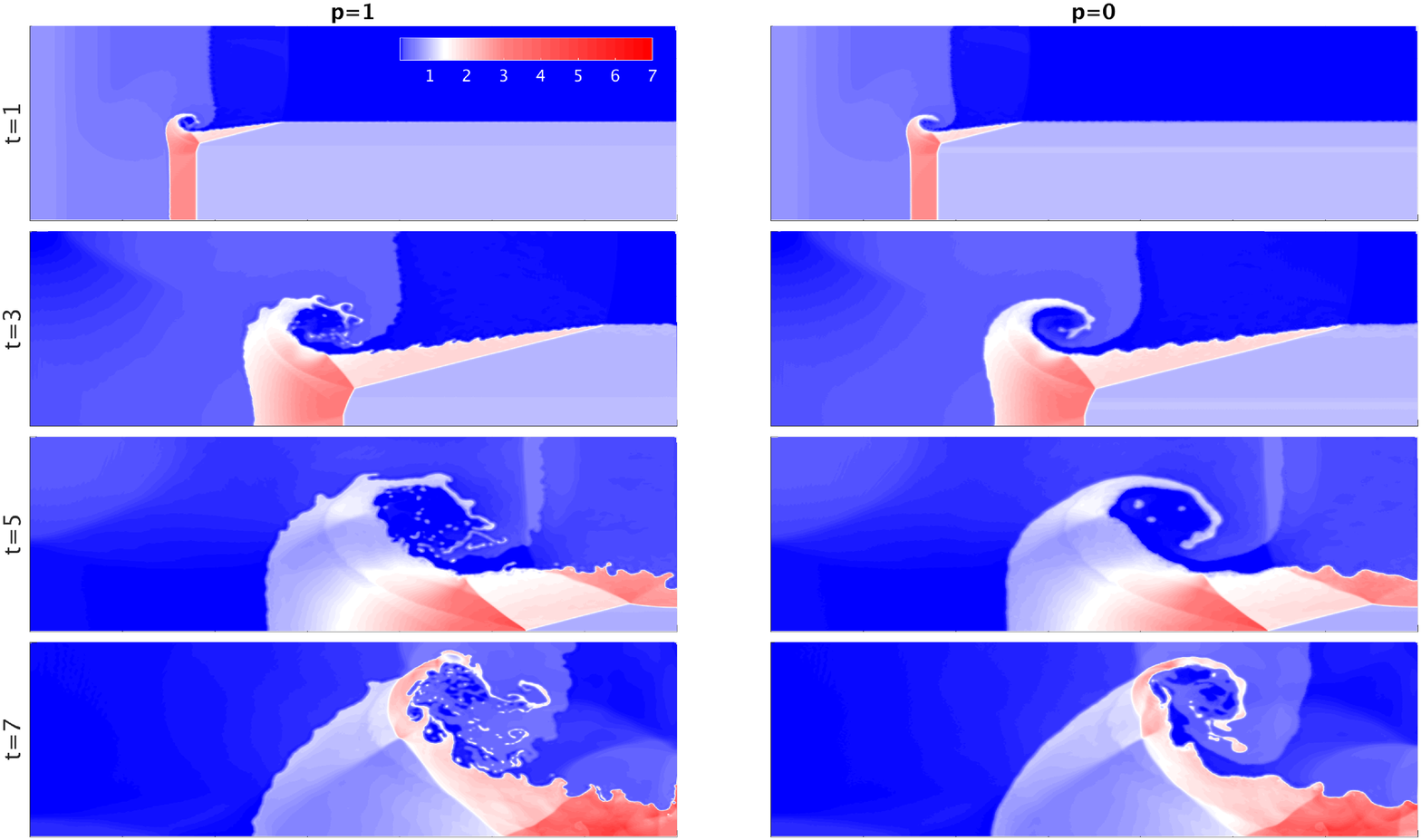}
\caption{Color-map of the mass density in the triple-point shock test. Each row shows a snapshot at a different time, while each column corresponds to a different VE. The column on the left was calculated with $X_a= \left(\langle m/\rho\rangle_a\right)^p$ and $p=1$, while the one on the right corresponds to $X_a=1$ (i.e., standard VE).}
\label{triplepoint}
\end{figure*}

\clearpage
\begin{figure*}
\includegraphics[angle=-90,width=\textwidth]{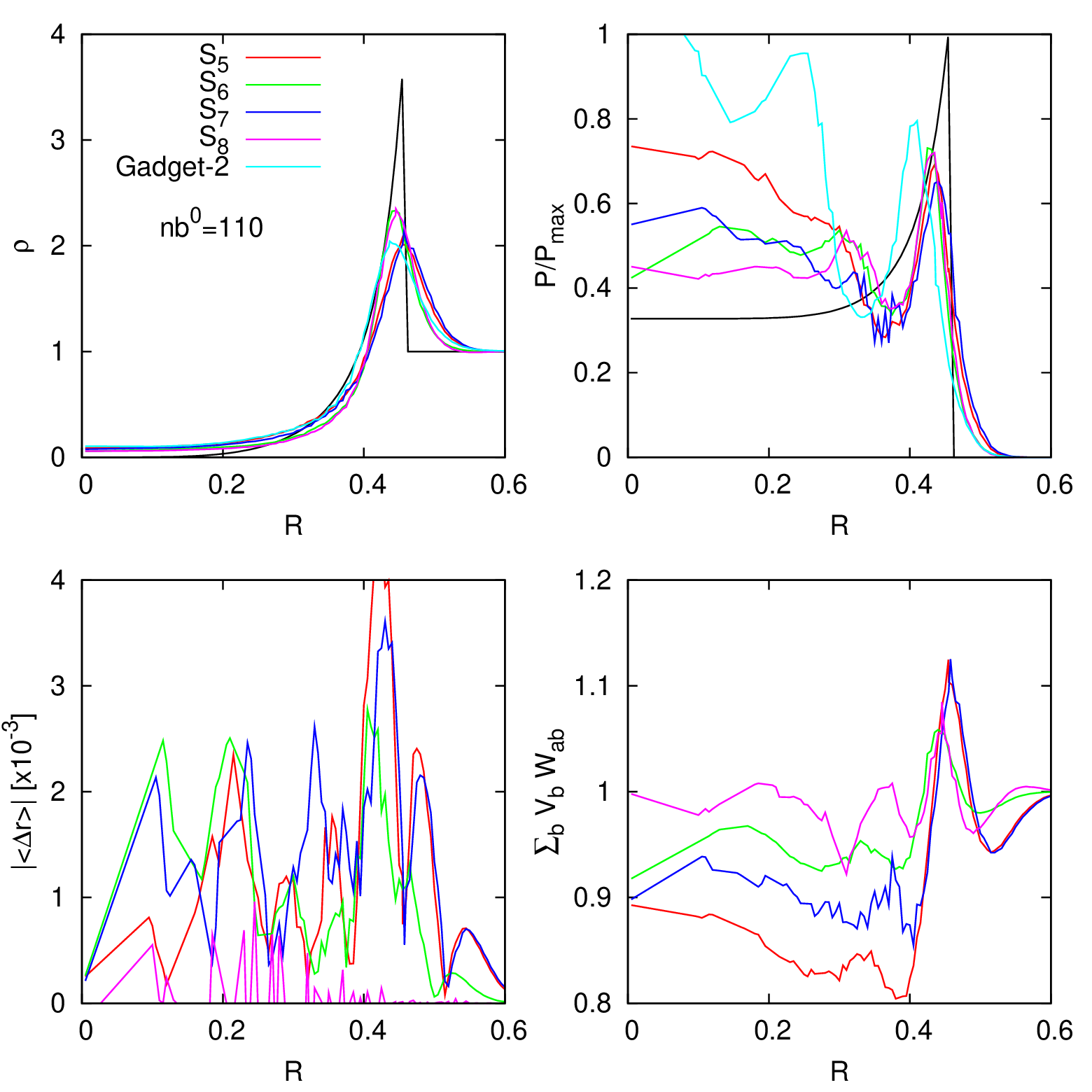}
\caption{Results for the 3D-Sedov test for models $S_5$, $S_6$, $S_7$, $S_8$ of Table~\ref{table3} and Gadget-2. We show here the radial profiles of density (upper-left), normalized pressure (upper-right), and the fulfillment of the conditions $|\langle\Delta\mathbf{r}\rangle|=0$ (bottom-left) and $\sum_b V_b W_{ab}=1$ (bottom-right). Solid black lines are the analytic solutions for density and normalized pressure profiles.}
\label{sedov_3}
\end{figure*}

\clearpage
\begin{figure*}
\includegraphics[angle=-90,width=\textwidth]{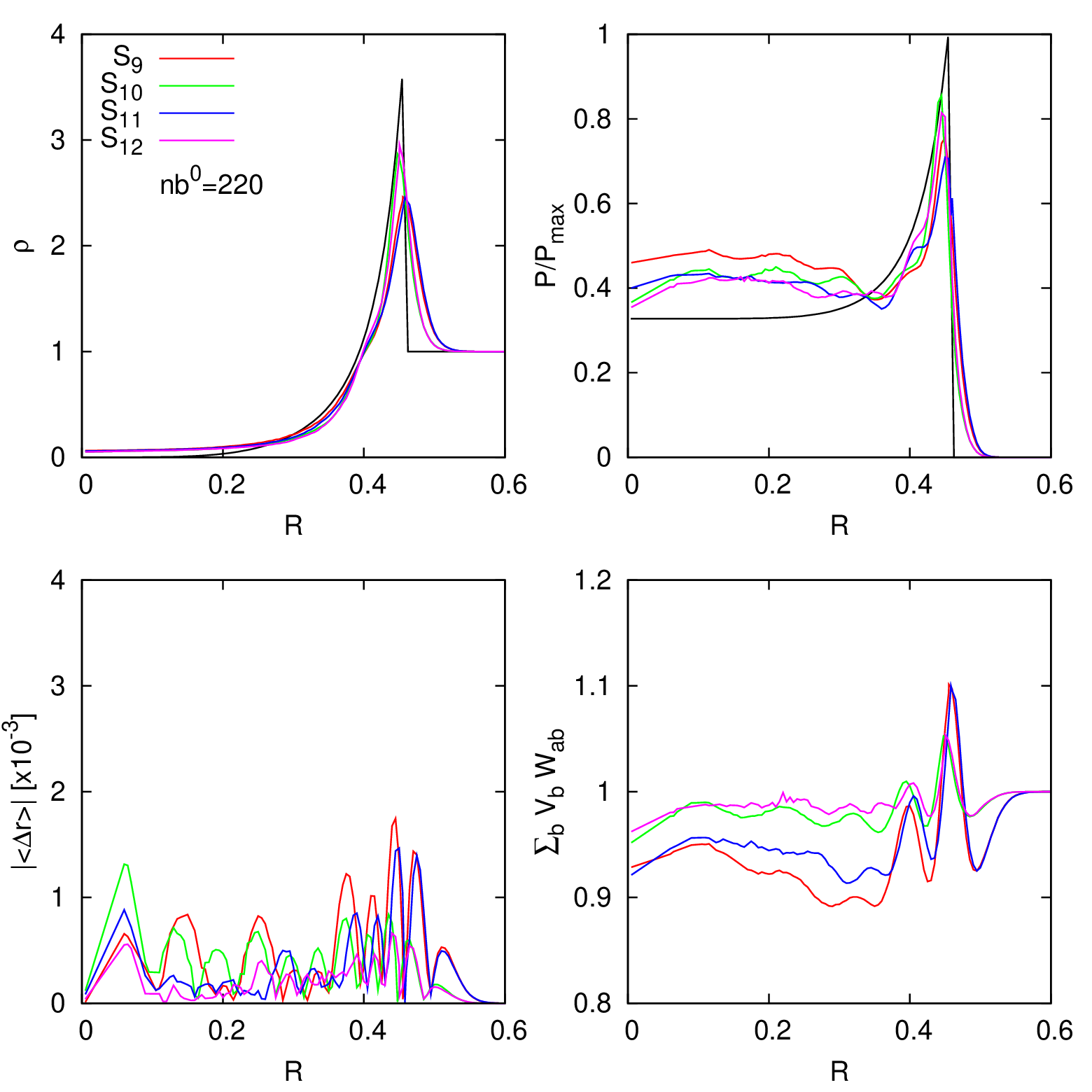}
\caption{Same as Fig.~(\ref{sedov_3}) but for models $S_9$, $S_{10}$, $S_{11}$, $S_{12}$, with $n_b=220$, of Table~\ref{table3}.}
\label{sedov_4}
\end{figure*}

\clearpage
\begin{figure*}
\includegraphics[angle=-90,width=\textwidth]{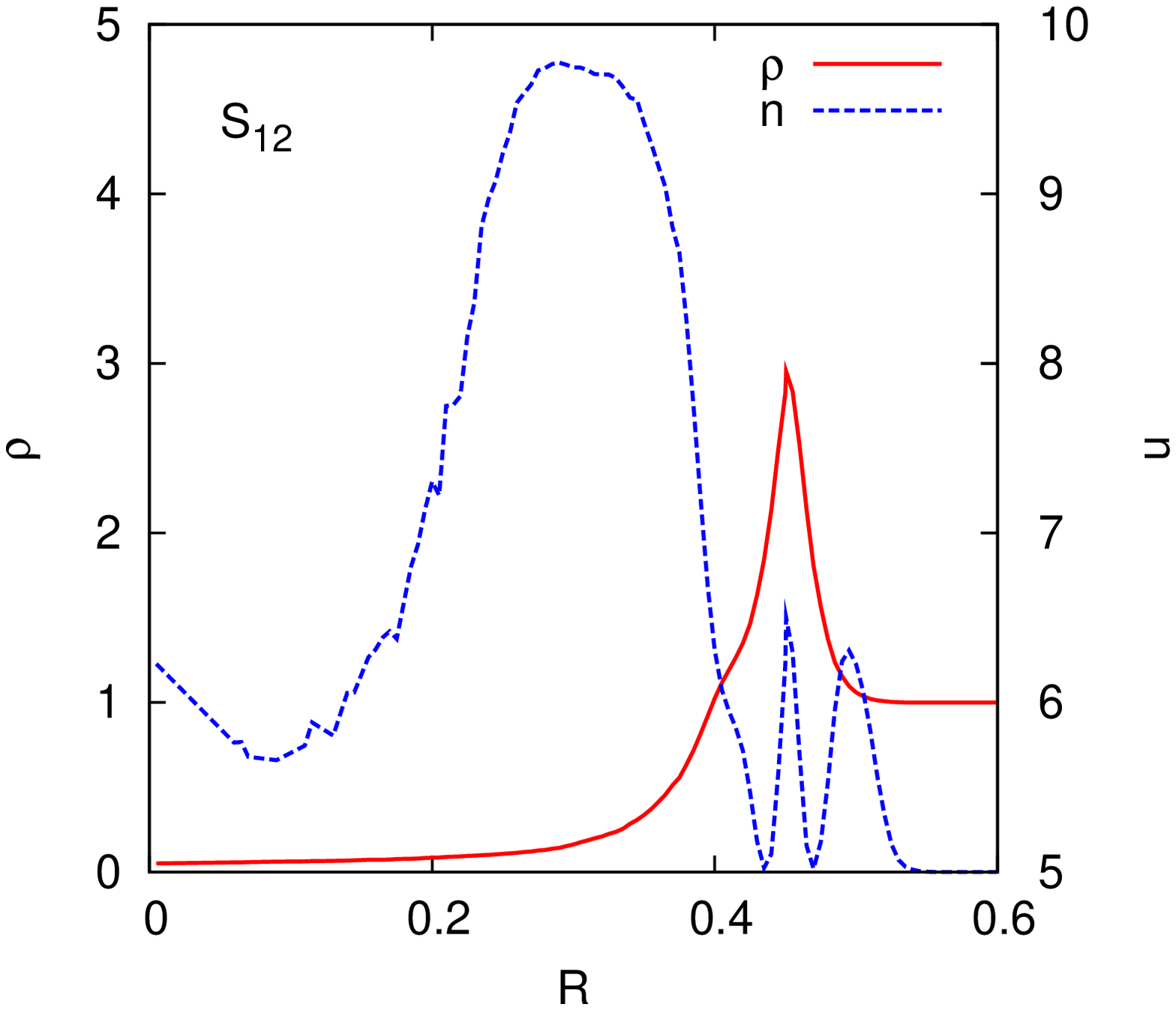}
\caption{Radial profiles of the $sinc$ kernel index, $n$, and density for model $S_{12}$ of Table~\ref{table3}. We note how the scheme increases the kernel index to improve the interpolations in the complicated regions of low- or of fast-changing density.}
\label{sedov_5}
\end{figure*}

\clearpage
\begin{figure*}
\includegraphics[angle=-90,width=\textwidth]{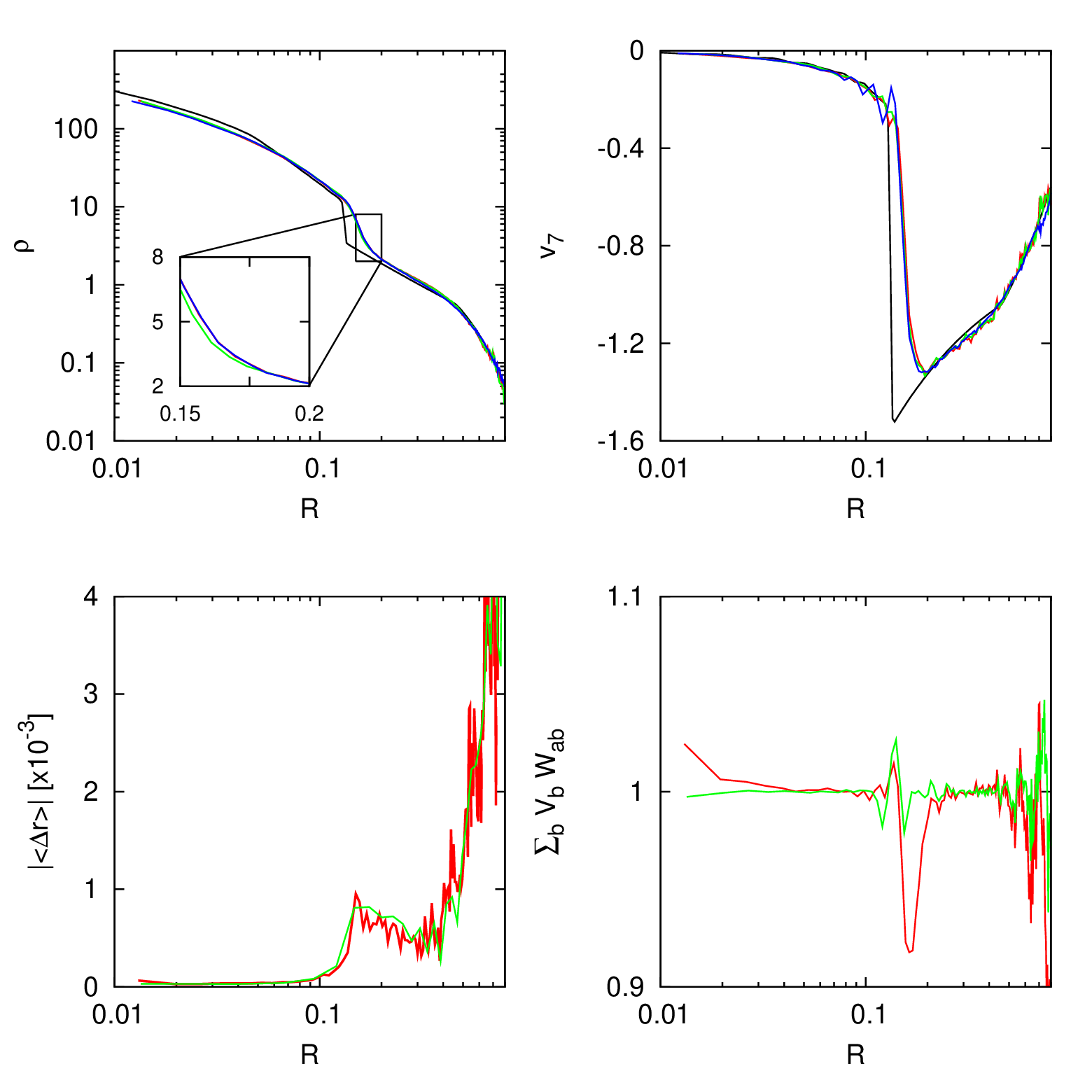}
\caption{Upper panels: density (left) and radial velocity profiles (right) for the collapse of an initially isothermal sphere of gas at $t=0.8$. The solid black line is the 1D-PPM solution. We show the SPHYNX calculations with $p=0$ (red) and $p=1$ (green) in Eq.~\ref{xrhobis}. Calculations using GADGET-2 are in blue. Lower panels: profiles of $\vert\langle\Delta\mathbf{r}\rangle\vert$ (left) and $\sum_b V_b~W_{ab}$ (right) for $p=0$ (red) and $p=1$ (green).}
\label{evrardfig1}
\end{figure*}
\end{document}